\documentclass[secnumarabic,amssymb,amsmath,natbib,graphicx,aps]{revtex4-2}
\usepackage{subcaption}
\usepackage{tikz}       
\usepackage{hyperref,comment}
\usepackage{booktabs}
\usepackage{float}
\usepackage{dcolumn}
\usepackage{xcolor}
\usepackage[normalem]{ulem}

\begin{document}

\title{Radiation properties and images of loop quantum Reissner-Nordström black hole with a thin accretion disk}
\author{Qian Li}
\author{Jia-Hui Huang}
\email{huangjh@m.scnu.edu.cn}
\affiliation{Key Laboratory of Atomic and Subatomic Structure and Quantum Control (Ministry of Education), Guangdong Basic Research Center of Excellence for Structure and Fundamental Interactions of Matter, School of Physics, South China Normal University, Guangzhou 510006, China}
\affiliation{Guangdong Provincial Key Laboratory of Quantum Engineering and Quantum Materials, Guangdong-Hong Kong Joint Laboratory of Quantum Matter, South China Normal University, Guangzhou 510006, China}


\begin{abstract}
	We investigate the characteristics of circular motion of charged particles around loop quantum Reissner-Nordström black hole (LQRNBH) and the radiation properties and observational appearance of a thin accretion disk around it. By calculating the shadow radius and utilizing observational data from M87* and Sgr A*, we derive constraints on the quantum parameter $\zeta$ and charge parameter $Q$. In order to support a accretion disk around the black hole, the charge parameter $Q$ is restricted to be tiny ($Q\lesssim10^{-21}$). 
The circular motion of charged particles around LQRNBH and the influence of the model parameters on the circular motion are studied. It is found that although the direct charge correction to the geometry is negligible, the electromagnetic coupling $\kappa$ substantially impacts the circular motion and the radiative properties.  
Then, by considering a thin accretion disk model, the local radiation fluxes, redshift factors and the observed fluxes are studied. With the ray-tracing method, the isoradial curves, direct images of redshift factors and the observed fluxes are illustrated. The influence of the model parameters on these quantities are systematically performed. 
It is found that the influence of the quantum parameter $\zeta$ is generally very small but whether its a positive or negative effect depends on the sign of $\kappa$, i.e. attractive or repulsive electromagnetic force. Although the charge of the black hole is tiny in our study, the electromagnetic interaction generally has significant influence on the physical quantities.  
\end{abstract}
\maketitle	

	\section{Introduction}
	
Black holes (BHs), predicted by Einstein's general relativity, are believed to be important astrophysical objects, which has successfully withstood the observation of gravitational waves since 2015 \cite{LIGOScientific:2016aoc} and the observation of shadow images in recent years \cite{EventHorizonTelescope:2019dse,EventHorizonTelescope:2022wkp}. However, it is known that black holes suffer from singularity problem which can not be avoided under certain common physical conditions in general relativity \cite{Penrose:1964wq,Hawking:1970zqf}, due to the geodesic incompleteness of spacetimes. 
Spacetime singularities may lead to the conflict between general relativistic dynamics and the unitarity of quantum evolution, i.e., BH information paradox \cite{Marolf:2017jkr,Buoninfante:2024oxl}. Thus, many efforts are made to resolve singularities of BHs, and typical approaches include string theory, loop quantum gravity, etc. (see \cite{Buoninfante:2024oxl} and references therein).

Symmetry-reduced cosmological models and BH models have been successfully constructed in loop quantum gravity framework \cite{Han:2005km,Giesel:2011idc,Thiemann:2007pyv,Ashtekar:2004eh,Addazi2022,Gambini:2020nsf,Ashtekar:2005qt,Li:2018opr,Assanioussi:2018hee,Yang:2009fp,
Ding:2008tq,Ashtekar:2006wn,Ashtekar:2006rx,Ashtekar:2003hd}, which resolve the classical singularities. A challenging issue in the effective theory of the quantum symmetry-reduced BH models is about the diffeomorphism covariance. 
Recently, this issue has been addressed through several approaches in spherically symmetric models, such as the quantum Oppenheimer-Snyder model \cite{Lewandowski:2022zce,Kelly:2020lec,Shi:2024vki} and loop quantum BHs (LQBHs) proposed in \cite{Zhang:2024khj,Zhang:2024ney} by constructing effective Hamiltonian constraints.

The LQBHs have been studied extensively from various aspects. The quasinormal modes of these spherical LQBHs have been explored by several recent works \cite{Yang:2022btw,Cao:2024oud,Zhang:2024svj,Skvortsova:2024atk,Zinhailo:2024kbq,Gong:2023ghh,Malik:2024nhy,Konoplya:2024lch}. The correspondence between grey-body factors and quasinormal modes of these LQBHs were also studied \cite{Skvortsova:2024msa}. The features of the shadows of these quantum black holes were studied in \cite{Zhang:2023okw,Ye:2023qks,Liu:2024soc,Peng:2020wun,He:2025hbu}, and the constraints on the parameter encoding the quantum gravity effect from Event Horizon Telescope (EHT) observations \cite{Konoplya:2024lch,Zhao:2024elr,Shu:2024tut,Afrin:2022ztr} and gravitational wave observation \cite{Zi:2024jla} were also discussed. Furthermore, the optical appearance of the accretion disk around LQBHs and the strong gravitational lensing effects of the LQBHs were explored in \cite{You:2024jeu, Liu:2024wal,Li:2024afr,Shu:2024tut}.
The investigation of primordial LQBH accounts for the dark matter was studied in \cite{Calza:2025mwn,Papanikolaou:2023crz}.

Recently, the framework developed in \cite{Zhang:2024khj,Zhang:2024ney} for constructing LQBHs has been extended to the electrovacuum setting with a cosmological constant \cite{Yang:2025ufs}. By solving the effective dynamics determined by the effective Hamiltonians, the authors derived three LQBH models. When the cosmological constant is taken zero, solution I is just a Reissner-Nordström (RN) black hole with loop quantum gravity modification which will be abbreviated to LQRNBH.  
In this paper, we study the radiation properties and optical appearance of the LQRNBH surrounded by a geometrically thin accretion disk, and explore the influence of the electric charges and the loop quantum parameter on various quantities and images of LQRNBH.

The remainder of this paper is organized as follows. In Section 2, we consider the shadow of the LQRNBH and constrain both the charge parameter and quantum parameter with the observational data of M87* and Sgr A*. In Section 3, taking into account of the electromagnetic interaction, we consider the circular motion of test charged particles around the LQRNBH and discuss the influence of model parameters on the motion. In Section 4, we consider various radiation properties of the LQRNBH surrounded by a geometrically thin and optically thick accretion disk. In Section 5, we numerically study the isoradial curves, redshift factors and observed fluxes of the direct images of the LQRNBH with the accretion disk. The final section is devoted to the conclusion.

\section{The LQRNBH metric and Constraints on model parameters}
The spherical and asymptotically flat LQRNBH is described by the following line element \cite{Yang:2025ufs},
\begin{align}
	ds^2 &= -f(r)dt^2 + \frac{dr^2}{f(r)} + r^2 d\Omega^2,\\
	f(r) &= \left(1 - \frac{2M}{r} + \frac{Q^2}{r^2}\right) \left(1 + \frac{\zeta^2}{r^2} \left(1 - \frac{2M}{r} + \frac{Q^2}{r^2}\right)\right),
\end{align}
where $d\Omega^2=d\theta^2+\sin^2\theta\,d\phi^2$. $M$ and $Q$ are respectively the mass and charge parameters of the black hole. $\zeta$ is the loop quantum correction parameter, which is expected to be proportional to the Planck length. However, here we relax this point and only treat $\zeta$ as a positive phenomenological effective parameter since this treatment is self-consistent in loop quantum gravity. The limits $Q\rightarrow0$ and $\zeta\rightarrow0$ give the loop quantum Schwarzschild solution \cite{Zhang:2024ney,Zhang:2024khj} and the classical RN metric, respectively. 

Then, we consider the event horizon of the LQRNBH. From the LQRNBH metric, we have  
\begin{equation}
	f(r)=A(r)F(r),\qquad
	A(r)=1-\frac{2M}{r}+\frac{Q^2}{r^2},\qquad
	F(r)=1+\frac{\zeta^2}{r^2}A(r).
\end{equation}
The roots of $A(r)=0$ give the usual RN-type horizons, $ r_{\pm}=M\pm\sqrt{M^2-Q^2}$.
The event horizon of LQRNBH is located at  $r_+$ which is not affected by the loop quantum parameter $\zeta$ and is the same as that of the RNBH. 

For the calculation of the shadow radius of a static and spherical black hole, many methods have been proposed and for a recent review we refer \cite{Perlick:2021aok}.
The photon-sphere radius $r_{ph}$ is the largest positive root of the following equation
	\begin{equation}
	rf'(r) - 2f(r) = 0,
	\end{equation}
and the shadow radius for an observer at infinity is
	\begin{equation}\label{Rsh}    
	R_{sh} = \frac{r_{ph}}{\sqrt{f(r_{ph})}}.
	\end{equation}
For the LQRNBH, the equation satisfied by $r_{ph}$ is 
	\begin{equation}
	    (r^2-3M r+2 Q^2) (r^4 + 2 Q^2 \zeta^2 - 4 M r \zeta^2 + 2 r^2 \zeta^2) = 0.
	\end{equation}
The first factor gives $r_{ph}$ of the LQRNBH since the roots of the second factor are always smaller than it. Hence, the photon-sphere radius of LQRNBH is the same as that of the RNBH, which is
\begin{equation}
r_{ph}=\frac{3M}{2}+\frac{1}{2}\sqrt{9M^2-8Q^2}.
\end{equation}
Plugging this into \eqref{Rsh}, we can obtain the shadow radius $R_{sh}$ of the LQRNBH,
	\begin{equation}
	    R_{sh} =\frac{r_{ph}^4}{\sqrt{(r_{ph}^2-2M r_{ph}+Q^2)(r_{ph}^4+\zeta^2(r_{ph}^2-2M r_{ph}+Q^2))}}.
	\end{equation}
Although the event horizon and photon sphere are independent of $\zeta$, the shadow radius depends on the parameter $\zeta$. When $Q \rightarrow 0$, the shadow radius reduces to $R_{sh} = \frac{27M^2}{\sqrt{27M^2 + \zeta^2}}$, which is the same as the previous result in \cite{Shu:2024tut}.

\begin{figure}[H]
	    \centering
	    \includegraphics[width=0.35\textwidth]{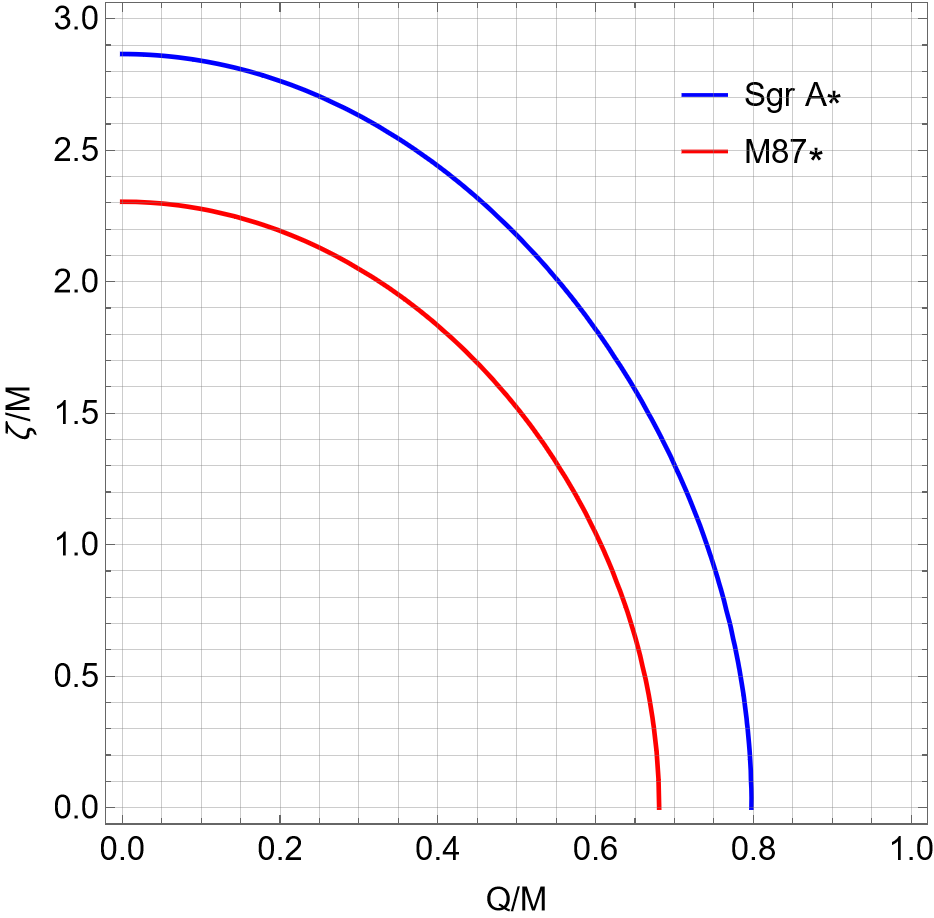}
	    \caption{Constraints on $\zeta$ and $Q$ from the observations of M87* and Sgr A*.}
	    \label{fig:zetaQCons}
	\end{figure}

The Event Horizon Telescope (EHT) collaboration announced the images of two supermassive black holes: M87* at the center of the Messier 87 galaxy and Sgr A* at the center of the Milky Way. For M87*, the angular diameter of its shadow is $\theta_{\text{sh}}^{\text{M87*}} =(42\pm 3)\mu as$, its distance to Earth is $ D_{\text{M87*}} = (16.8_{-0.7}^{+0.8})~\text{Mpc} $ and its estimated mass is $ M_{\text{M87*}} = (6.5 \pm 0.7) \times 10^9~M_\odot $ \cite{EventHorizonTelescope:2019dse, EventHorizonTelescope:2019ths, EventHorizonTelescope:2019pgp, EventHorizonTelescope:2019ggy}. With these data, the dimensionless shadow diameter of the black hole can be computed as $ d_{\text{sh}} = \frac{D\theta_{\text{sh}}}{M}$, which is
$
d_{\text{sh}}^{\text{M87*}} = \frac{D_{\text{M87*}}~\theta_{\text{sh}}^{\text{M87*}}}{M_{\text{M87*}}} \simeq 11.0 \pm 1.5.
$
The corresponding shadow radius of the black hole, $R_{\text{sh}}^{\text{M87*}}$, is
$
	   4.75 \lesssim R_{\text{sh}}^{\text{M87*}} \lesssim 6.25 \, (1\sigma)
	    \label{M87},
$
which leads to following constraints on the model parameters of LQRNBH,
	\begin{align}
	    0 \le \zeta/M \lesssim 2.304 \, (1\sigma), ~~
	    0 \le Q/M \lesssim 0.681 \, (1\sigma).
	    \label{M87_cons_eq}
	\end{align}
 For Sgr A*, the observed shadow angular diameter is $\theta_{\text{sh}}^{\text{Sgr A*}} =(48.7\pm 7.0)\mu as$, its distance to earth is $ D_{\text{Sgr A*}} = (8.15\pm 0.15)~\text{kpc} $ and its mass is $ M_{\text{Sgr A*}} = (4_{-0.6}^{+1.1}) \times 10^6~M_\odot $ \cite{EventHorizonTelescope:2022wkp, EventHorizonTelescope:2022wok, EventHorizonTelescope:2022exc, EventHorizonTelescope:2022urf, EventHorizonTelescope:2022xqj}. Besides these data, the EHT also provide the fractional deviation between the observed shadow radius and the shadow radius of a Schwarzschild BH, $\delta\equiv\frac{R_{sh}}{3\sqrt{3}M}-1$. Taking the
average of the Keck- and VLTI-based estimates of $\delta$, the $1\sigma$ interval for $\delta$ is
$
-0.125\lesssim\delta\lesssim 0.005~(1\sigma)
$\cite{Vagnozzi2023},
which leads to the following constraints on the model parameters of LQRNBH,
\begin{align}
	    0 \le \zeta/M \lesssim 2.875 \, (1\sigma), ~~ 0 \le Q/M \lesssim 0.799 \, (1\sigma).
	    \label{Sgr_cons_eq}
	\end{align} 
These two constraints on the parameters $\zeta$ and $Q$ are shown in Fig.\ref{fig:zetaQCons}. The red and blue curves are the boundaries of the constraints from M87* and Sgr A* respectively. When $Q=0$, the constraints on the quantum parameter $\zeta$ are consistent with the previous results \cite{Konoplya:2024lch, Shu:2024tut}. 

It should be noted that the discussion of the constraints on the parameters from the shadow observations is a rough theoretical investigation. The astrophysical black holes are believed to be neutral, or at least their charge can be negligible. So, we focus on black holes with tiny charge parameters ($Q<10^{-20}$) in the following sections. 

\section{Circular Motion of charged massive test particles}

In this section we consider the circular motion of a charged test particle with mass $m$ and charge $q$ in the LQRNBH spacetime. This is not a geodesic motion since the particle is subjected to both gravitational and electromagnetic forces. For simplicity, we set $M=1$ in the subsequent discussion and the other parameters are rescaled as  
\begin{equation}
		\frac{r}{M}\rightarrow r,\qquad 
		\frac{Q}{M}\rightarrow Q,\qquad 
        \frac{\zeta}{M}\rightarrow\zeta.
\end{equation}

The Lagrangian of the charged test particle is \cite{Chen:2023bao,Pugliese2011charged}
\begin{equation}\label{lag}
	\mathcal{L}
	=
	\frac{1}{2}g_{\mu\nu}\dot{x}^{\mu}\dot{x}^{\nu}
	+\eta A_{\mu}\dot{x}^{\mu},
\end{equation}
where $\eta=q/m$ is the charge-to-mass ratio or specific charge of the particle and the dot means derivative with respect to the proper time. The electromagnetic field of the LQRNBH is described by
\[
A_\mu dx^\mu=\frac{Q}{r}dt.
\]
The equations of motion of the test particle can be derived from the Lagrangian by using the Euler-Lagrange equation \cite{Chen:2023bao,Pugliese2011charged}. In fact, since the Lagrangian is independent of the variables $t$ and $\phi$,  we have following two conserved quantities for the particle  
\begin{equation}
	E=-p_t=-\frac{\partial \mathcal{L}}{\partial \dot{t}},
	\qquad
	L=p_\phi=\frac{\partial \mathcal{L}}{\partial \dot{\phi}},
\end{equation}
where $E$ and $L$ are the conserved specific energy and angular momentum. Due to the spherical symmetry of the LQRNBH spacetime and the electromagnetic field, without loss of generality we restrict ourselves to the study of the equatorial motion with $\theta=\frac{\pi}{2}$. Then, we have
\begin{equation}
	E=f(r)\dot{t}-\frac{\kappa}{r},
	\qquad
	L=r^2\dot{\phi},
	\label{eq:charged-conserved-quantities}
\end{equation}
where $\kappa$ describes the electromagnetic coupling between the charged particle and LQRNBH, which is defined as 
\begin{equation}
	\kappa\equiv\eta Q.
	\label{eq:kappa-definition}
\end{equation}
Using $g_{\mu\nu}u^\mu u^\nu=-1$, the first-order equations of equatorial motion can be written as
\begin{align}
	\dot{t}
	&=
	\frac{E+\kappa/r}{f(r)},
	\\
	\dot{\phi}
	&=
	\frac{L}{r^2},
	\\
	\dot{r}^{\,2}
	&=
	\left(E+\frac{\kappa}{r}\right)^2
	-
	f(r)\left(1+\frac{L^2}{r^2}\right)
	\equiv R(r).
	\label{eq:charged-radial-equation}
\end{align}
We can see that the electromagnetic interaction impact the dynamics of the particle through the combination \(E+\kappa/r\).

The circular equatorial orbits of the test particle are determined by the conditions $R(r) = 0$ and $\partial_r R(r) = 0$. From these conditions, we can obtain the specific energy $E$, the specific angular momentum $L$ and the angular velocity $\Omega$ of the particle on a circular equatorial orbit with radius $r$,
\begin{subequations}
	\label{eq:charged-circular-quantities}
	\begin{align}
		E(r)
		&=
		\frac{
			f(r)\left[\kappa+\mathcal{S}(r)\right]
		}{
			r\mathcal{D}(r)
		}
		-\frac{\kappa}{r},
		\label{eq:charged-energy}
		\\
		L(r)
		&=
		r\sqrt{
		\frac{\left(E(r)+\kappa/r\right)^2}{f(r)}
		-1},
		\label{eq:charged-angular-momentum}
		\\
		\Omega(r)
		&=
		\frac{f(r)L(r)}
		{r^2\left(E(r)+\kappa/r\right)},
		\label{eq:charged-angular-velocity}
	\end{align}
\end{subequations}
where for compactness, we define
	\begin{equation}
		\mathcal{D}(r)=2f(r)-rf'(r),
		\qquad
		\mathcal{S}(r)
		=
		\sqrt{\kappa^2+2r^2\mathcal{D}(r)} .
	\end{equation}

Before we numerically examine how the parameters $\zeta$, $Q$, and $\eta$ affect these three physical quantities, we provide an explanation for the selection of parameters in all the subsequent numerical studies. In the accretion disk around an astrophysical black hole, the main charged particles are negatively charged electrons and positively charged ions. The specific charge of electron is approximately $-2*10^{21}$ and the specific charges of the ions are several orders of magnitude smaller than that of the electron. Thus, in the subsequent studies we choose $|\eta|\leq 2\times 10^{21}$. Since $\kappa=\eta Q$ has remarkable impact on the circular motion of the charged particle, in order to support an reasonable ISCO and an accretion disk around the black hole, we choose $-1\leq\kappa\leq3$. Then, the charge parameter of the black hole is required to be very small, which is $Q\lesssim 10^{-21}$ (i.e., a solar mass black hole carries electric charges less than 1 C.). The circular motion of a particle with $\kappa>0$ is significantly different from that with $\kappa<0$. Without loss of generality, the charge of the black hole is always chosen to be positive, and the charge of the particle may be positive or negative.

%
%

\begin{figure}
	\centering
	
	\begin{subfigure}[b]{0.32\textwidth}
		\centering
		\includegraphics[width=\textwidth]{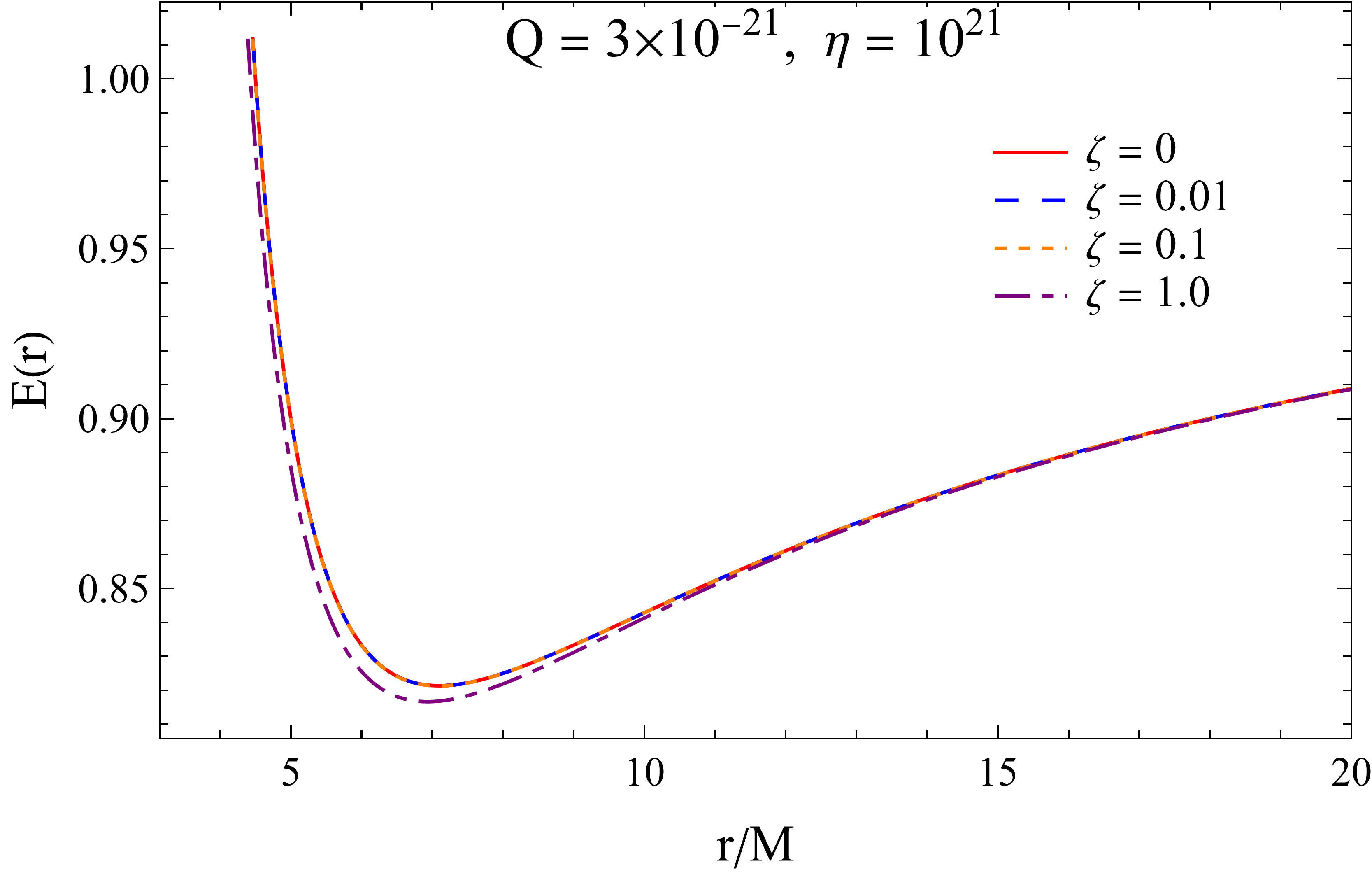}
		\caption{}
		\label{fig:energy-zeta}
	\end{subfigure}
	\begin{subfigure}[b]{0.32\textwidth}
		\centering
		\includegraphics[width=\textwidth]{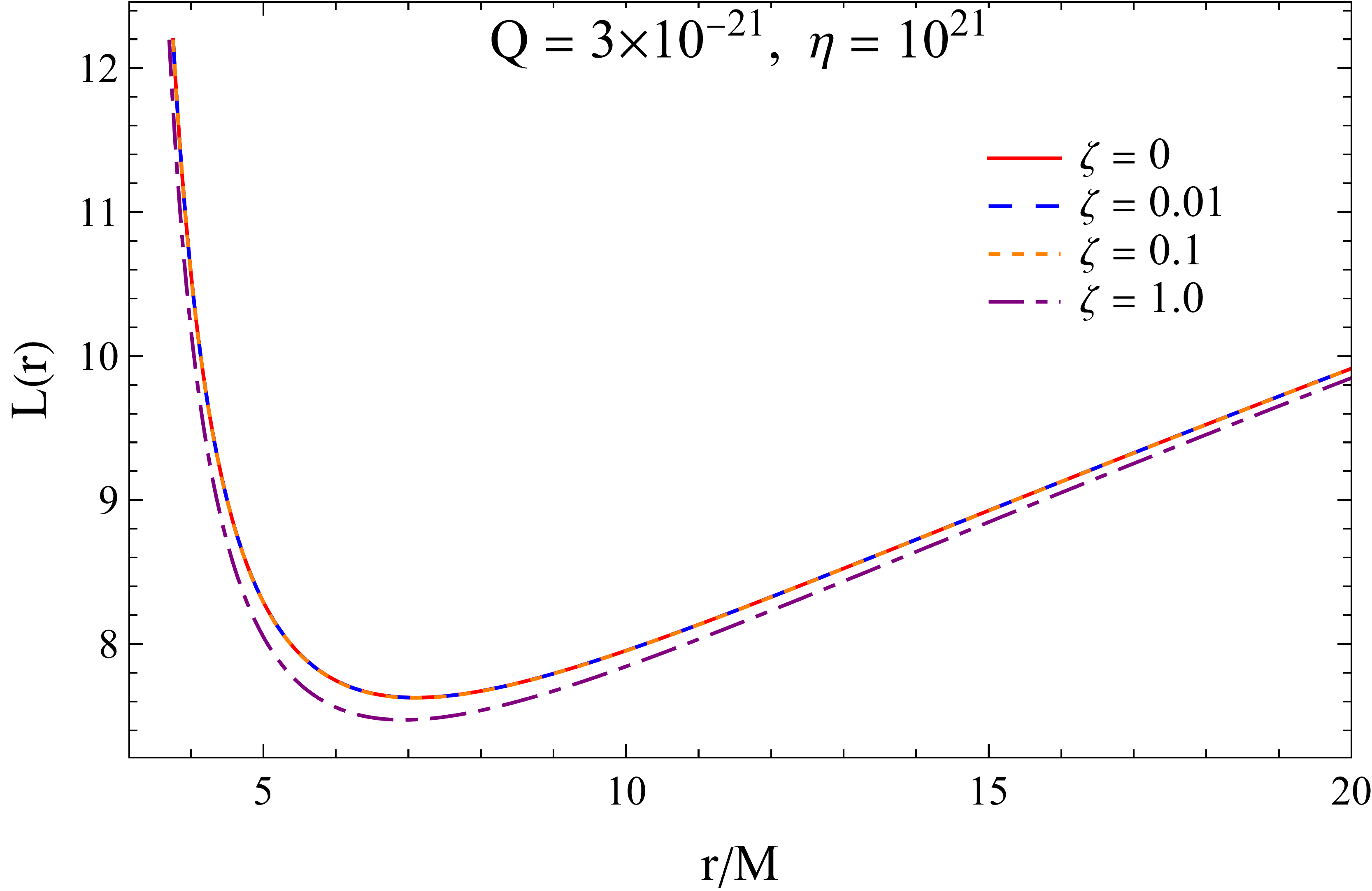}
		\caption{}
		\label{fig:angular-momentum-zeta}
	\end{subfigure}
	\begin{subfigure}[b]{0.32\textwidth}
		\centering
		\includegraphics[width=\textwidth]{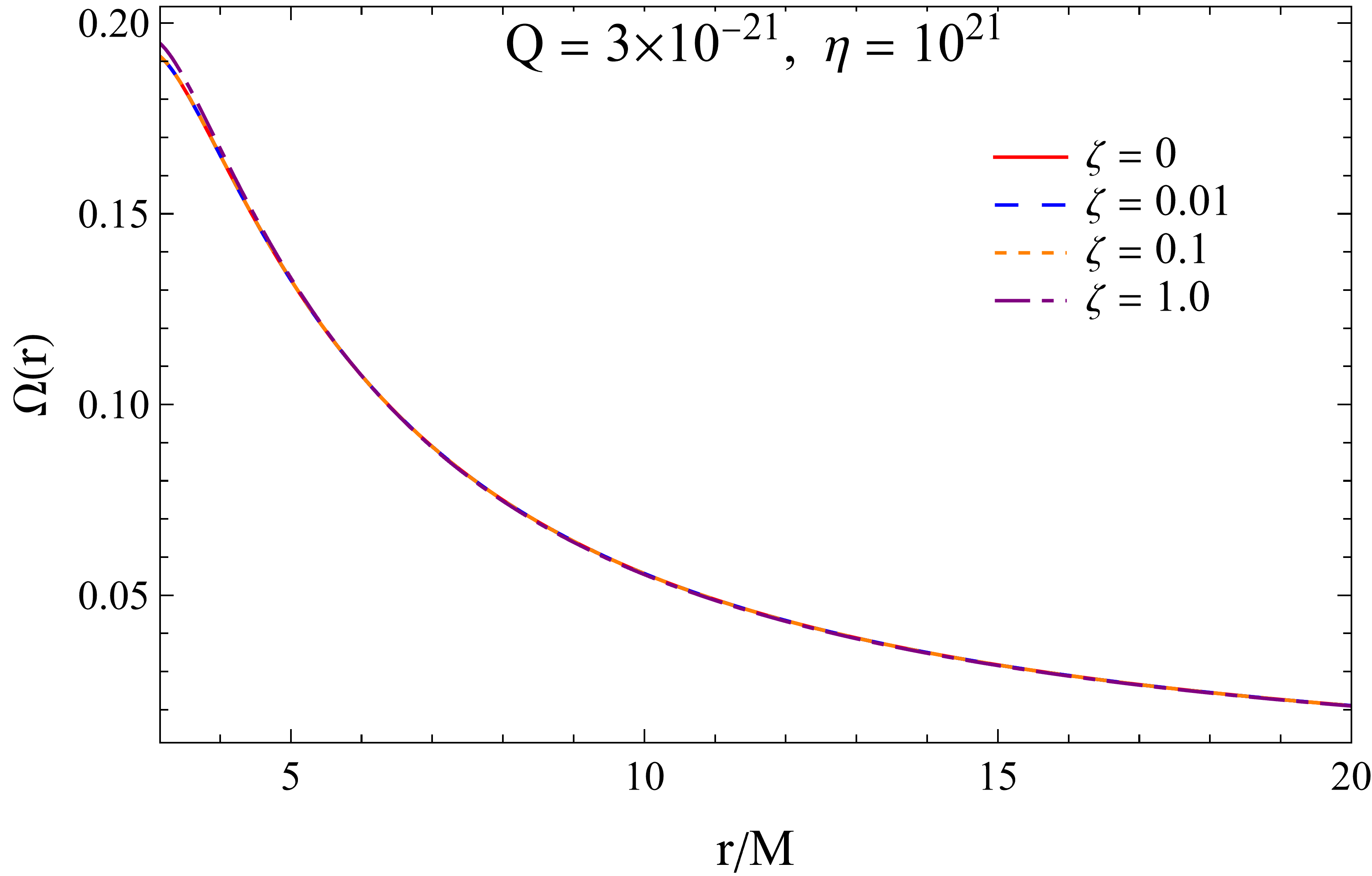}
		\caption{}
		\label{fig:omega-zeta}
	\end{subfigure}

	\begin{subfigure}[b]{0.32\textwidth}
		\centering
		\includegraphics[width=\textwidth]{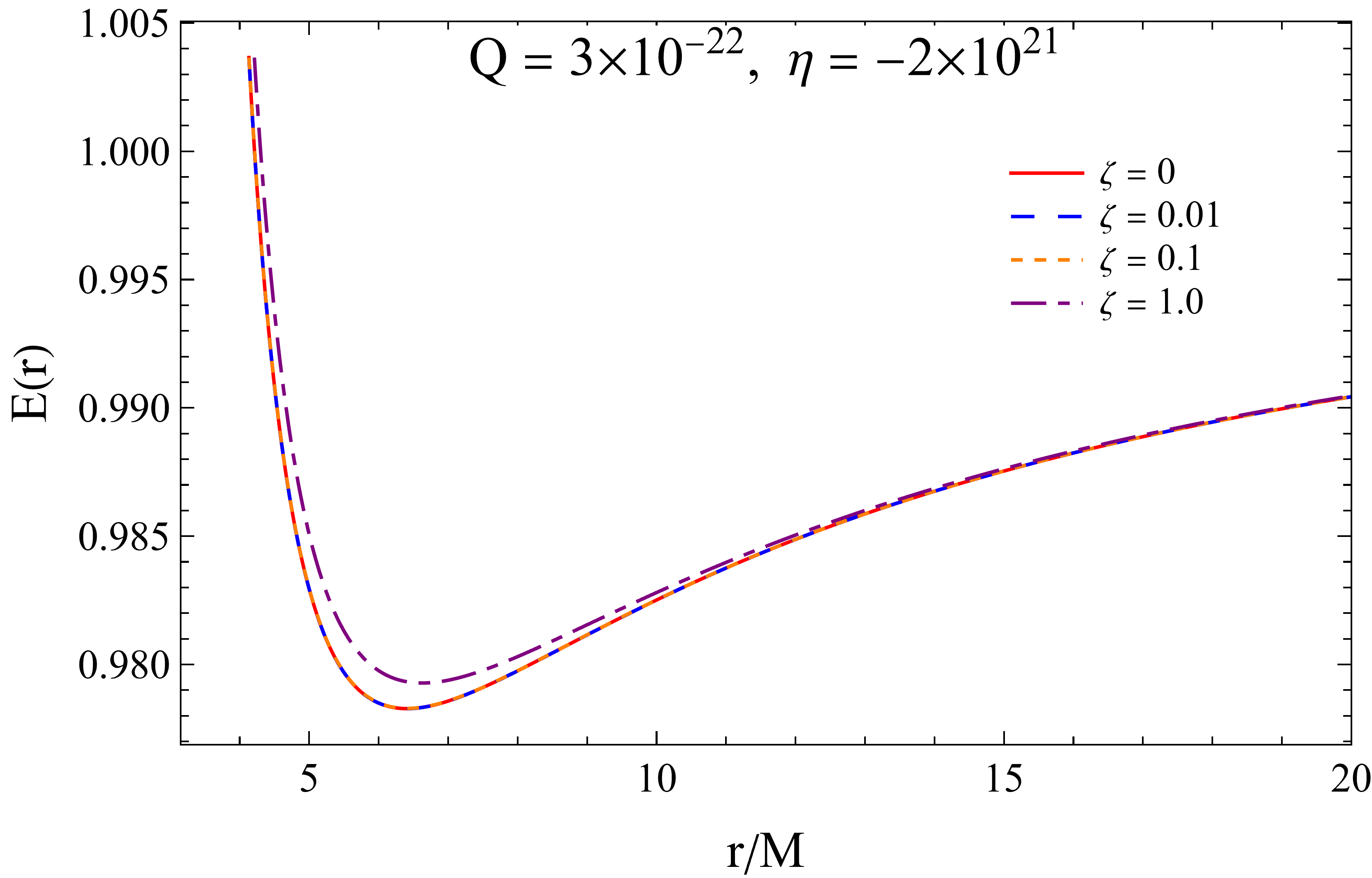}
		\caption{}
		\label{fig:energy-zeta-Negative}
	\end{subfigure}
	\begin{subfigure}[b]{0.32\textwidth}
		\centering
		\includegraphics[width=\textwidth]{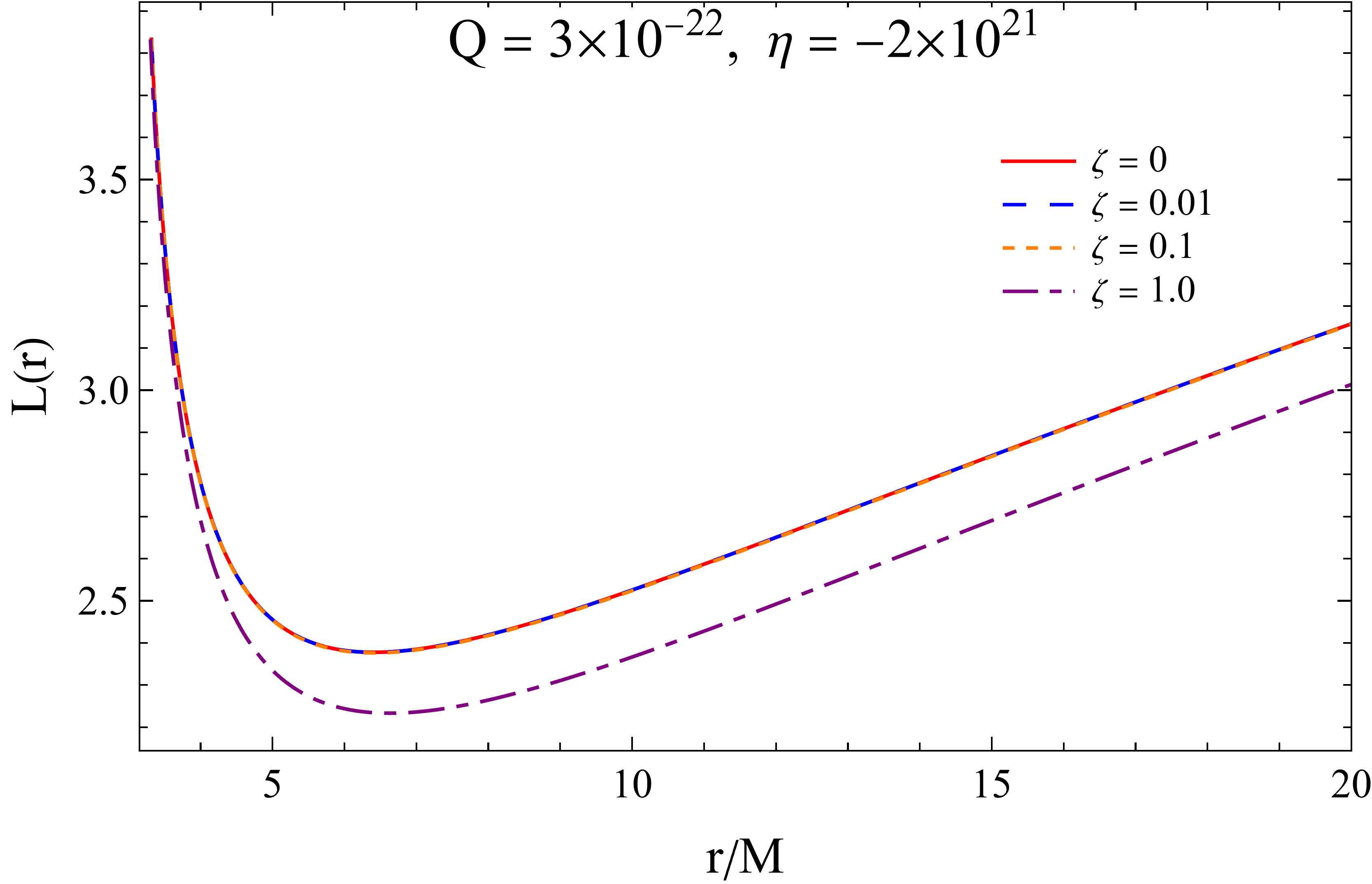}
		\caption{}
		\label{fig:angular-momentum-zeta-Negative}
	\end{subfigure}
	\begin{subfigure}[b]{0.32\textwidth}
		\centering
		\includegraphics[width=\textwidth]{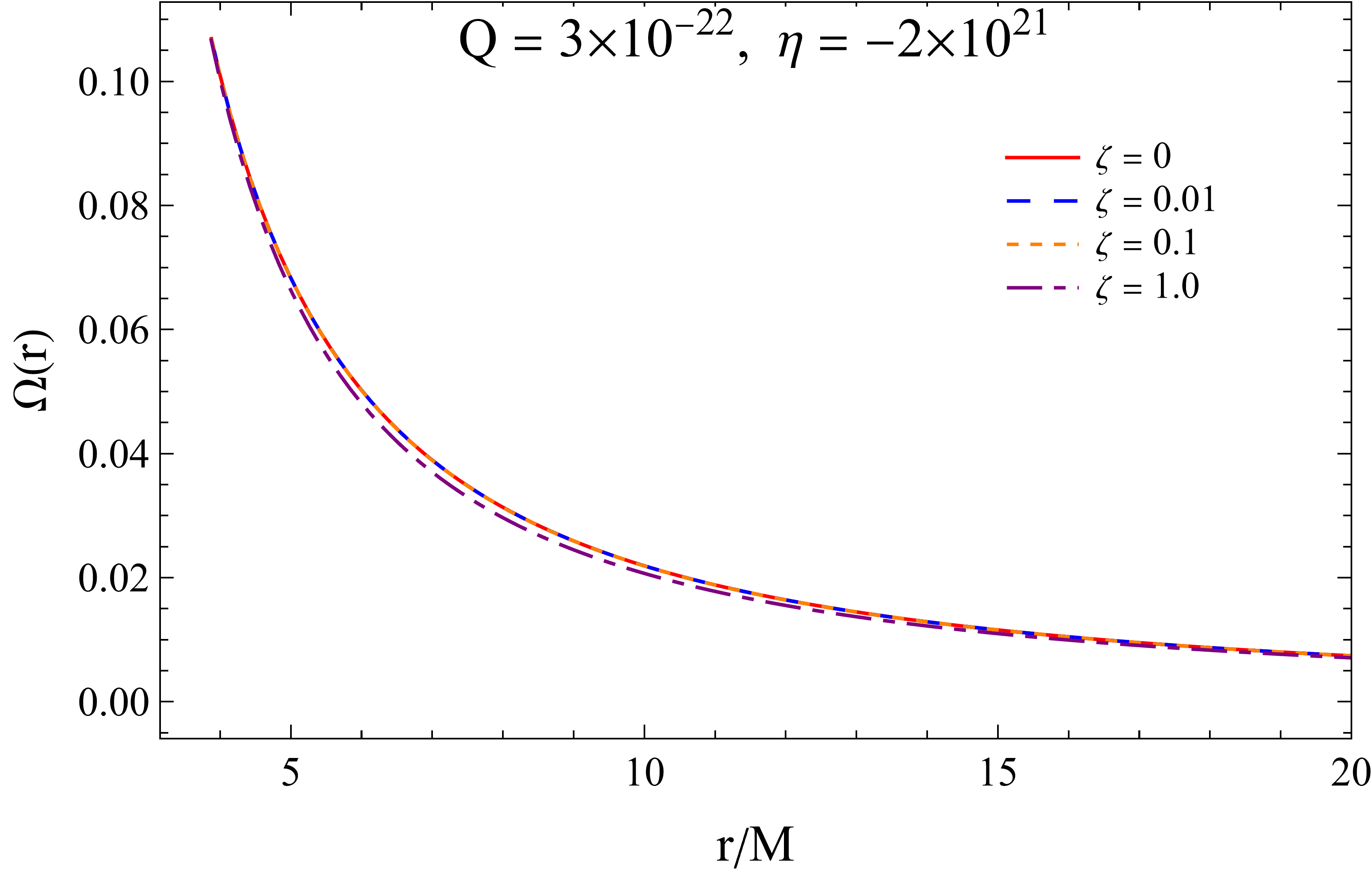}
		\caption{}
		\label{fig:omega-zeta-Negative}
	\end{subfigure}
	
	\begin{subfigure}[b]{0.32\textwidth}
		\centering
		\includegraphics[width=\textwidth]{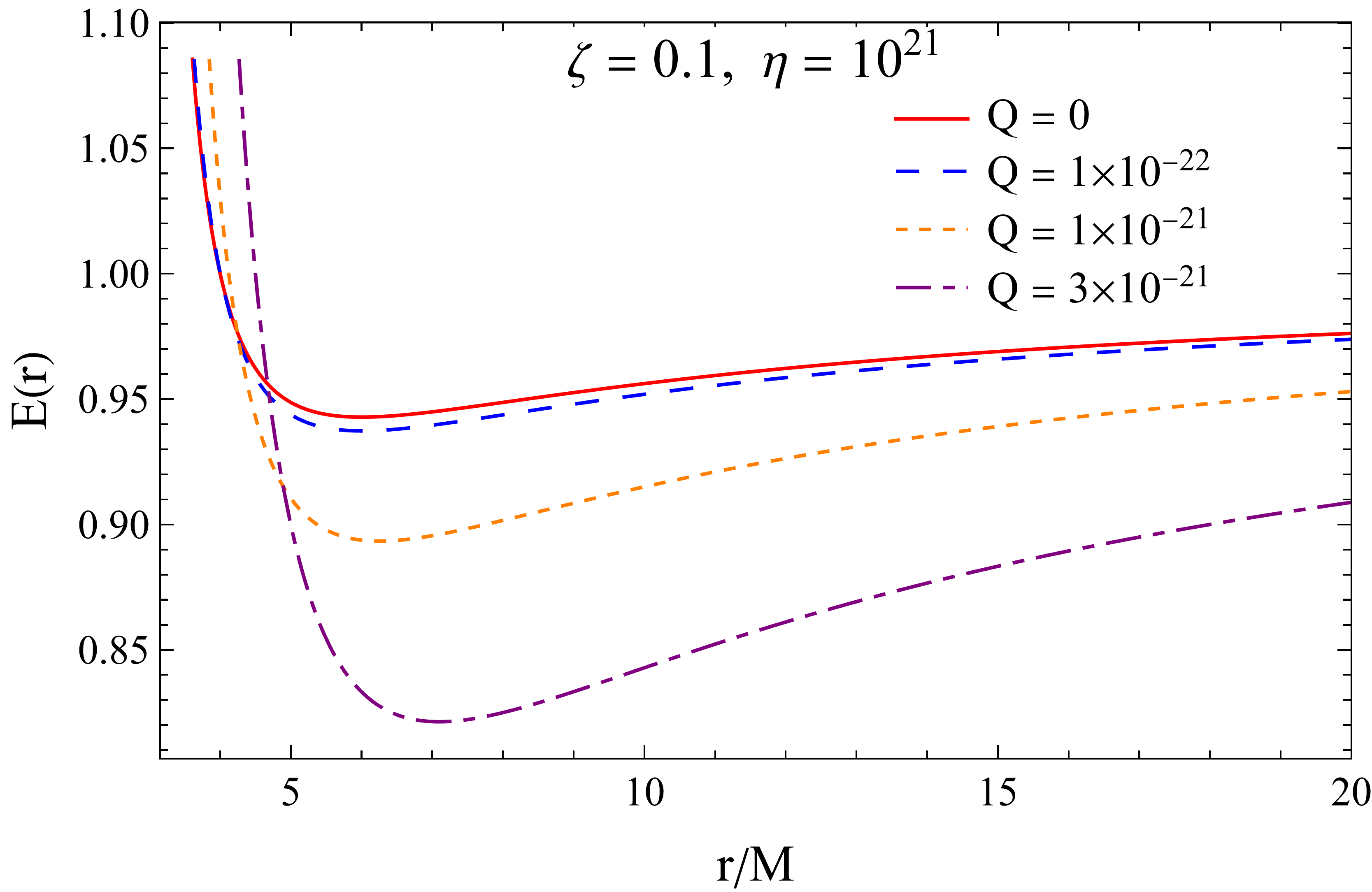}
		\caption{}
		\label{fig:energy-Q}
	\end{subfigure}
	\begin{subfigure}[b]{0.32\textwidth}
		\centering
		\includegraphics[width=\textwidth]{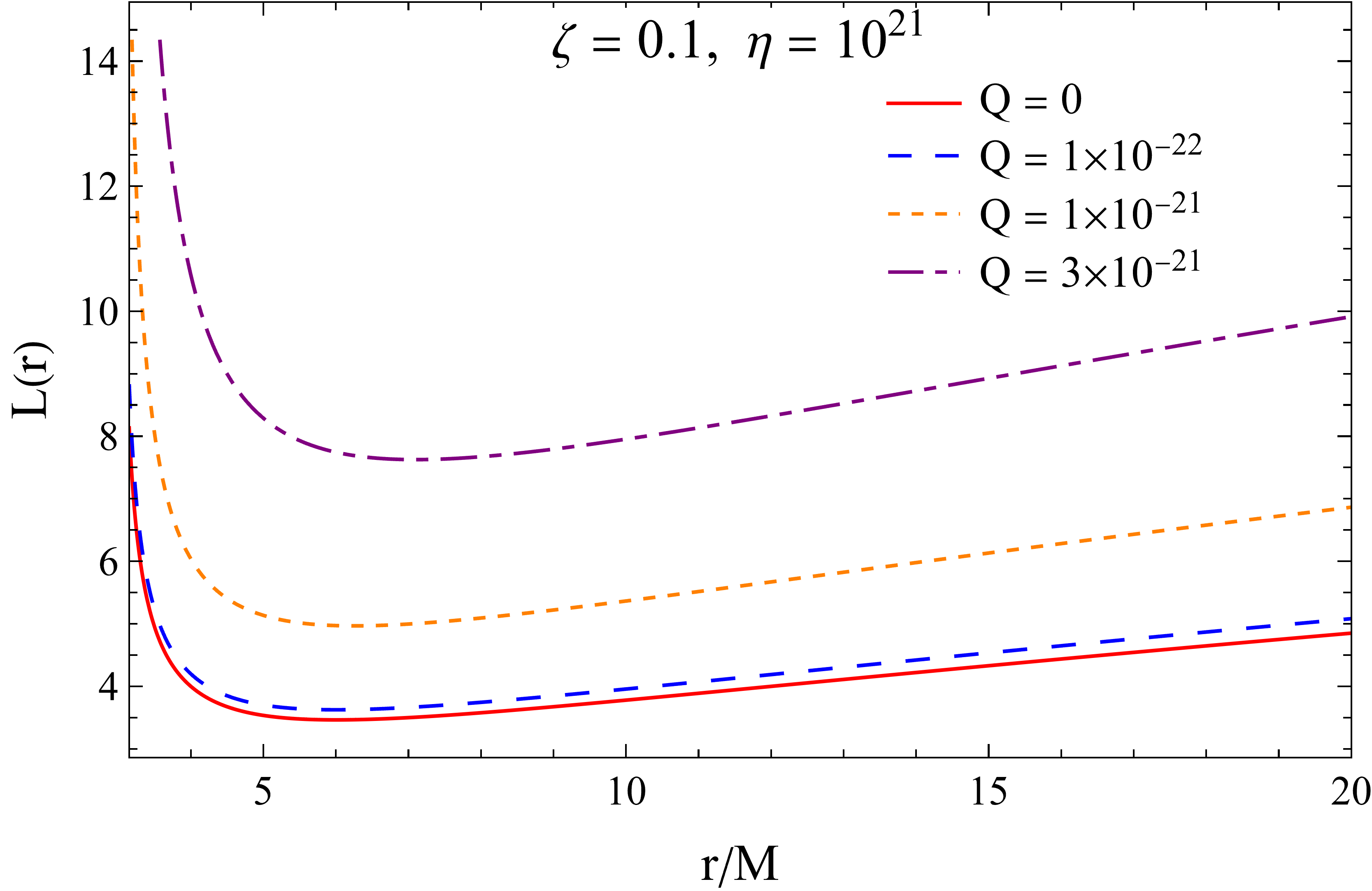}
		\caption{}
		\label{fig:angular-momentum-Q}
	\end{subfigure}
	\begin{subfigure}[b]{0.32\textwidth}
		\centering
		\includegraphics[width=\textwidth]{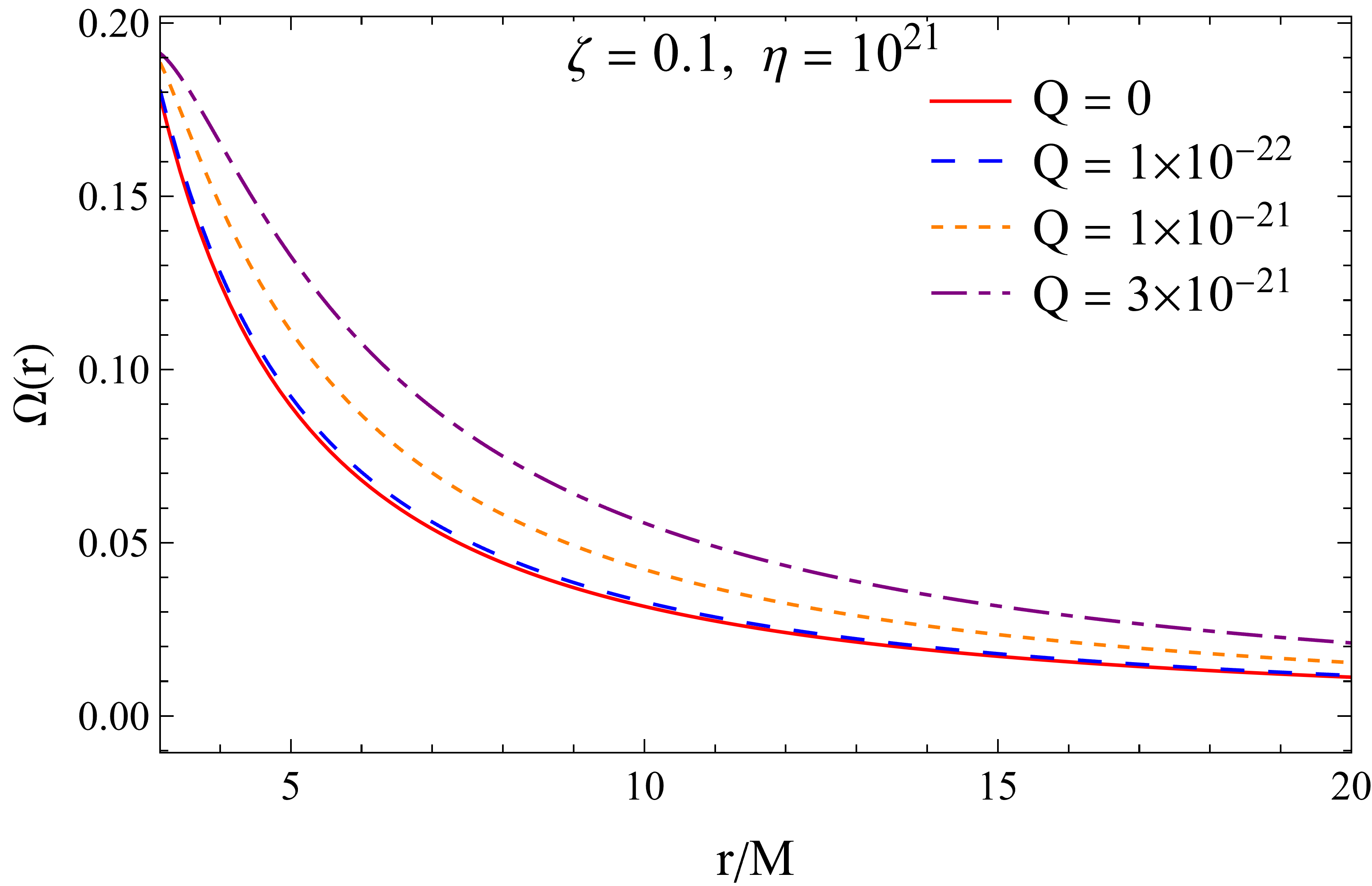}
		\caption{}
		\label{fig:omega-Q}
	\end{subfigure}
	
	\begin{subfigure}[b]{0.32\textwidth}
		\centering
		\includegraphics[width=\textwidth]{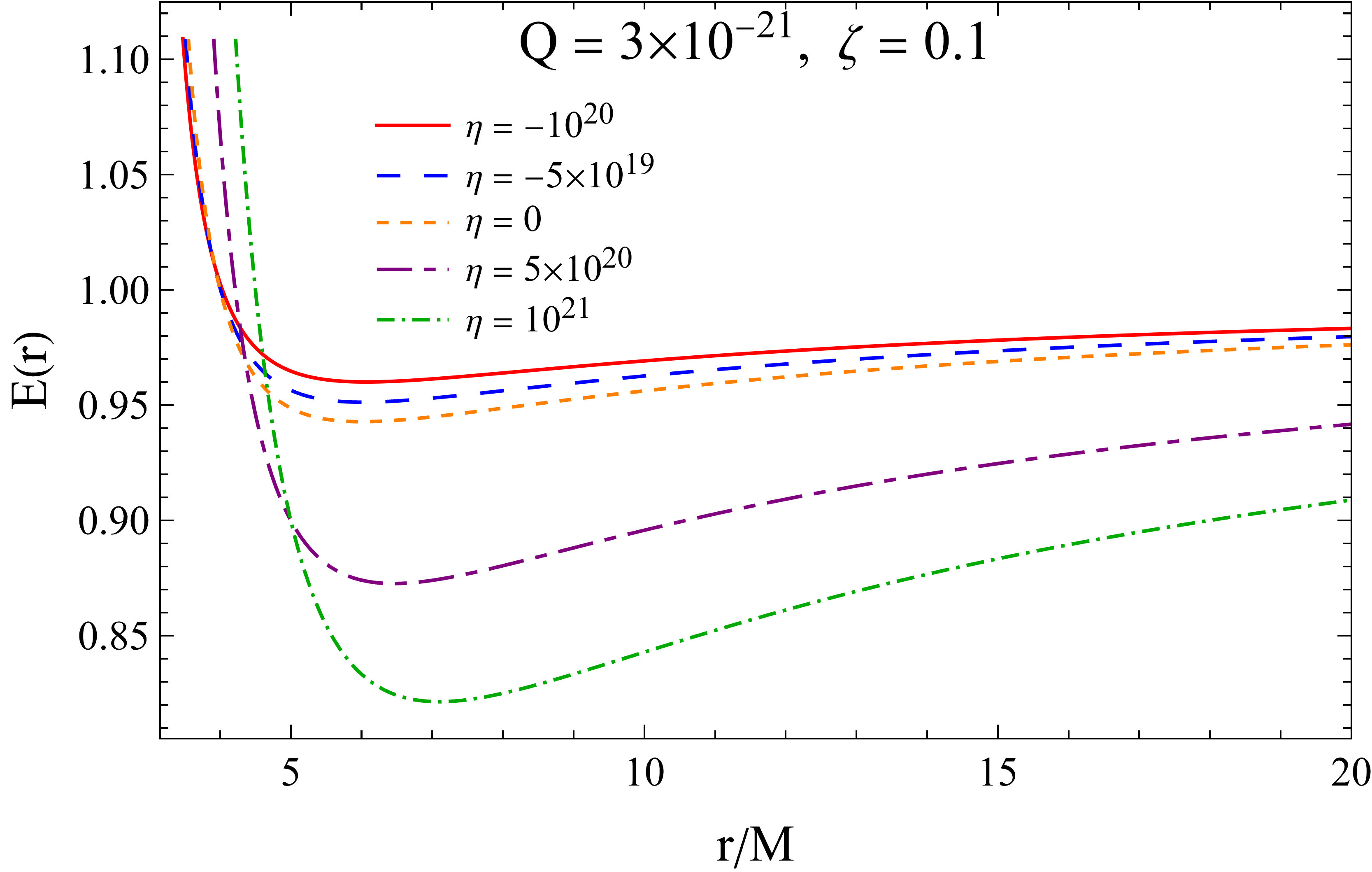}
		\caption{}
		\label{fig:energy-eta}
	\end{subfigure}
	\begin{subfigure}[b]{0.32\textwidth}
		\centering
		\includegraphics[width=\textwidth]{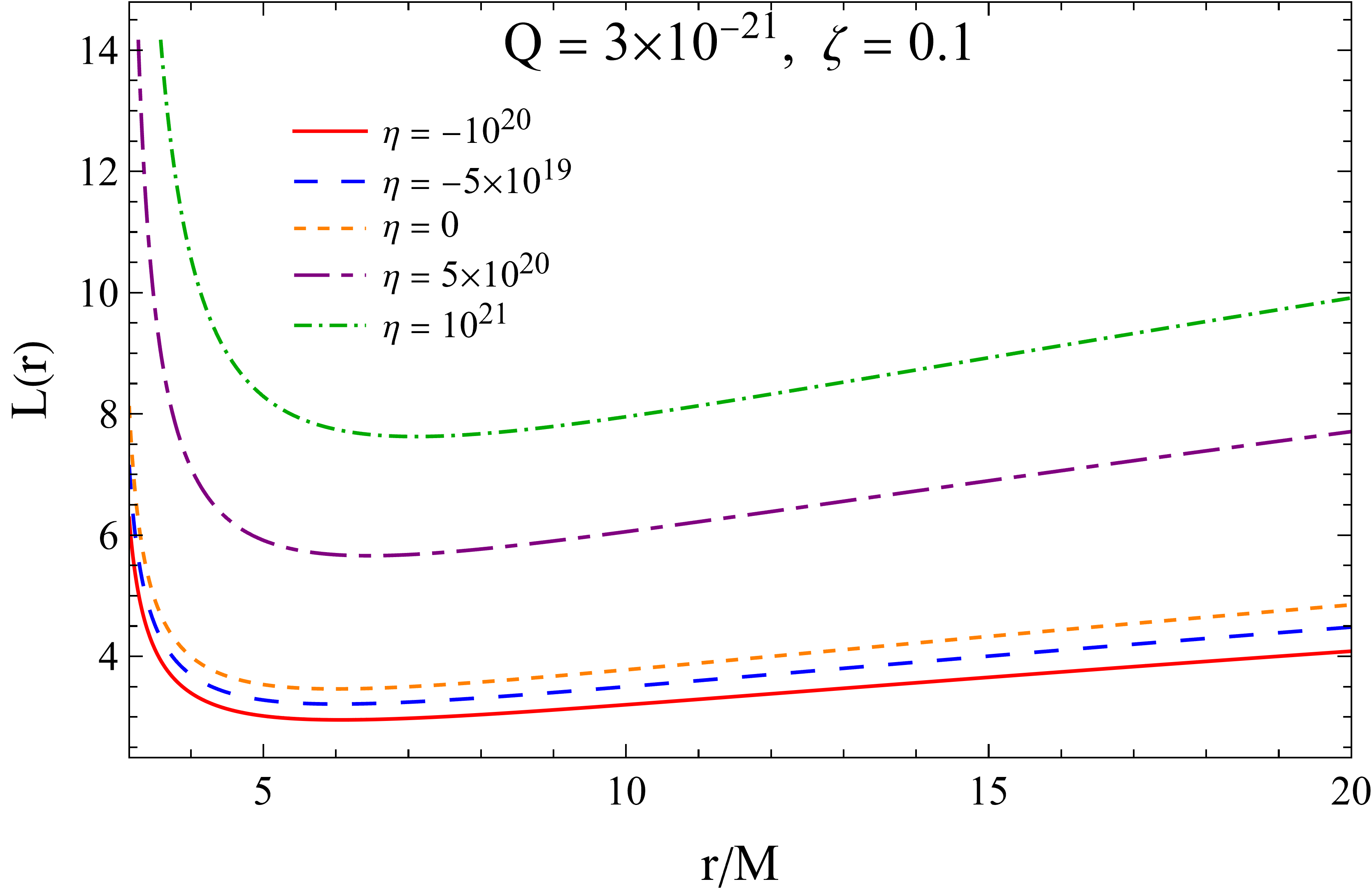}
		\caption{}
		\label{fig:angular-momentum-eta}
	\end{subfigure}
	\begin{subfigure}[b]{0.32\textwidth}
		\centering
		\includegraphics[width=\textwidth]{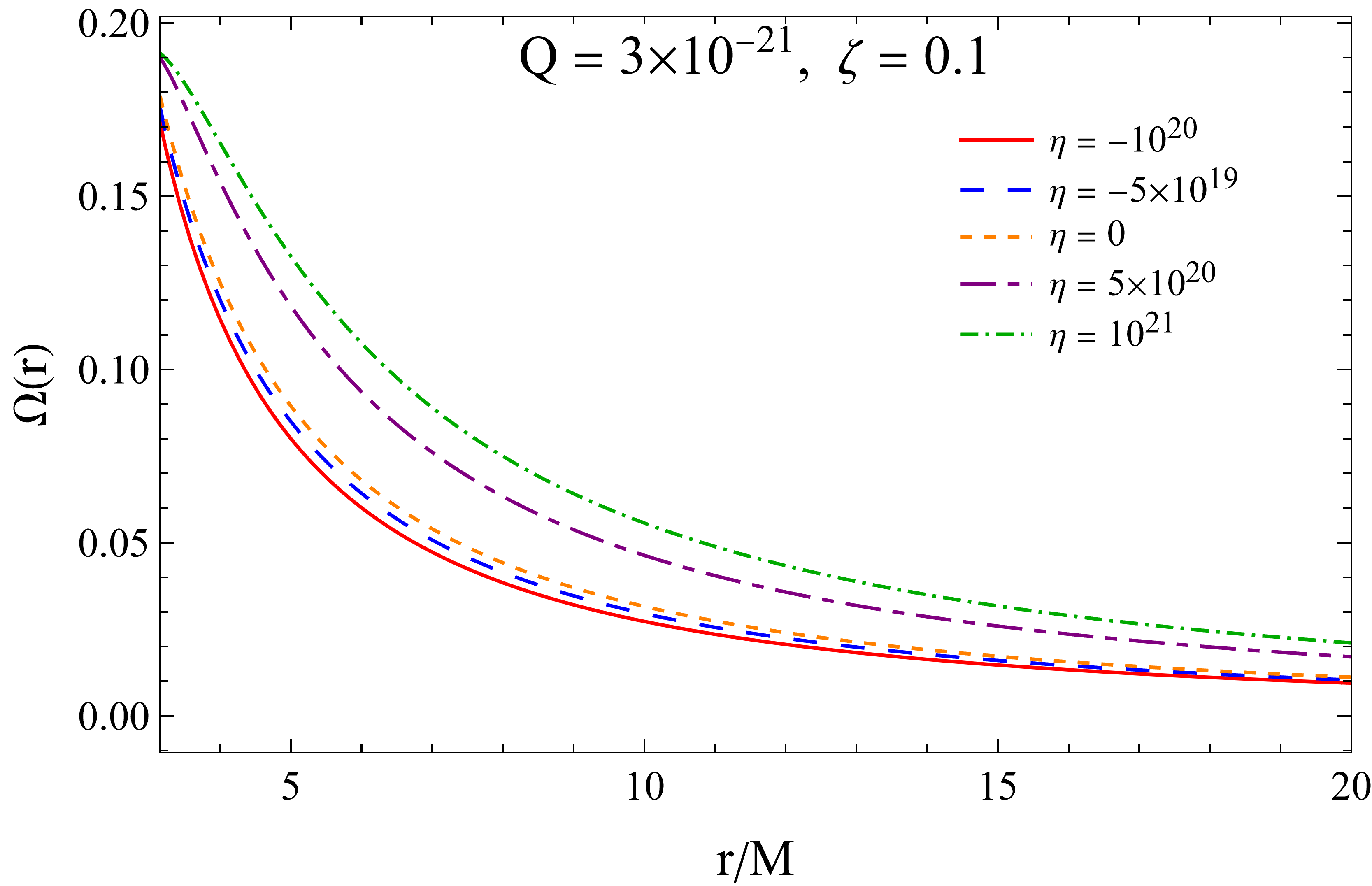}
		\caption{}
		\label{fig:omega-eta}
	\end{subfigure}
	
	\caption{Energy \(E\), angular momentum \(L\), and angular velocity \(\Omega\) of charged particles on equatorial circular orbits around the LQRNBH with various model parameters. }
	\label{fig:charged-circular-parameters}
\end{figure}
In Fig. \ref{fig:charged-circular-parameters}, we plot the energy $E$, the angular momentum $L$, and the angular velocity $\Omega$ of the particles on different circular orbits around LQRNBHs with various parameters. The first and second rows show the influence of $\zeta$ on these three quantities for positive and negative $\kappa$, respectively. The third row shows the influence of black hole charge parameter $Q$ on them, and the last row shows the influence of charge-to-mass ratio of the particles $\eta$ on them. In the first row $\kappa=3$ and in the second row $\kappa=-0.6$. 
From these two rows, we can observe that relatively pronounced effects on $E$ and $L$ curves can only be produced when the loop quantum parameter $\zeta\sim 1$, and the influence of the loop quantum parameter on the angular velocity $\Omega$ remains almost negligible. The loop quantum parameter slightly decreases the particles' energy for positive $\kappa$ while slightly increases it for negative $\kappa$. The loop quantum parameter $\zeta$ decreases the angular momentum  of the particles for both positive and negative $\kappa$. An interesting observation is that the loop quantum effect on the angular momentum is significantly more obvious when $\kappa$ is negative. 

Another interesting observation is that the influence of the quantum parameter $\zeta$ on $E$ is relatively obvious only for particles on orbits in the  near region, while the influence of the quantum parameter $\zeta$ on the angular momentum $L$ is much more obvious for particles in the far region. These two interesting observation can be understood by series expansions of $E$ and $L$ with respect to $r$. As $r \rightarrow \infty$, $E \sim 1 - \frac{1+\kappa}{2 r} + \frac{3+4\kappa+\kappa^2}{8 r^2}+\frac{27-8Q^2(2+\kappa)-8  (-3 + \zeta^2)  \kappa + 5  \kappa^2}{16r^3}+O(\frac{1}{r^4})$ and $L \sim \sqrt{r(1+\kappa)} - \frac{6 -2 Q^2 - 2\zeta^2+5\kappa+\kappa^2}{4\sqrt{1+\kappa}} \frac{1}{\sqrt{r}}+ O(r^{-\frac{3}{2}})$.
We can see that the leading order influences of $\zeta$ on $E$ and $L$ are $\frac{1}{r^3}$ and $\frac{1}{\sqrt{r}}$ respectively. Although $L$ is more sensitive to $\zeta$ in the region far from the event horizon, the influence of $\zeta$ on $L$ is combined with $Q$ in the leading order and next to leading order is needed to distinguish the influences from $Q$ and $\zeta$.  

From the third and fourth rows of Fig. \ref{fig:charged-circular-parameters}, we can see that the electromagnetic interaction between the black hole and the charged particles has significantly impact on the energy $E$, angular momentum $L$ and the angular velocity $\Omega$ of the particles. The electromagnetic interaction always decreases the energy while increases the angular momentum and the angular velocity. In fact, the electromagnetic coupling constant $\kappa$ plays the key role in the influence on these quantities.

\begin{figure}[htbp]
	\centering
	\begin{subfigure}[b]{0.48\textwidth}
		\centering
		\includegraphics[width=\textwidth]{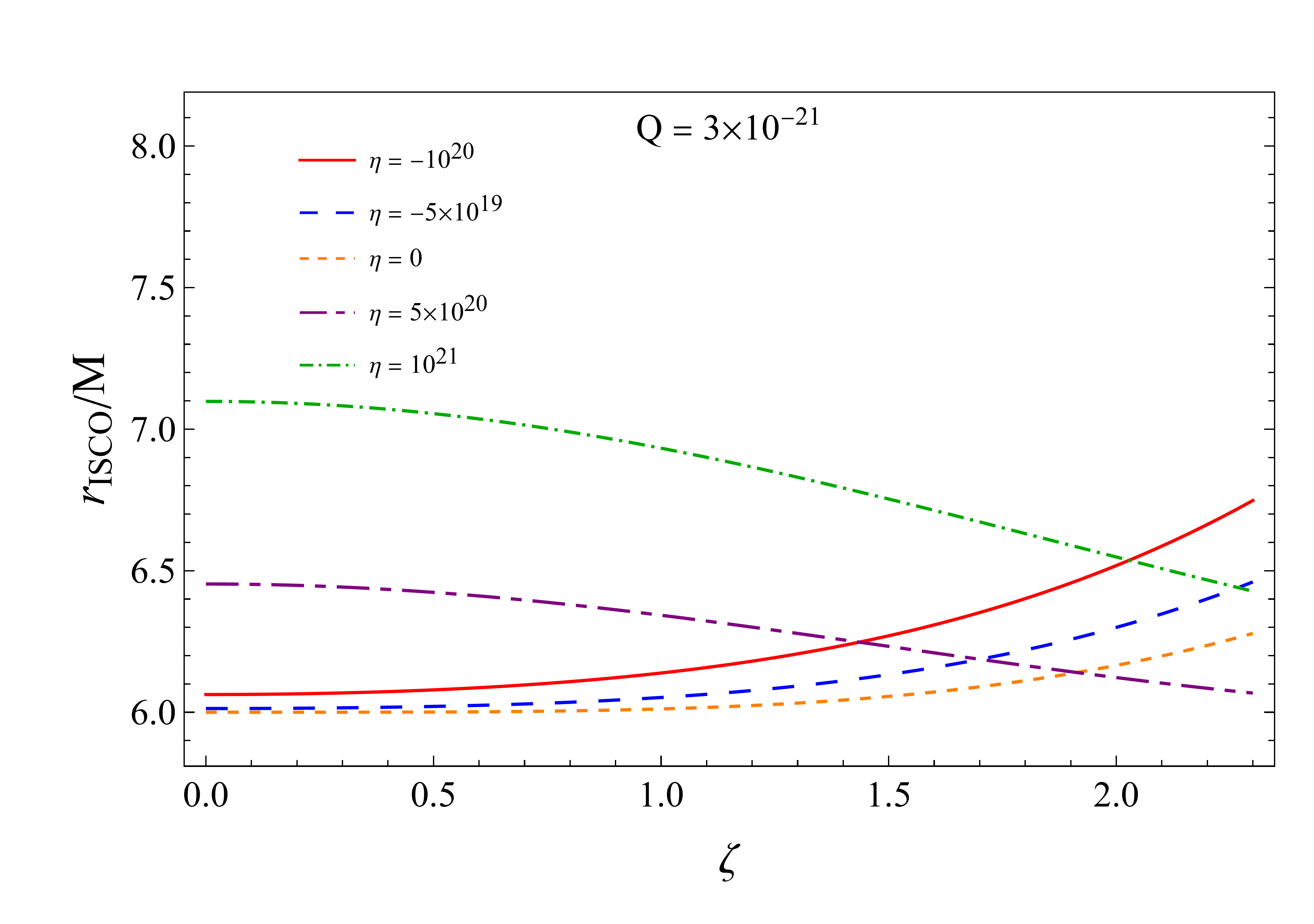}
		\caption{}
		\label{fig:isco-zeta}
	\end{subfigure}
	\hfill
	\begin{subfigure}[b]{0.48\textwidth}
		\centering
		\includegraphics[width=\textwidth]{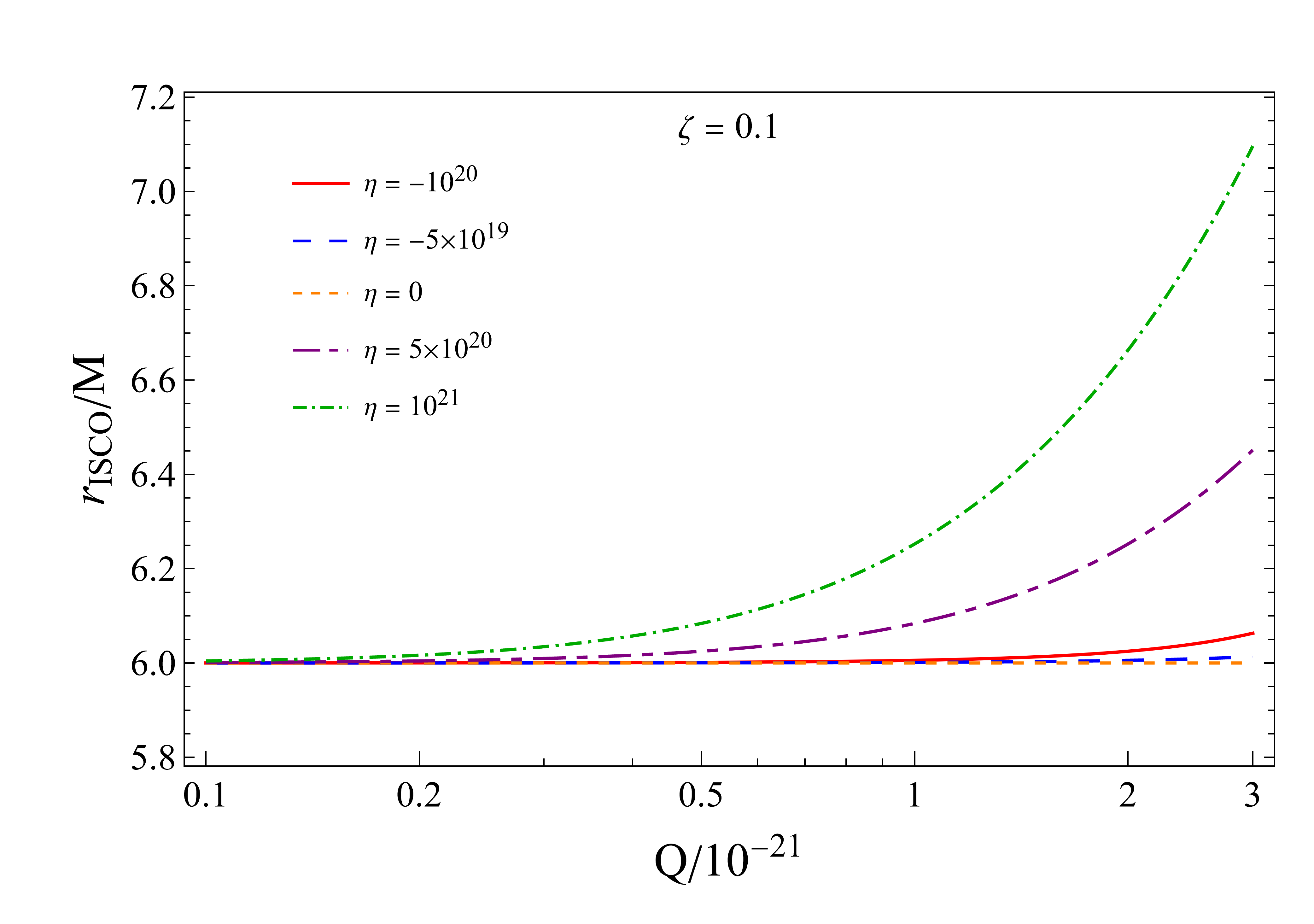}
		\caption{}
		\label{fig:isco-Q}
	\end{subfigure}
	\caption{The ISCO radius of charged particles around LQRNBHs as functions of $\zeta$ and $Q$.}
	\label{fig:charged-isco-q-zeta}
\end{figure}

Among the circular orbits, there usually exists an innermost stable circular orbit (ISCO) which is the inner boundary of the stable circular orbits. The ISCO is determined by an additional condition  
	\begin{equation}
	  -\frac{1}{2}  \partial_r^2  R(r) = -\frac{2 E \kappa}{r^3} - \frac{3 \kappa^2}{r^4}+\frac{3 L^2 f(r)}{r^4} - \frac{2 L^2 f'(r)}{r^3} 
+ \frac{1}{2} \left(1 + \frac{L^2}{r^2}\right) f''(r) = 0. 
	    \label{isco_sol}
	\end{equation}
Plugging Eq.\ref{eq:charged-energy} and Eq.\ref{eq:charged-angular-momentum} into the above equation, we can obtain the equation satisfied by the radius of ISCO. 
By numerically solving this equation, we plot the dependence of the ISCO radius on the loop quantum parameter $\zeta$ and black hole charge parameter $Q$ in Fig.\ref{fig:charged-isco-q-zeta}. In Fig.\ref{fig:isco-zeta}, we can see that a visible effect on the ISCO radius emerges only when the quantum parameter $\zeta$ is of order unity. When $\zeta$ increases, the ISCO radius increases for negative $\kappa$ while decreases for positive $\kappa$. In Fig.\ref{fig:isco-Q}, we can see that the ISCO radius increases monotonically as the increase of the absolute value of the electromagnetic coupling $\kappa$, and in particular, when $|\kappa|$ is greater than 1, the ISCO radius increases remarkably.

\section{Radiation Properties of a Thin Accretion Disk around LQRNBH}

In this section, we consider various radiation properties when the LQRNBH is surrounded by a accretion disk, which is modelled here by the geometrically thin and optically thick Novikov–Page–Thorne accretion disk \cite{Page:1974he,Alloqulov:2025pex}. We will focus on the influence of the model parameters on these properties.

In the process of a test particle falling from infinity to the black hole, the efficiency of converting its gravitational energy into radiation can be used as a key indicator to quantify the energy conversion capacity of the central black hole. Assuming all the radiation energy reaches to infinity and the falling particle at infinity is at rest ($E_{\infty} \approx 1$), the radiation efficiency $\epsilon$ of a black hole for a test particle falling from spatial infinity to the ISCO are defined as \cite{Novikov1973,Liu:2024brf}
	\begin{equation}
	    \epsilon= \frac{E_{\infty }-E_{\text{ISCO}}}{E_{\infty }} \approx  1 - E_{\text{ISCO}}.
	\end{equation}
In Table \ref{tab:isco_params_q}, we calculate the radiation efficiency of LQRNBH for various values of the model parameters. It is obvious that 
the influence of the quantum parameter $\zeta$ on the radiation efficiency is very small. However, the electromagnetic interaction has significant impact on it, and this impact depends on whether the interaction is attractive or repulsive. When the interaction is repulsive (i.e. $\kappa>0$), the radiation efficiency increases significantly as the increase of $\kappa$. And, the radiation efficiency decreases remarkably as the increase of $|\kappa|$ when $\kappa<0$.  

\begin{table}[!htbp]
	\centering
	\small
	\caption{
		 The ISCO radius $r_{\mathrm{ISCO}}$, ISCO energy
		$E_{\mathrm{ISCO}}$, and radiative efficiency $\epsilon$ for various model parameters.
		}
	\label{tab:isco_params_q}
	\begingroup
	\setlength{\tabcolsep}{4pt}
	\renewcommand{\arraystretch}{1.08}
	\begin{tabular}{@{}ccccccc@{}}
		\toprule
		$\zeta$ & $Q$ & $\eta$ & $\kappa$
		& $r_{\mathrm{ISCO}}$ & $E_{\mathrm{ISCO}}$ & $\epsilon$ \\
		\midrule
		0.00 & $3\times10^{-21}$ & $10^{21}$ & 3.00
		& 7.09808 & 0.821367 & 17.863\,\% \\
		0.01 & $3\times10^{-21}$ & $10^{21}$ & 3.00
		& 7.09806 & 0.821367 & 17.863\,\% \\
		0.10 & $3\times10^{-21}$ & $10^{21}$ & 3.00
		& 7.09632 & 0.821320 & 17.868\,\% \\
		1.00 & $3\times10^{-21}$ & $10^{21}$ & 3.00
		& 6.93275 & 0.816671 & 18.333\,\% \\
		\midrule
		0.10 & $0$ & $10^{21}$ & 0.00
		& 6.00000 & 0.942809 & 5.719\,\% \\
		0.10 & $1\times10^{-22}$ & $10^{21}$ & 0.10
		& 6.00442 & 0.937317 & 6.268\,\% \\
		0.10 & $1\times10^{-21}$ & $10^{21}$ & 1.00
		& 6.25243 & 0.893371 & 10.663\,\% \\
		0.10 & $3\times10^{-21}$ & $10^{21}$ & 3.00
		& 7.09632 & 0.821320 & 17.868\,\% \\
		\midrule
		0.10 & $3\times10^{-21}$ & $-10^{20}$ & $-0.30$
		& 6.06318 & 0.960062 & 3.994\,\% \\
		0.10 & $3\times10^{-21}$ & $-5\times10^{19}$ & $-0.15$
		& 6.01336 & 0.951292 & 4.871\,\% \\
		0.10 & $3\times10^{-21}$ & $0$ & 0.00
		& 6.00000 & 0.942809 & 5.719\,\% \\
		0.10 & $3\times10^{-21}$ & $5\times10^{20}$ & 1.50
		& 6.45182 & 0.872590 & 12.741\,\% \\
		0.10 & $3\times10^{-21}$ & $10^{21}$ & 3.00
		& 7.09632 & 0.821320 & 17.868\,\% \\
		\bottomrule
	\end{tabular}
	\endgroup
\end{table}

In our chosen thin disk model, the local radiative flux emitted from the disk surface is \cite{Page:1974he}
\begin{equation}
	\tilde{F}(r) = -\frac{\dot{M}}{4\pi \sqrt{-g / g_{\theta\theta}}}
	\frac{\Omega_{,r}}{(E - \Omega L)^2}
	\int_{R_I}^{r} (E - \Omega L) L_{,r} \, dr,
\end{equation}
where $\dot{M}$ is the mass accretion rate, $g$ is the metric determinant, and $R_I=r_{\mathrm{ISCO}}$. The functions $E(r)$, $L(r)$, and $\Omega(r)$ are those in Eq.~\eqref{eq:charged-circular-quantities}. The contribution of the electromagnetic interaction to the flux through these orbital quantities and the ISCO which is regarded as the inner edge of the disk. 

For the convenience of comparison with astrophysical observation, we recover dimensions in the following discussion. The relation between the radiation flux $\tilde{F}$ and the dimensional radiation flux $F$ is \cite{Liu:2025hhg,Collodel:2021gxu}
\begin{eqnarray}
F = \frac{c^{6}}{G^2 M^2} \tilde{F}.
\end{eqnarray}
The values of physical parameters used here are: $c = 2.99792458 \times 10^{10}$ cm s$^{-1}$, $M_\odot = 1.989 \times 10^{33}$ g, $G=6.67430\times 10^{-8}$ cm$^3$ g$^{-1}$ s$^{-1}$.

\begin{figure}[H]
	\centering
	\begin{subfigure}[t]{0.32\textwidth}
		\centering
		\includegraphics[
		width=\linewidth,
		height=5.2cm,
		keepaspectratio
		]{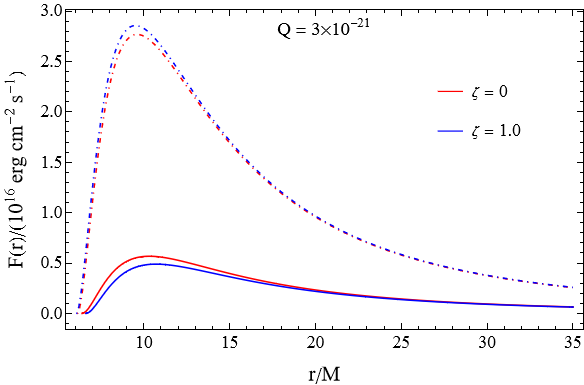}
		\caption{}
		\label{fig:Fvsz}
	\end{subfigure}
	\begin{subfigure}[t]{0.32\textwidth}
		\centering
		\includegraphics[
		width=\linewidth,
		height=5.2cm,
		keepaspectratio
		]{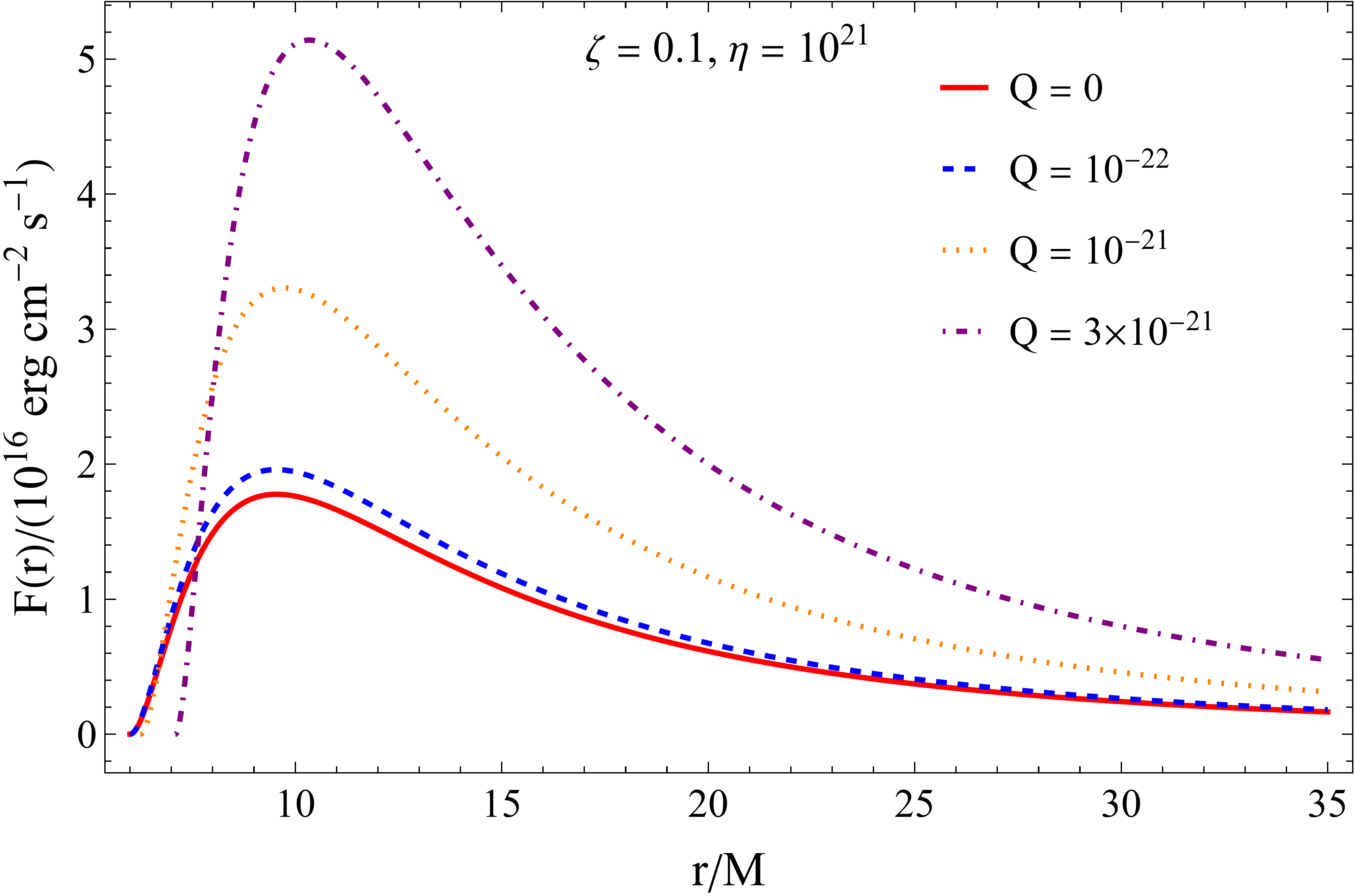}
		\caption{}
		\label{fig:FvsQ}
	\end{subfigure}
	\begin{subfigure}[t]{0.32\textwidth}
		\centering
		\includegraphics[width=\textwidth,height=5.2cm,
		keepaspectratio]{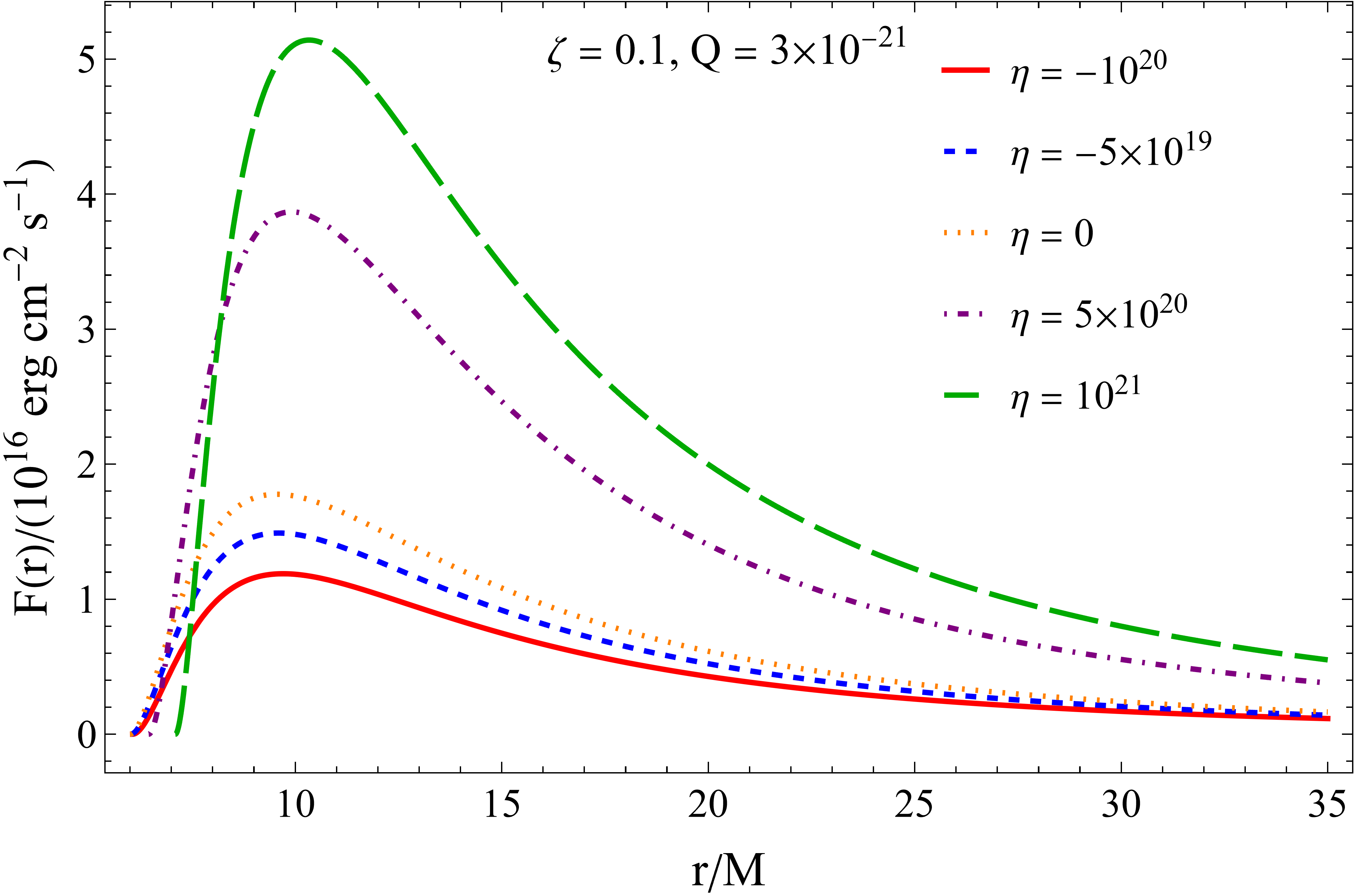}
		\caption{}
		\label{fig:Fvseta}
	\end{subfigure}
	\caption{
		Local radiation flux $F(r)$ emitted from the disk around the LQRNBH for
		various model parameters. In the left panel, $\eta=2\times10^{20}$ ($\kappa=0.6$) for the dashed-dotted lines and  $\eta=-2\times10^{20}$ ($\kappa=-0.6$) for the solid lines.
		}
	\label{fig:F}
\end{figure}

In Fig.\ref{fig:F}, we show the local radiation flux distribution of the accretion disk surrounding the black hole and the influence of parameters $\zeta$, $Q$, and $\eta$ on the flux. The mass accretion rate $\dot{M}$ is assumed to be $2\times10^{-3}M_\odot/\text{yr}$  and the black hole mass is taken as $M=2\times 10^6 M_\odot$. 
For each curve, the flux rises from zero at the ISCO, reaches its maximum, and then decreases. From the left panel, 
we can see that compared with the $\zeta=0$ case, the $\zeta=1$ case exhibits a visible but still moderate increase of the flux when $\kappa>0$; however, when $\kappa<0$, the $\zeta=1$ case exhibits a decrease of the flux. These results are consistent with the previous observation about the influence of $\zeta$ on the radiation efficiency. Similarly, from the middle and right panels, we can see that the electromagnetic coupling constant $\kappa$ exerts an important influence on the flux. The local flux increases significantly as the increases of $\kappa$ when $\kappa>0$ and decreases remarkably as the increases of $|\kappa|$ when $\kappa<0$.   
This result is also consistent with previous observation of influence on the radiation efficiency.

When the accretion disk around the black hole is under local thermodynamic equilibrium, the local effective temperature $T_{\mathrm{eff}}$ of the accretion disk and the radiation flux satisfy the Stefan-Boltzmann relation $F(r) = \sigma T_{\mathrm{eff}}^4(r)$, the Stefan-Boltzmann constant $\sigma = \frac{2\pi^5 k^4}{15 c^2 h^3}=5.67 \times 10^{-5}$. 
To quantitatively illustrate the difference in these quantities between LQRNBH and the corresponding black holes in general relativity, we calculate two examples and summarize the data in Table~\ref{tab:flux_temperature_peaks}.

 For the first example, we refer to observational data for the black hole M87* as our model parameters. We take the black hole mass $M = 6.5 \times 10^9 M_\odot$, the accretion rate $\dot{M} = 10^{-3} M_\odot/\mathrm{yr}$ \cite{EventHorizonTelescope:2019ggy,Drew:2025euq, EventHorizonTelescope:2021srq}. For the second example,  we adopt a mass of $M = 4.3 \times 10^6 M_\odot$ and an accretion rate of $7.0\times10^{-9} M_\odot/\text{yr}$, which refer the supermassive black hole at the center of our galaxy, Sgr A* \cite{EventHorizonTelescope:2022wkp, EventHorizonTelescope:2022wok, EventHorizonTelescope:2022exc, EventHorizonTelescope:2022urf, EventHorizonTelescope:2022xqj}.
In each case, the other parameters of LQRNBH are chosen as $Q=3\times10^{-21}$, $\zeta=0.1$, and $\eta=10^{21}$, the RNBH has the same $Q$ and $\eta$ but vanishing $\zeta$, and the Schwarzschild black hole has $Q=\zeta=0$.  

\begin{table}[!htbp]
	\centering
	\small
	\caption{
		Peak values of the local radiation flux and the local effective temperature for two black hole models (M87* and Sgr A*).
		$r_{\mathrm{peak}}$ is the location of the peak value.
	}
	\label{tab:flux_temperature_peaks}
	\begingroup
	\setlength{\tabcolsep}{4pt}
	\renewcommand{\arraystretch}{1.12}
	\begin{tabular}{@{}llcccc@{}}
		\toprule
		BHs & Spacetime
		& $r_{\mathrm{ISCO}}/M$
		& $r_{\mathrm{peak}}/M$
		& $F_{\max}\,(\mathrm{erg\,cm^{-2}\,s^{-1}})$
		& $T_{\mathrm{eff}}^{\max}\,(\mathrm{K})$ \\
		\midrule
		M87* & LQRN
		& 7.09632 & 10.33664
		& $2.43354\times10^{9}$ & 2559.55 \\
		M87* & RN
		& 7.09808 & 10.33980
		& $2.43124\times10^{9}$ & 2558.95 \\
		M87* & Schwarzschild
		& 6.00000 & 9.55093
		& $8.41144\times10^{8}$ & 1962.56 \\
		\midrule
		Sgr A* & LQRN
		& 7.09632 & 10.33664
		& $3.89791\times10^{10}$ & 5120.50 \\
		Sgr A* & RN
		& 7.09808 & 10.33980
		& $3.89423\times10^{10}$ & 5119.29 \\
		Sgr A* & Schwarzschild
		& 6.00000 & 9.55093
		& $1.34730\times10^{10}$ & 3926.18 \\
		\bottomrule
	\end{tabular}
	\endgroup
\end{table}
The dimensionless ISCO radii and location of the peak values of the fluxes are the same for the two BHs when the spacetime and particle parameters are fixed, whereas the dimensional fluxes and temperatures differ significantly because of the chosen masses and accretion rates.  
The difference of the data between LQRNBH and RNBH remains below $0.1\%$ and the loop quantum effect is still negligible when $\zeta=0.1$. In contrast, the peak value of the flux of LQRNBH  is about $2.9$ times of that of Schwarzschild black hole, which further confirms the electromagnetic coupling has significant influence on the properties of accretion disk.

In actual observation, due to the moving of gas in the disk and the strong gravitational effect of the black hole,  the observed radiation flux by a far observer is affected by both Doppler and gravitational redshifts. The observed radiation flux can be expressed as \cite{Page:1974he,Luminet:1979nyg, Huang:2023ilm,Wu:2024sng}
	\begin{equation}
	F_{\text{obs}} = \frac{F}{(1 + z)^4},
	\end{equation}
where the redshift factor $1+z$ is defined as
	\begin{equation}\label{redshft}
	    1+z= \frac{1 + \Omega b \sin\theta_0 \sin\alpha}{\sqrt{-g_{tt} - \Omega^2 g_{\phi\phi}}},
	\end{equation}
where $\theta_0$ is the inclination angle of the observer, $\alpha$ is the polar angle on the observer’s imaging plane, and $b$ is the impact parameter of the photon\cite{Luminet:1979nyg, Shu:2024tut}. 

\begin{figure}[t]
	\centering
	\begin{subfigure}[t]{0.32\textwidth}
		\centering
		\includegraphics[width=\textwidth]{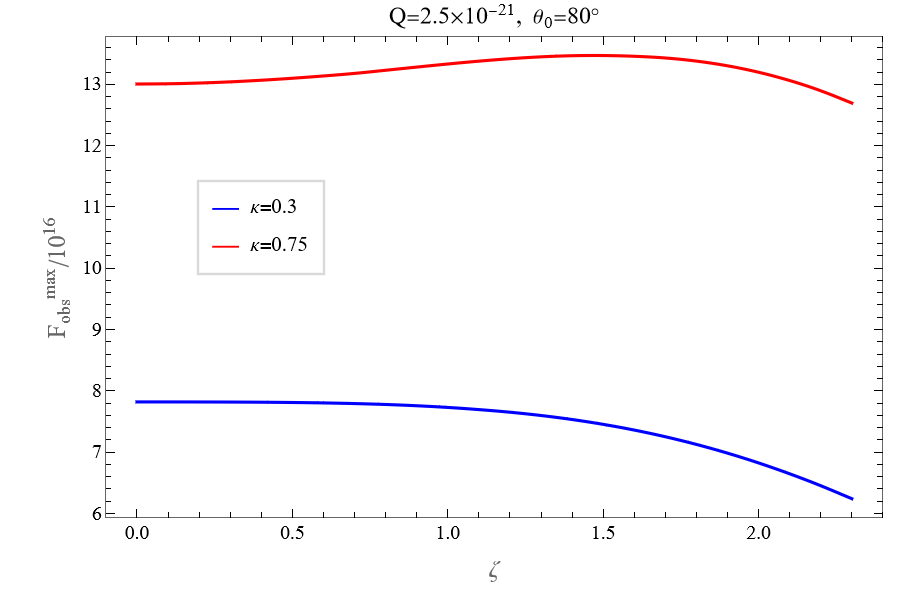}
		\caption{}
		\label{fig:Fobsmax-zeta}
	\end{subfigure}
	\hfill
	\begin{subfigure}[t]{0.32\textwidth}
		\centering
		\includegraphics[width=\textwidth]{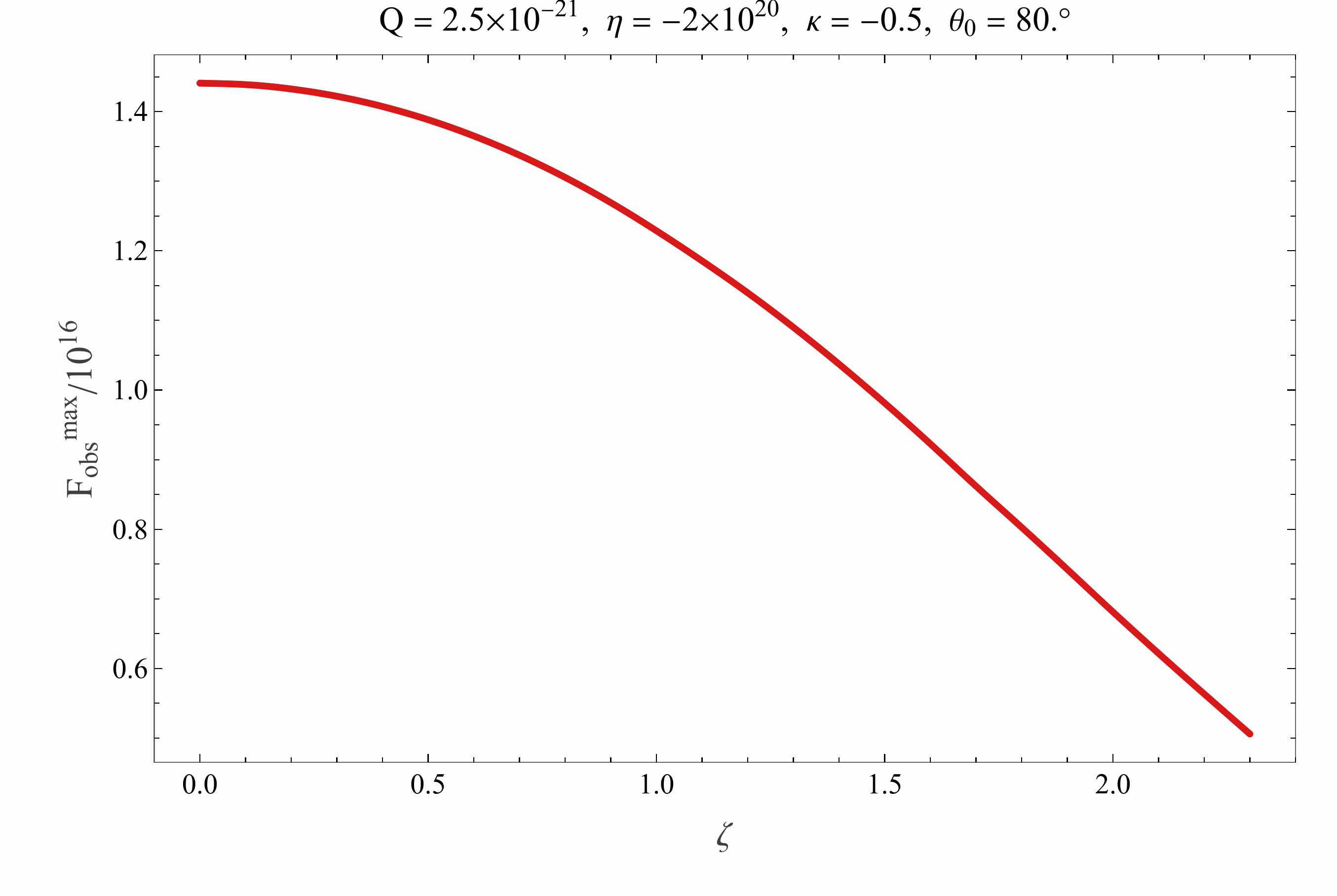}
		\caption{}
		\label{fig:Fobsmax-Q}
	\end{subfigure}
	\hfill
	\begin{subfigure}[t]{0.32\textwidth}
		\centering
		\includegraphics[width=\textwidth]{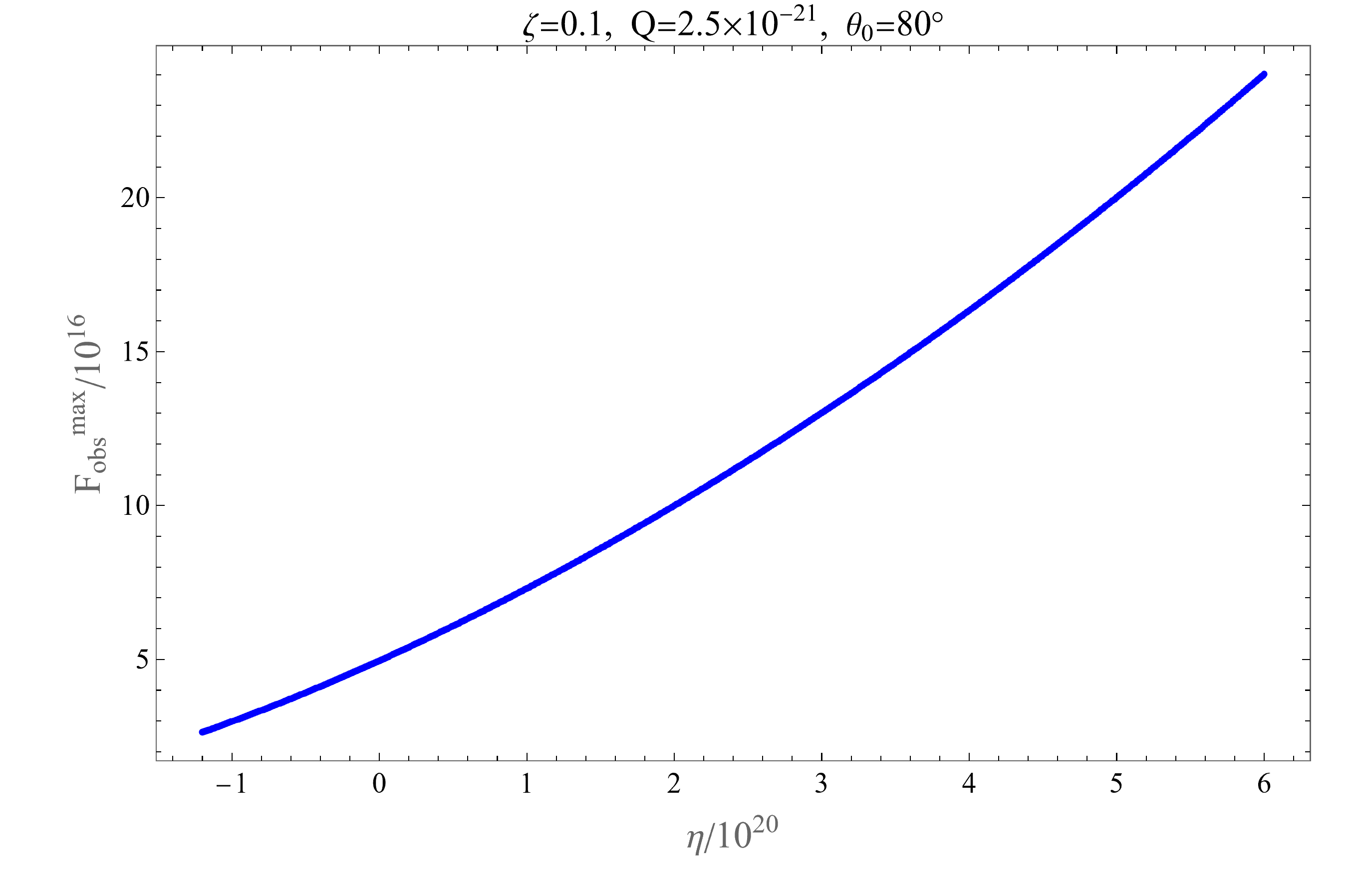}
		\caption{}
		\label{fig:Fobsmax-eta}
	\end{subfigure}
	\caption{
		The dependence of the maximum value of the observed flux $F_{\mathrm{obs}}^{\max}$ on various model parameters.
		}
	\label{fig:Fobspart}
\end{figure}

In Fig. \ref{fig:Fobspart}, we show the  maximum values of the observed flux $F_{\text{obs}}^{\max}$ of the LQRNBH accretion disk and their dependence on $\zeta$ and $\eta$. The mass accretion rate $\dot{M}$ is chosen as $2\times10^{-3}M_\odot/\text{yr}$  and the black hole mass is taken as $M=2\times 10^6 M_\odot$. The unit of the flux is $\mathrm{erg\,cm^{-2}\,s^{-1}}$. The observation angle is taken as $\theta_0=80^\circ$. 

The results in left and middle panels show that the monotonicity of $F_{\text{obs}}^{\max}$ as a function of $\zeta$ depends on the value of $\kappa$. We find that this function decreases monotonically for $-1<\kappa<0.33$, while for $\kappa>0.33$ it becomes non-monotonic, increasing first and then decreasing. 
The result in the right panel shows that the maximum of the observed radiation flux increases monotonically as the increase of $\kappa $ when the electromagnetic interaction is repulsive, i.e. $\kappa>0$.
 Compared with the $\kappa=0$ case, the maximum value of the observed flux increases by approximately $ 100\%$ when $\kappa= 0.5 $. When the electromagnetic interaction is attractive, i.e. $\kappa<0$, the maximum of the observed flux decreases monotonically as the increase of $|\kappa| $. 
Compared with the $\kappa=0$ case, the maximum value of the observed flux decreases by approximately $ 40\%$ when $\kappa=-0.25$. Since the value of $Q$ is very small in our discussion, the direct $Q^2$ correction to the metric is negligible, and thus the remarkable influence on the observed flux arises from the electromagnetic coupling. 
These results further confirm that the electromagnetic coupling $\kappa$ has significant impact on the physics of the accretion disk.

	\section{Optical Appearance and Images of the LQRNBH}

In this section, we investigate the observational appearance of the LQRNBH surrounded by the accretion disk and discuss the influence of the model parameters on the images. 
As we introduced before, our chosen accretion disk model is optically thick and thus we just consider the direct images of redshift factors and observed fluxes in this section.  
Since the LQRNBH is spherically symmetric, the method used in this section is the backward ray-tracing approach which follows Refs. \cite{Luminet:1979nyg,Tian:2019yhn,Shu:2024tut}.

We first provide a brief review of the method.  We assume the observer is far from the black hole and the observation plane is a flat plane with polar coordinates $(b, \alpha)$ which are introduced under Eq.\eqref{redshft}. The coordinates of a point emitter on the accretion disk are $(r,\phi)$. The deflection angle of light traveling from its emitter to its image point on the observation plane is $\gamma$. Then, the trigonometric relation between $\alpha$ and $\gamma$ is
	\begin{equation}
	\cot\gamma=\cos\alpha \tan\theta_0,
	\label{angle_relation}
	\end{equation}
	where $\alpha\in[0,2\pi]$, $\theta_0\in[0,\pi/2]$, and $\gamma\in[\pi/2-\theta_0,\pi/2+\theta_0]$. On the trajectory plane of the light, its trajectory equation is 
	\begin{equation}
	    \left(\frac{\mathrm{d} u}{\mathrm{d} \varphi } \right)^2=\frac{1}{b^2} -u^2(1-2 M u+Q^2 u^2)\left(1+\zeta^2 u^2(1-2 M u+Q^2 u^2)\right),
	    \label{geo_eq}
	\end{equation}
where $u=\frac{1}{r}$, and $\varphi$ is the polar angle coordinate of a point on the trajectory. Here, $b$ is also the impact parameter of the light in the backward ray-tracing method. In order to obtain the image of a circular orbit in the disk (i.e. the isoradial curve in \cite{Luminet:1979nyg}), we need the relation between the coordinates $b$ and $\alpha$ of an image point in the isoradial curve on the observation plane, which can be numerically derived from the above trigonometric equation and trajectory equation with the deflection angle $\gamma$ serving as a bridge quantity.

\begin{figure}[htbp]
	\centering
	\includegraphics[width=0.35\textwidth]
	{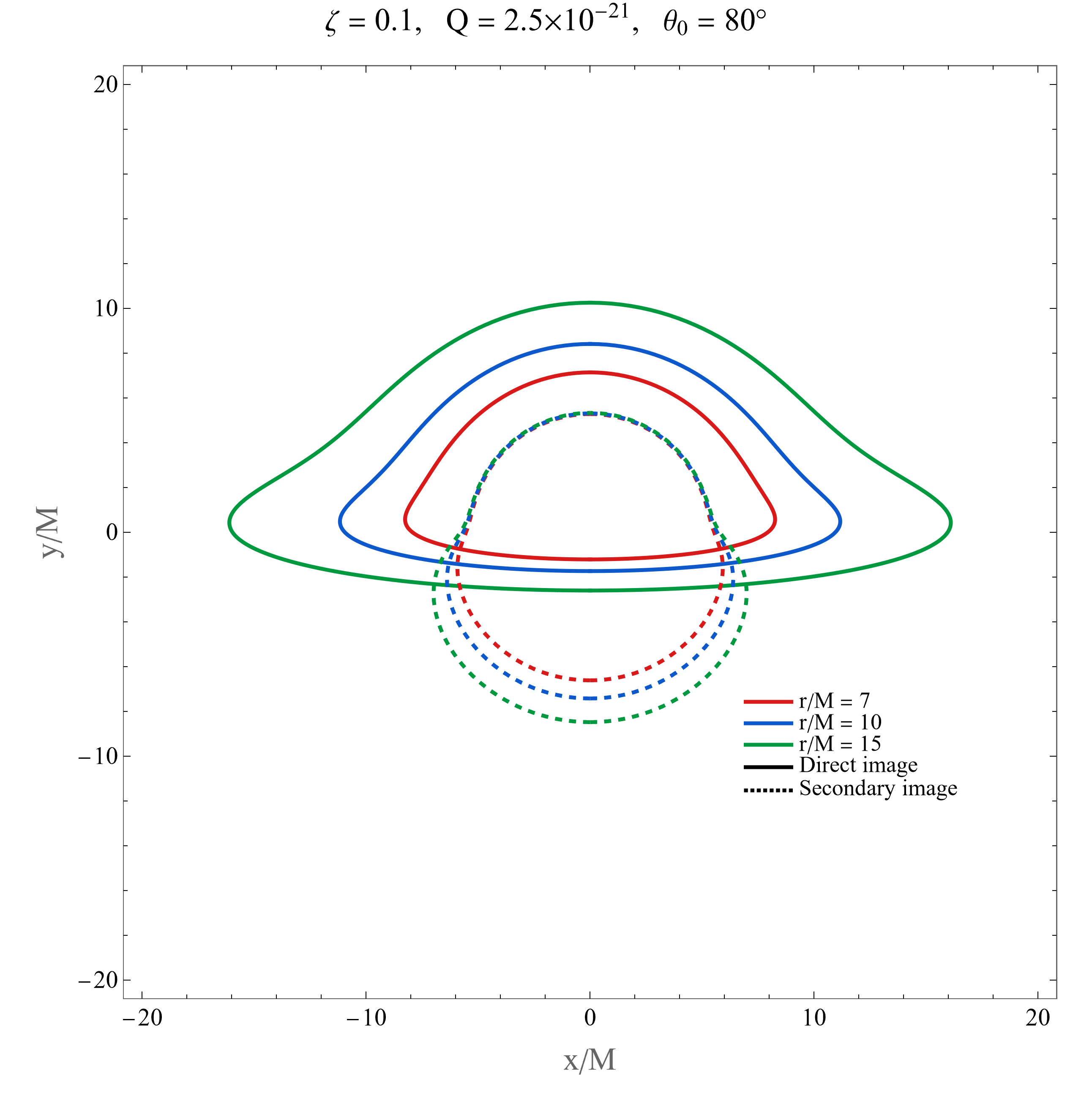}
	\caption{Representative isoradial curves for the case with $\theta_0=80^\circ$, $\zeta=0.1$,
		and $Q=2.5\times10^{-21}$.}
	\label{fig:isoradius}
\end{figure}
In Fig.\ref{fig:isoradius}, we illustrate the direct and secondary images of three typical circular orbits on the accretion disk around the LQRNBH with fixed model parameters.
Here we do not show the influence of charge parameter $Q$ on these curves since it is very tiny in our considered parameter ranges ($Q\lesssim 10^{-21}$).

\begin{figure}[htbp]
	\centering
	\begin{subfigure}[b]{0.31\textwidth}
		\centering
		\includegraphics[width=\textwidth]{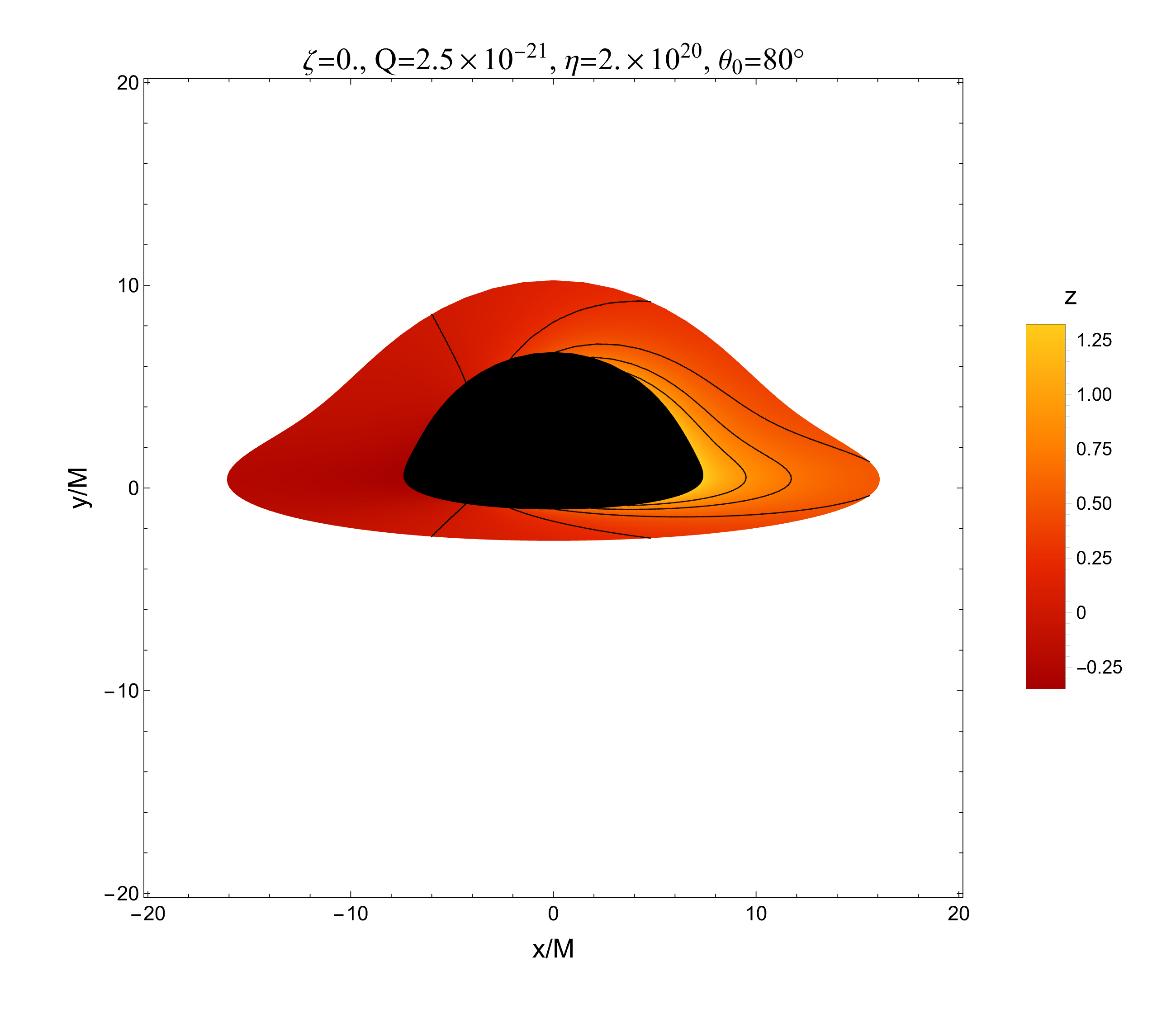} 
		\label{fig:80redz0Q05}
	\end{subfigure}
	\hfill
	\begin{subfigure}[b]{0.31\textwidth}
		\centering
		\includegraphics[width=\textwidth]{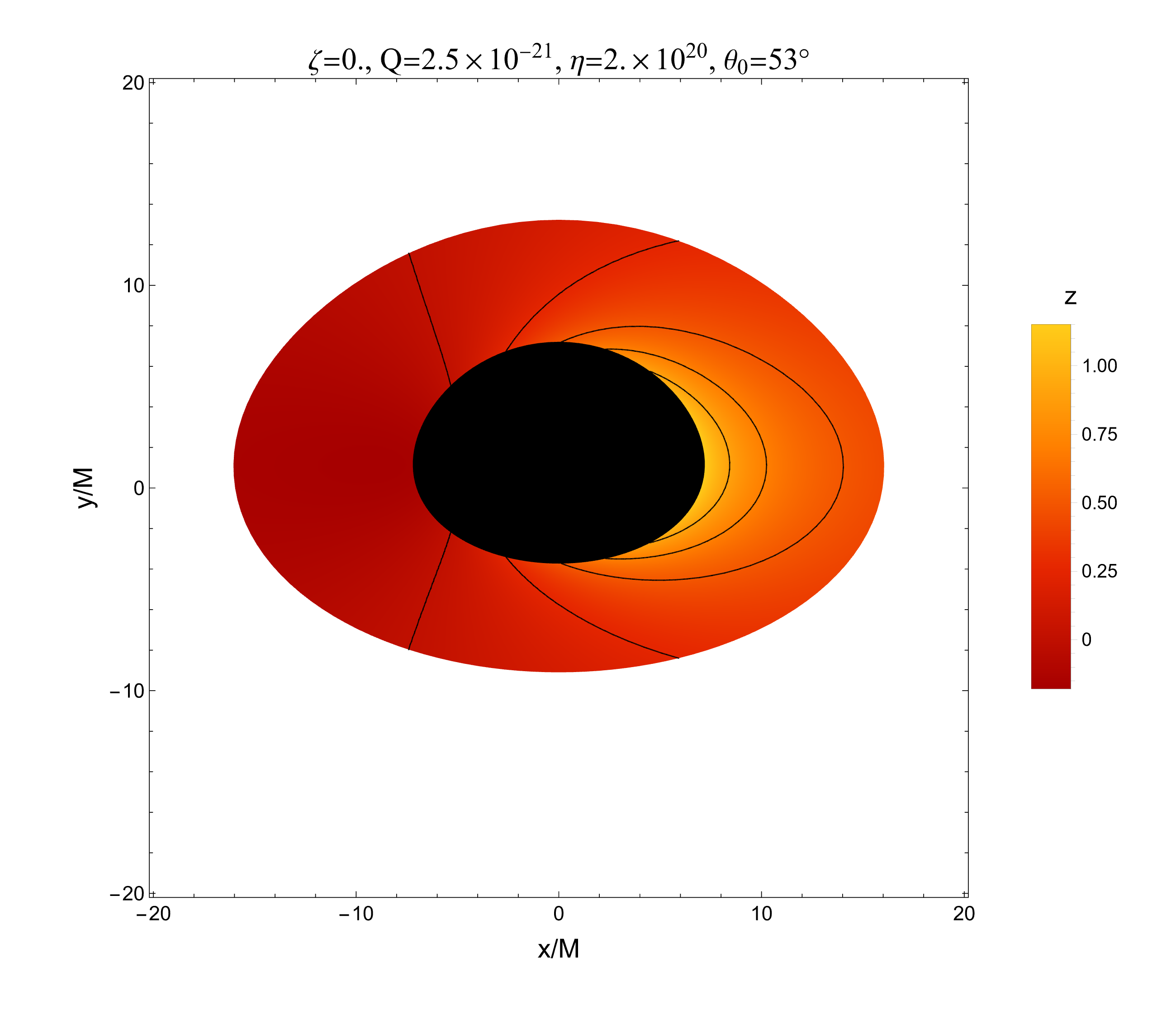} 
		\label{fig:53redz0Q05}
	\end{subfigure}
	\hfill
	\begin{subfigure}[b]{0.31\textwidth}
		\centering
		\includegraphics[width=\textwidth]{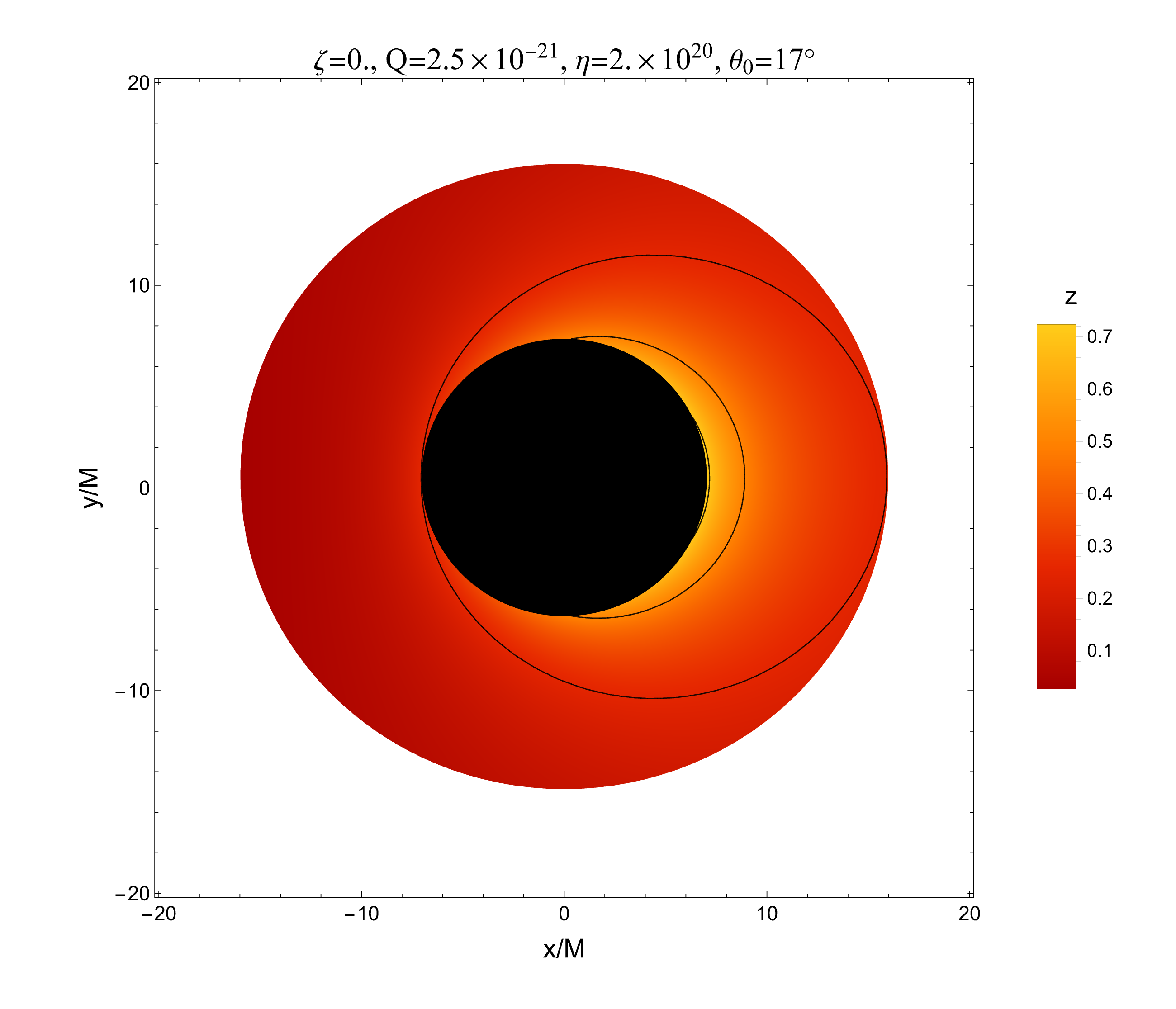}
		\label{fig:17redz0Q05}
	\end{subfigure}
	\par\medskip
	
	\begin{subfigure}[b]{0.31\textwidth}
		\centering
		\includegraphics[width=\textwidth]{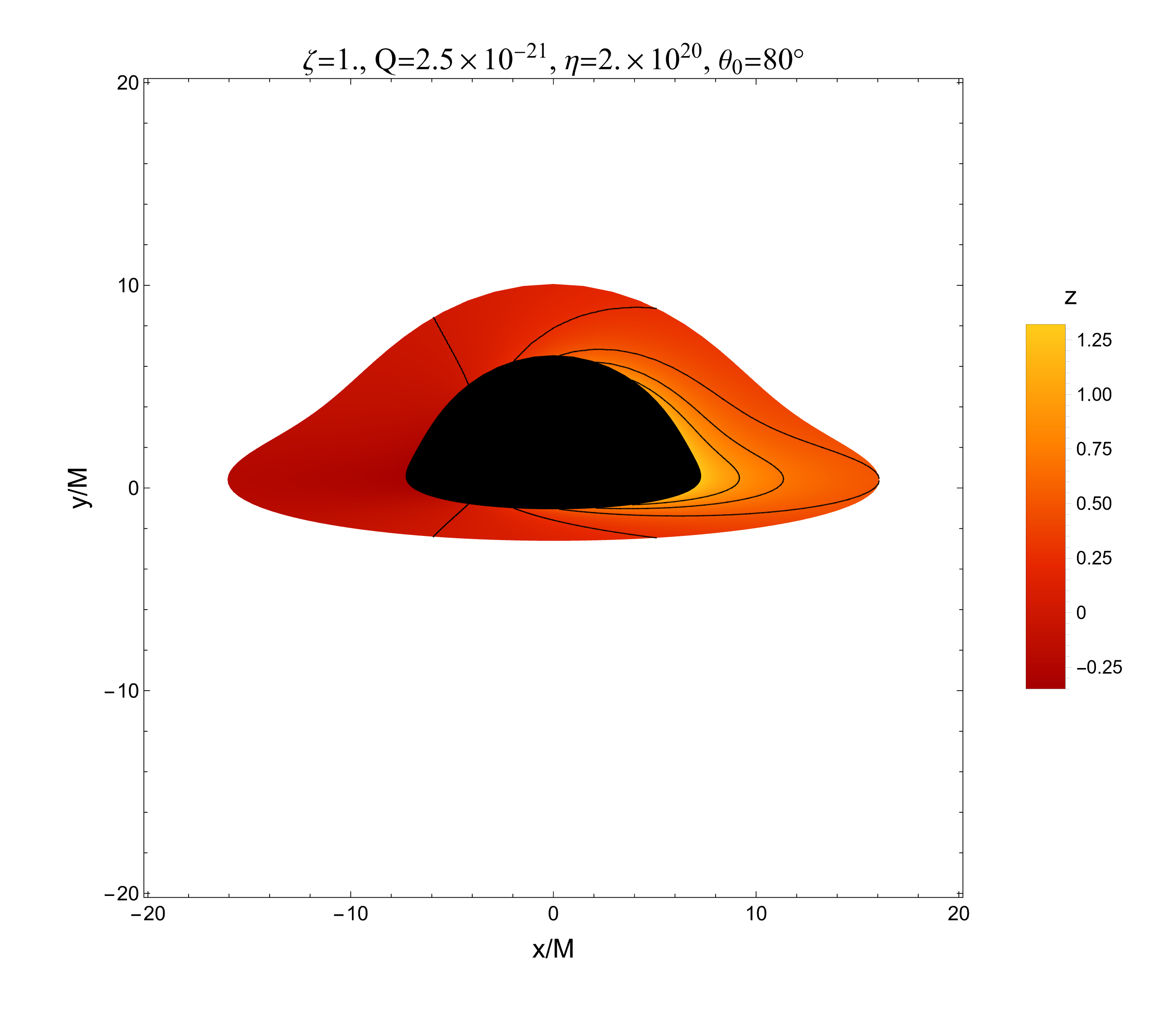}
		\label{fig:80redz15Q05a}
	\end{subfigure}
	\hfill
	\begin{subfigure}[b]{0.31\textwidth}
		\centering
		\includegraphics[width=\textwidth]{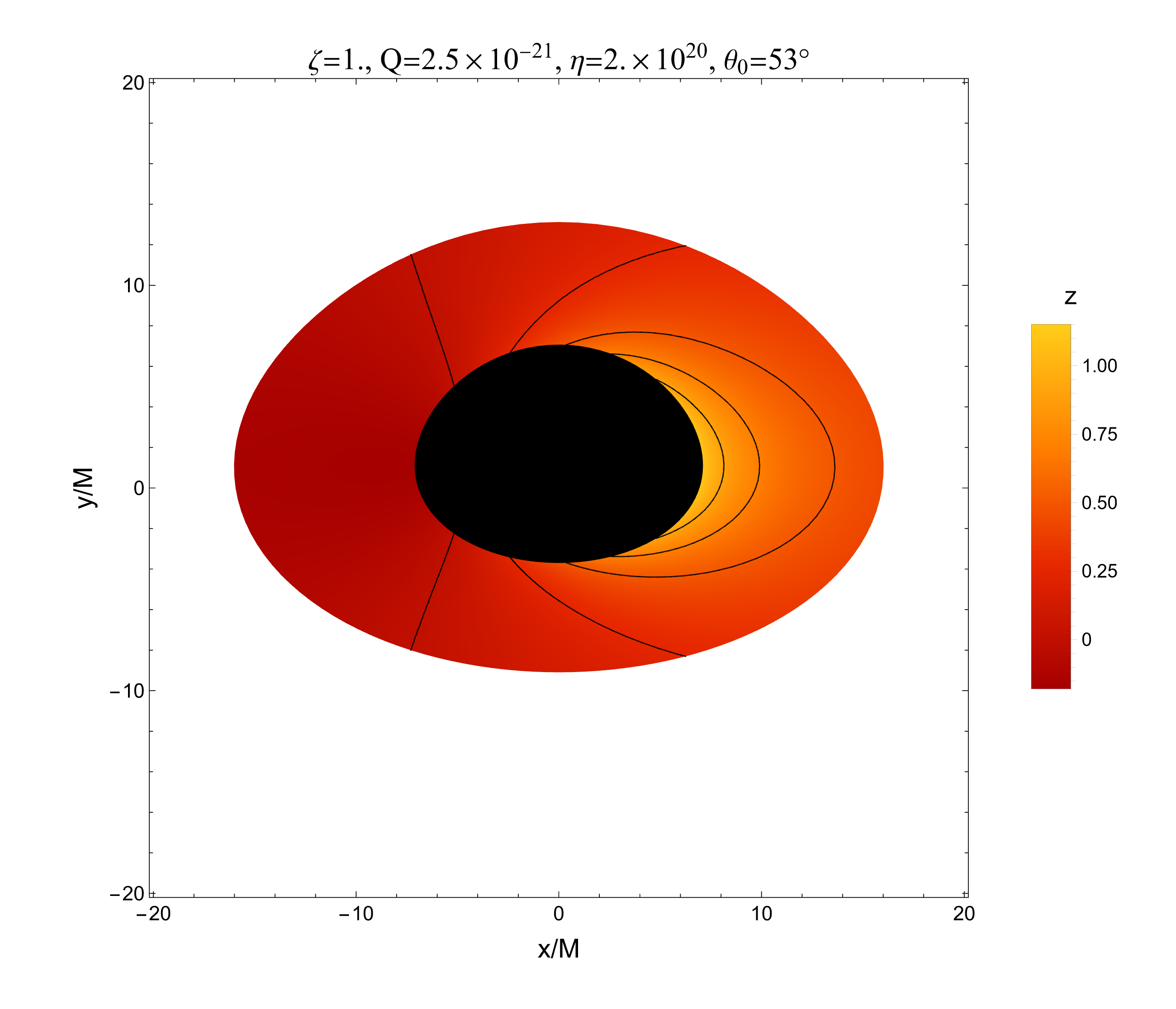} 
		\label{fig:53redz15Q05a}
	\end{subfigure}
	\hfill
	\begin{subfigure}[b]{0.31\textwidth}
		\centering
		\includegraphics[width=\textwidth]{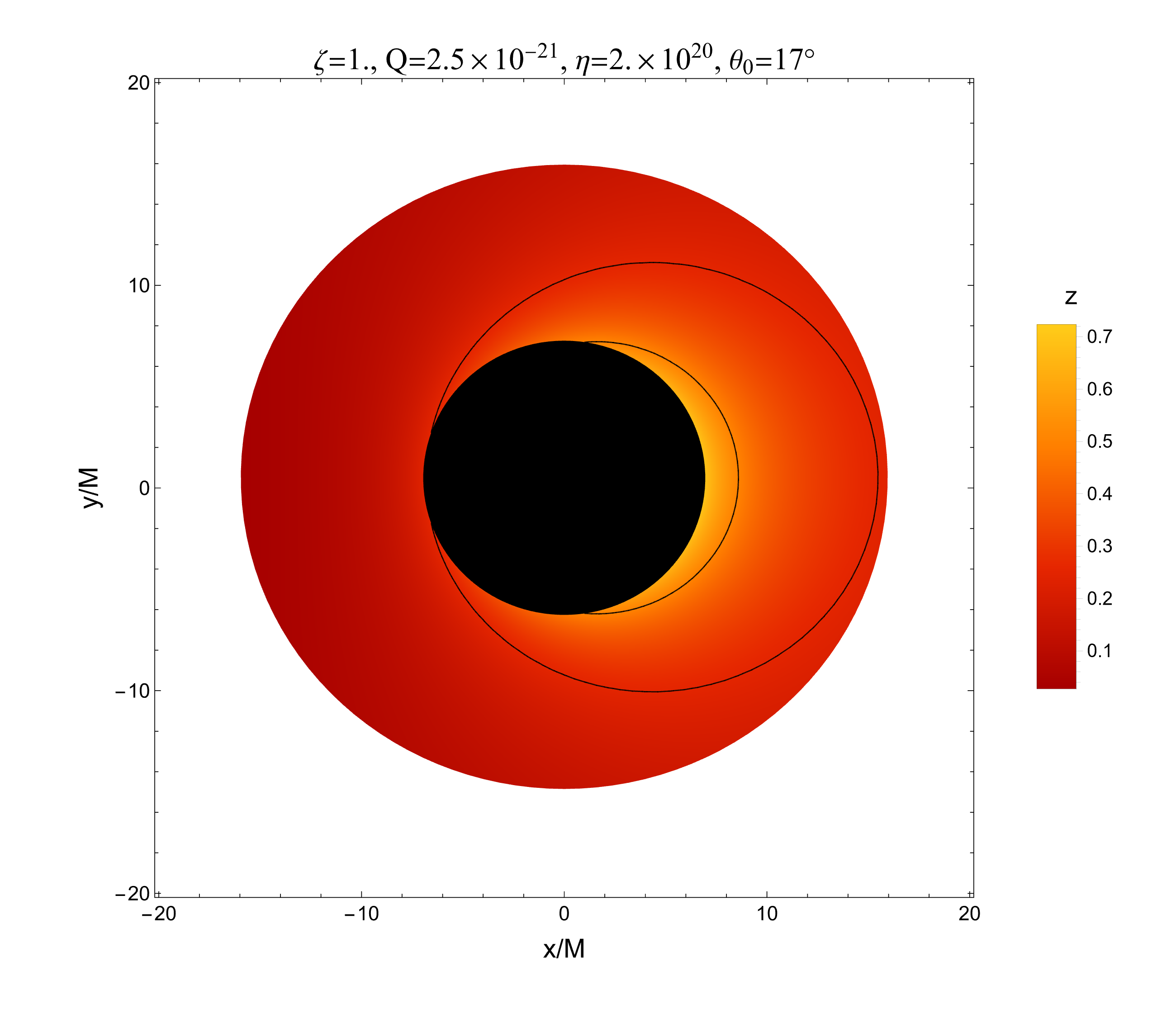}
		\label{fig:17redz15Q05a}
	\end{subfigure}
	
	\caption{Influence of $\zeta$ on the direct images of the redshift factor $z$ at different observation angles.}
	\label{fig:Redshift diffz}
\end{figure}

The redshift factor $z$ is defined in Eq.~\eqref{redshft}. It contains both the gravitational redshift caused by the black hole and the Doppler redshift caused by the relative motion of the particles in the accretion disk with respect to the observer. In Fig. \ref{fig:Redshift diffz}, we show the images of the redshift factors at different observation angles and the influence of the quantum parameter $\zeta$ on them, and we also plot the redshift contours with $z=0,0.25,0.5,0.7,0.9$.
The electromagnetic coupling is fixed at $\kappa=0.5$. $\zeta=0$ for the images in the first row and  $\zeta=1$ for the images in the second row. 
We see that the effects of $\zeta$ on the images with different inclination angles are almost negligibly small. However, we notice that compared with the $\zeta=0$ case, the $z$ contours move inward to the black hole when $\zeta=1$, which means the quantum parameter reduces the maximum values of the redshift factor.

\begin{figure}[htbp]
	\centering
	\begin{subfigure}[b]{0.31\textwidth}
		\centering
		\includegraphics[width=\textwidth]{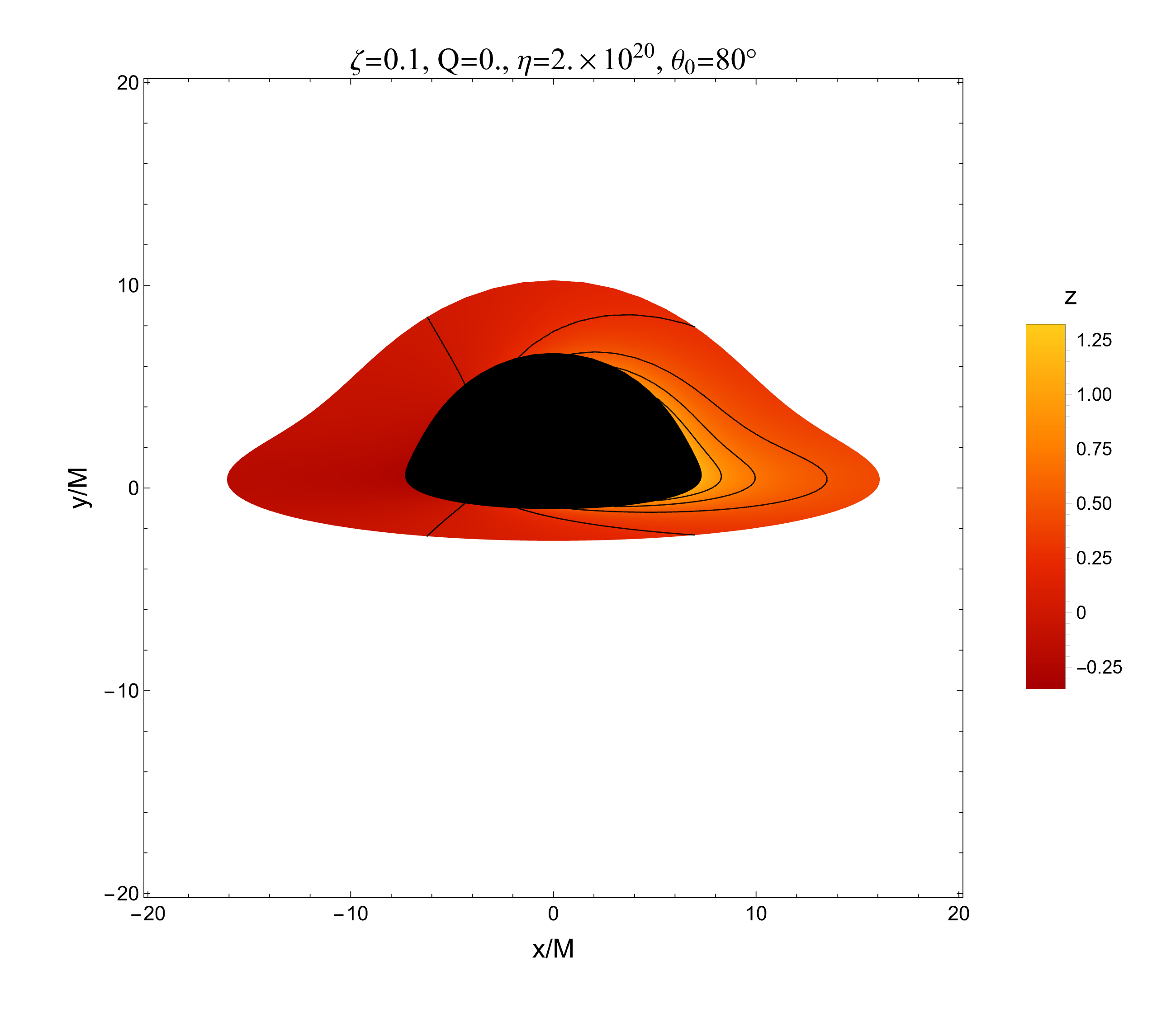} 
		\label{fig:80redz15Q03}
	\end{subfigure}
	\hfill
	\begin{subfigure}[b]{0.31\textwidth}
		\centering
		\includegraphics[width=\textwidth]{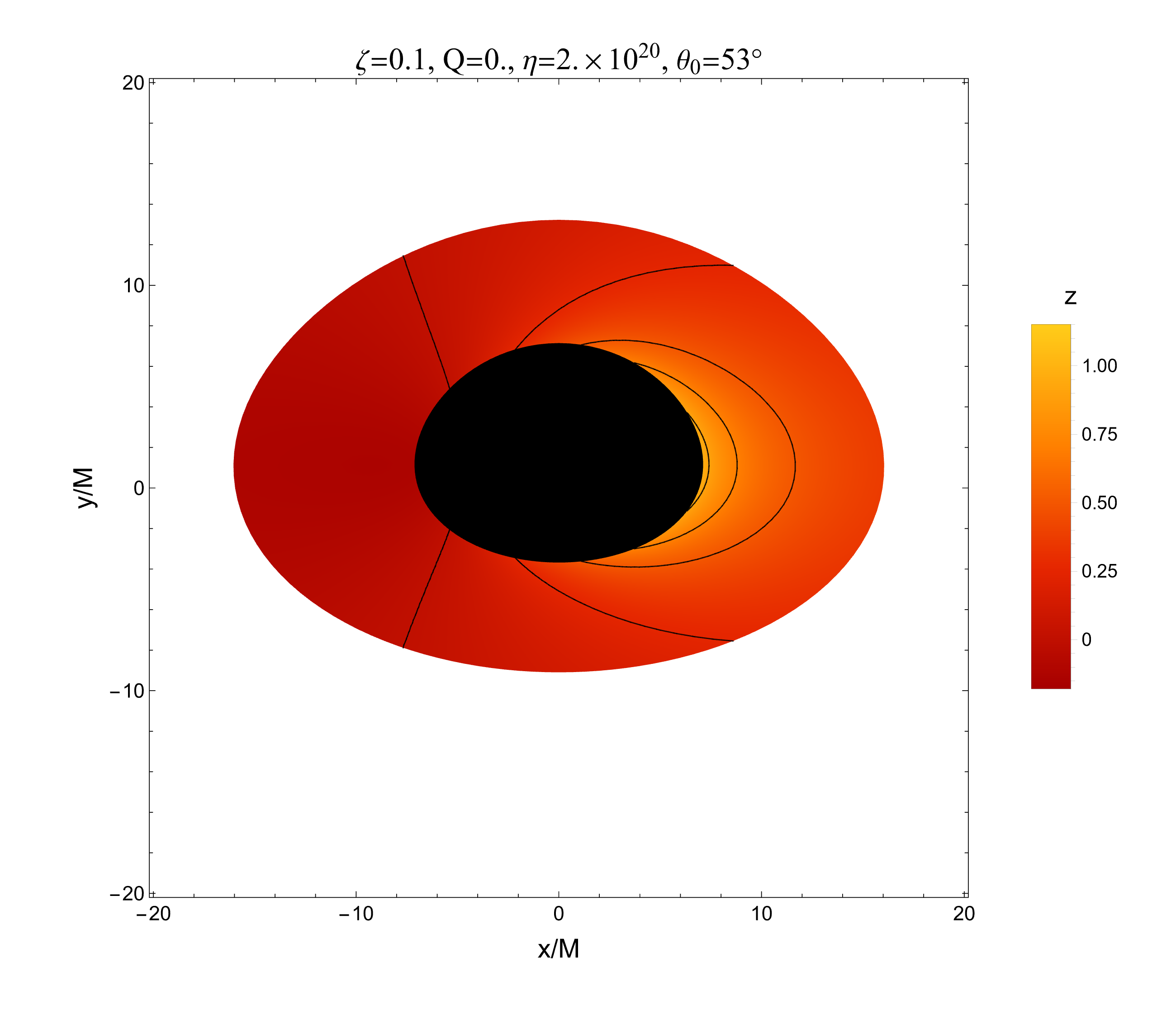}
		\label{fig:53redz15Q03}
	\end{subfigure}
	\hfill
	\begin{subfigure}[b]{0.31\textwidth}
		\centering
		\includegraphics[width=\textwidth]{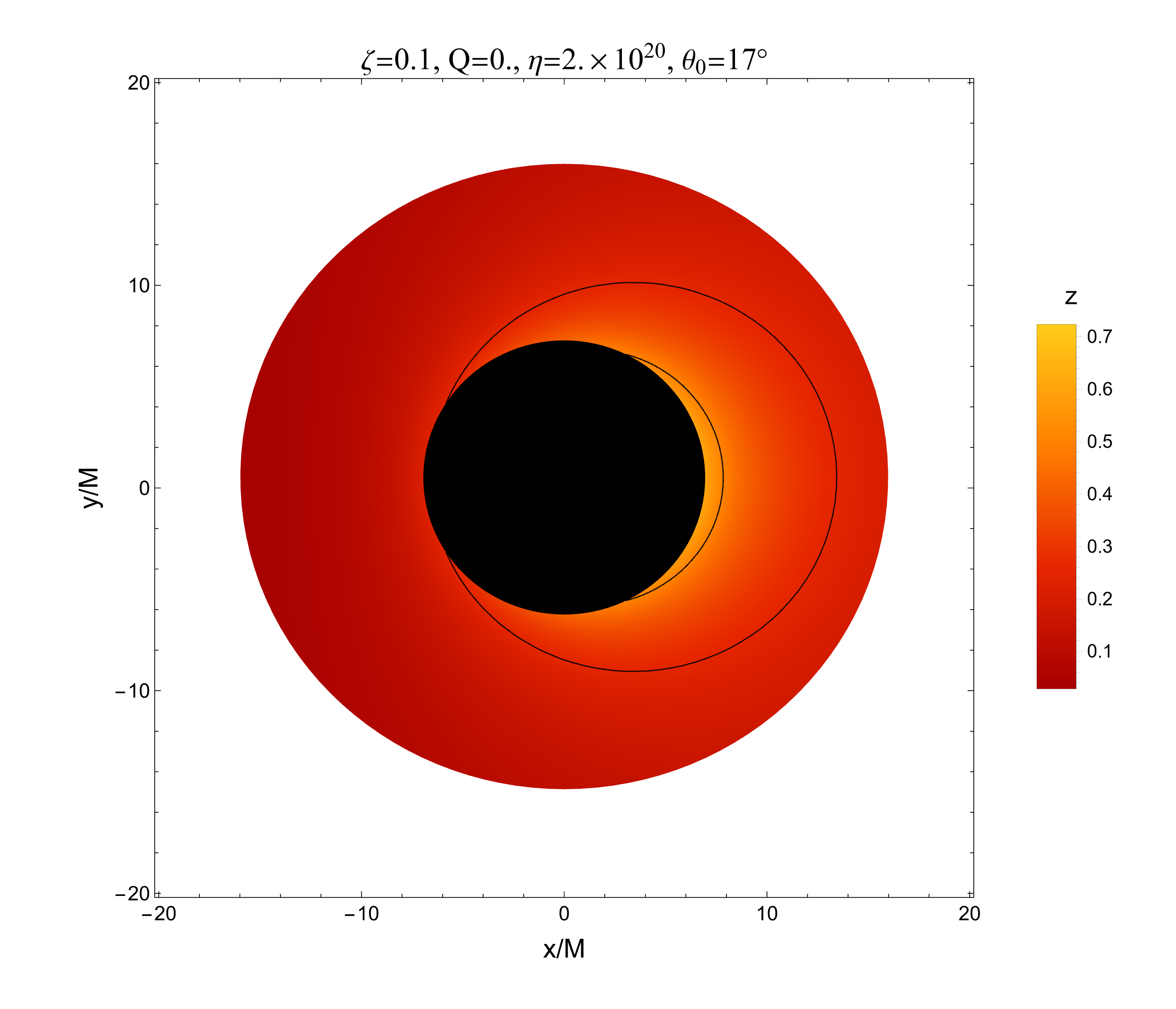}
		\label{fig:17redz15Q03}
	\end{subfigure}
	\par\medskip
	
	\begin{subfigure}[b]{0.31\textwidth}
		\centering
		\includegraphics[width=\textwidth]{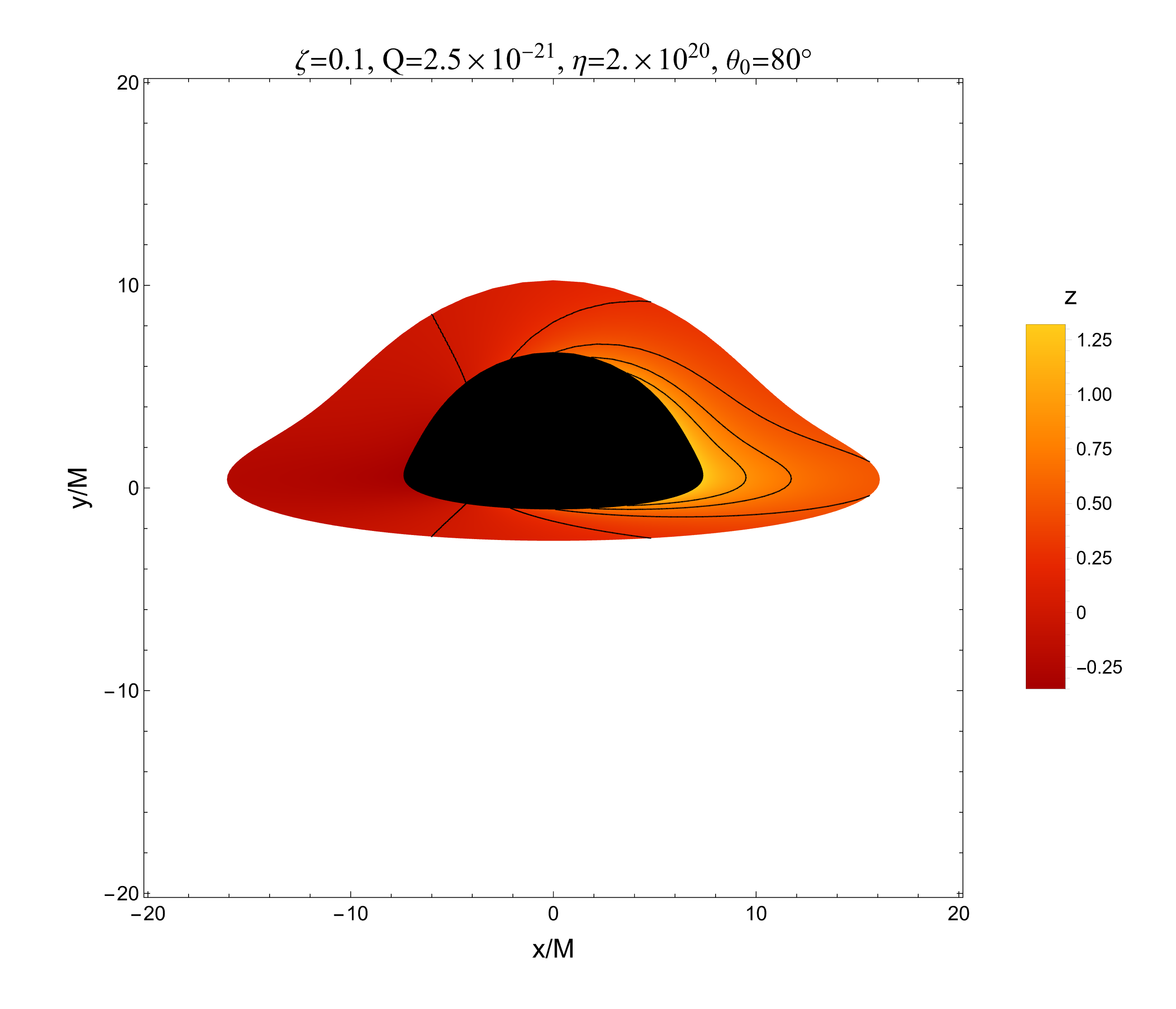}
		\label{fig:80redz15Q05}
	\end{subfigure}
	\hfill
	\begin{subfigure}[b]{0.31\textwidth}
		\centering
		\includegraphics[width=\textwidth]{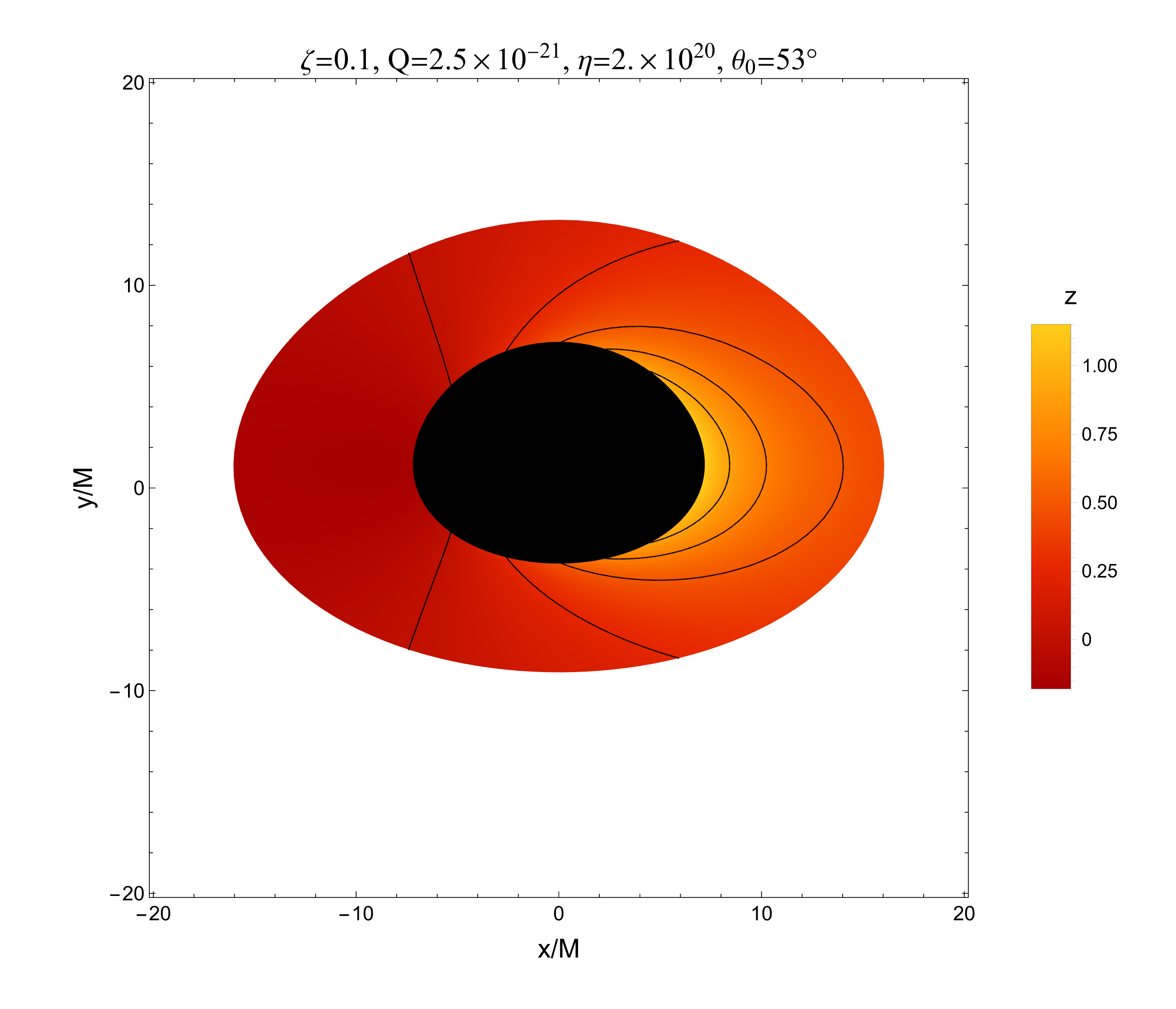}
		\label{fig:53redz15Q05}
	\end{subfigure}
	\hfill
	\begin{subfigure}[b]{0.31\textwidth}
		\centering
		\includegraphics[width=\textwidth]{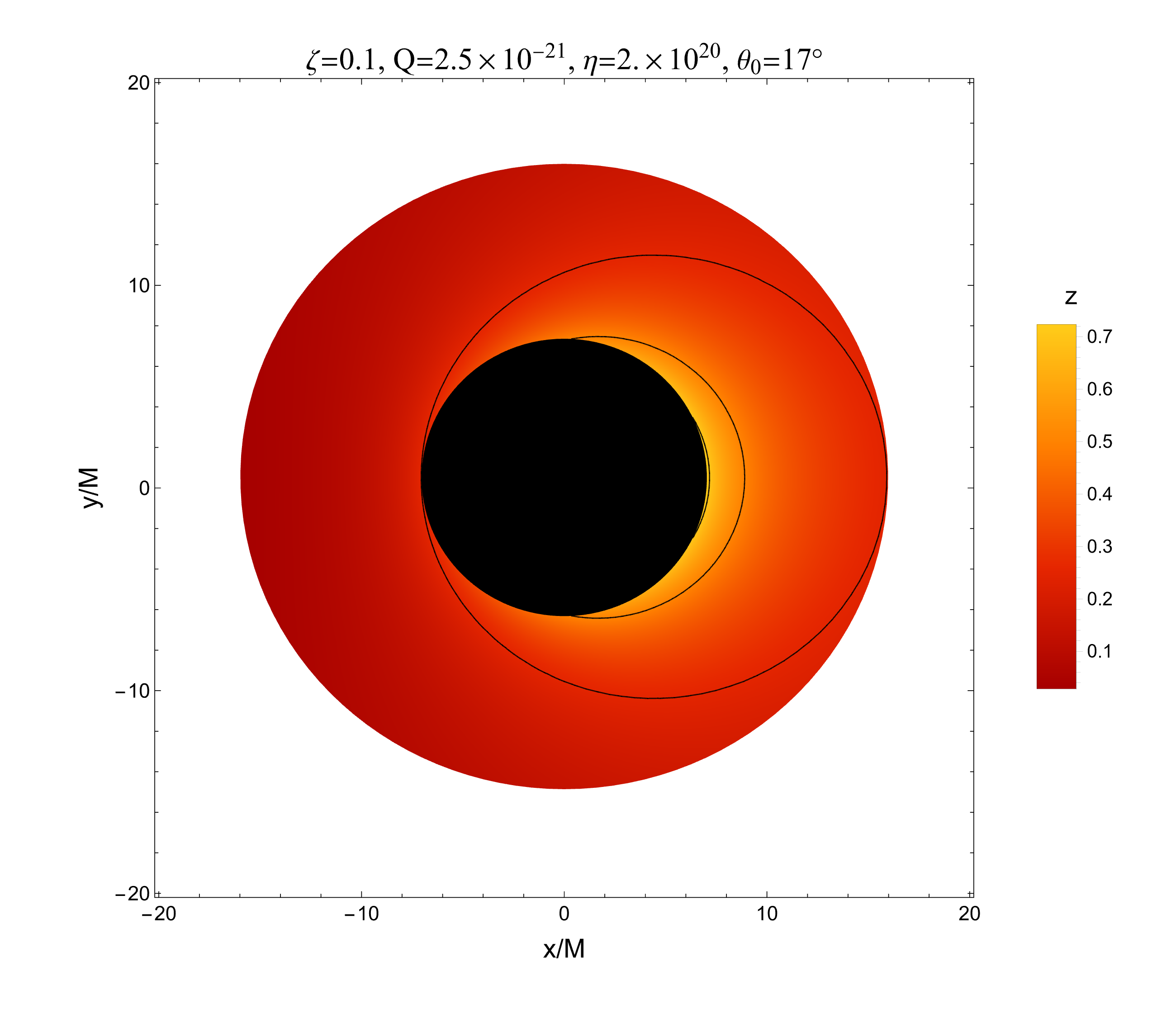}
		\label{fig:17redz15Q05}    
	\end{subfigure}
	
	\caption{Influence of $Q$ on the direct images of the redshift factor $z$ at different observation angles when $\kappa>0$.}
	\label{fig:Redshift diffQ}
\end{figure}

In Fig. \ref{fig:Redshift diffQ}, we show the images of the redshift factors and the influence of charge parameter $Q$ on them. $Q=0$ ($\kappa=0$) for the images in the first row and  $Q=2.5\times10^{-21}$ ( $\kappa=0.5$) for the images in the second row. The redshift contours are $z=0,0.25,0.5,0.7,0.9$. We can see that the increase of $Q$ has remarkable effect on the redshift factors. The high-redshift region extends obviously and the redshift contours move outward remarkably. The redshift factor increases as the increase of $Q$. 

\begin{figure}[htbp]
	\centering
	\begin{subfigure}[b]{0.31\textwidth}
		\centering
		\includegraphics[width=\textwidth]{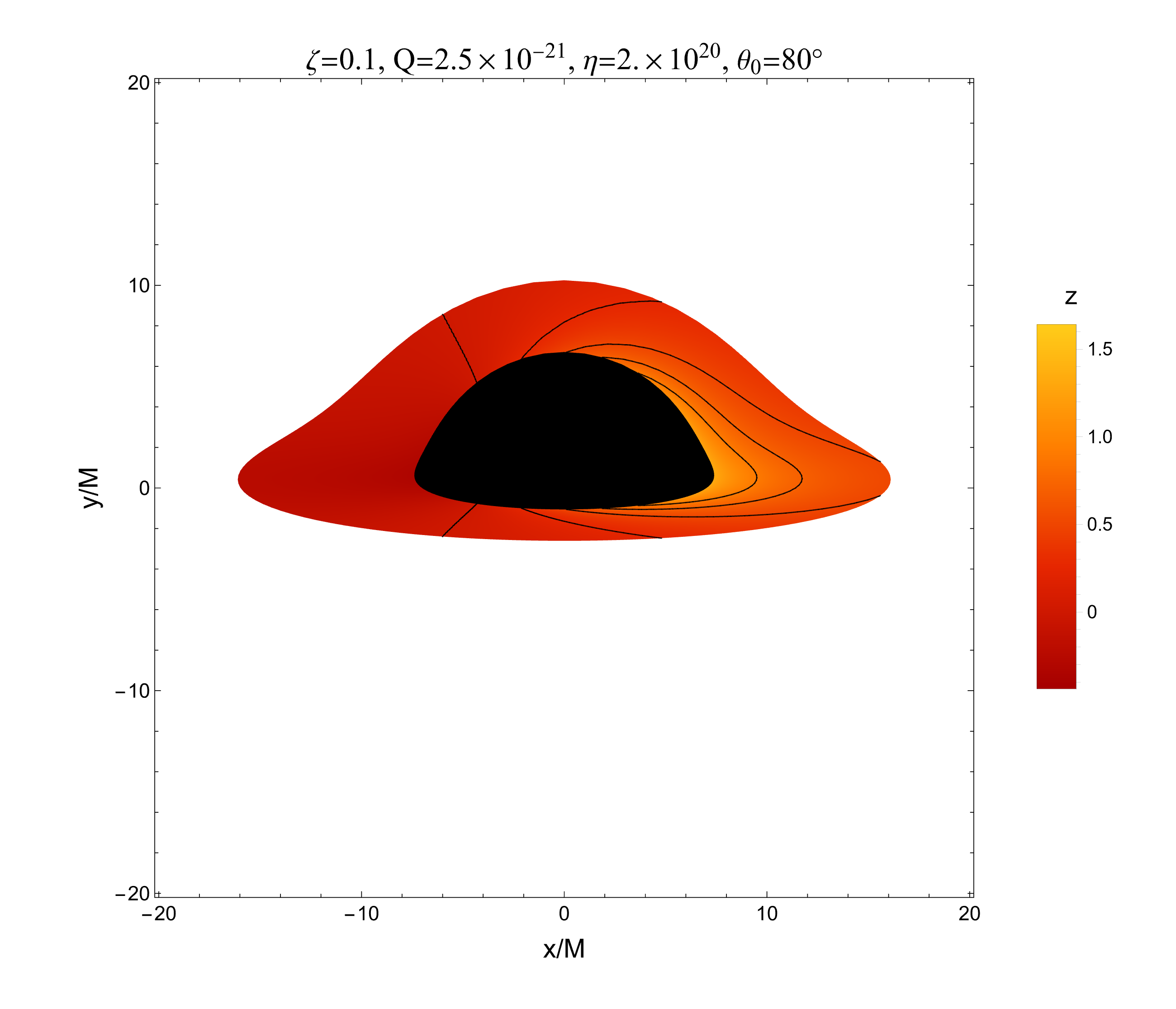}
	\end{subfigure}
	\hfill
	\begin{subfigure}[b]{0.31\textwidth}
		\centering
		\includegraphics[width=\textwidth]{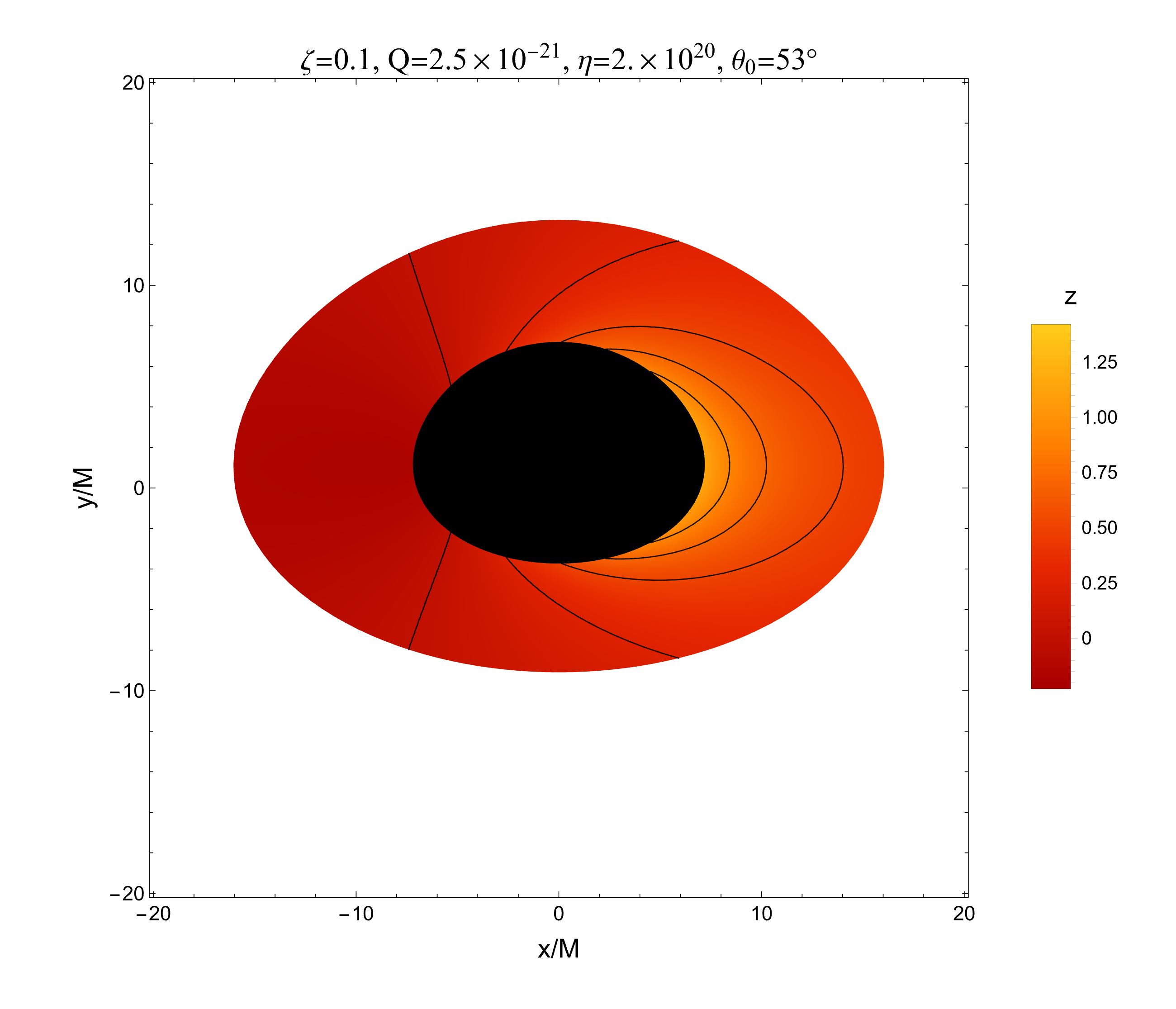}
	\end{subfigure}
	\hfill
	\begin{subfigure}[b]{0.31\textwidth}
		\centering
		\includegraphics[width=\textwidth]{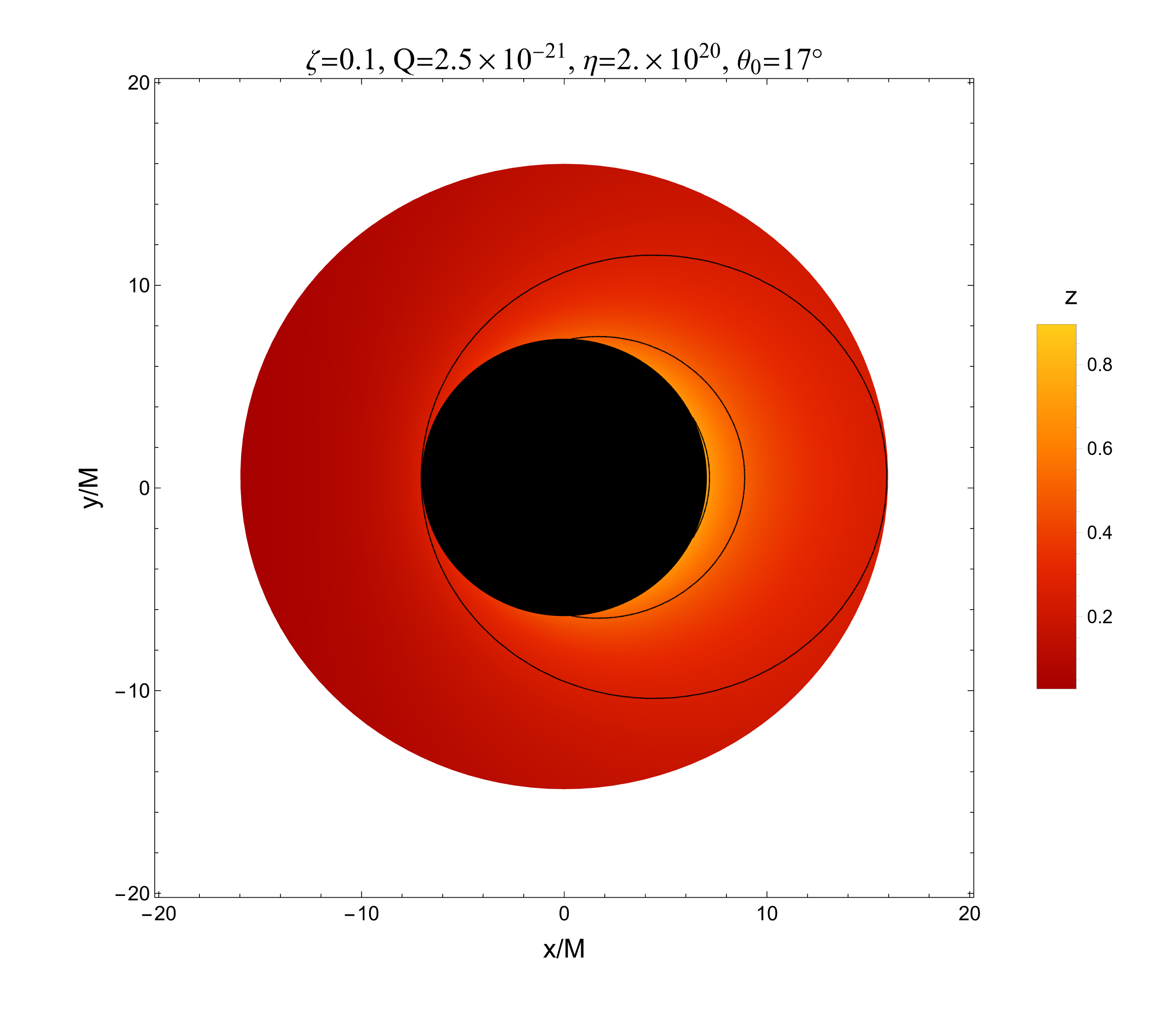}
	\end{subfigure}
	\par\medskip
	
	\begin{subfigure}[b]{0.31\textwidth}
		\centering
		\includegraphics[width=\textwidth]{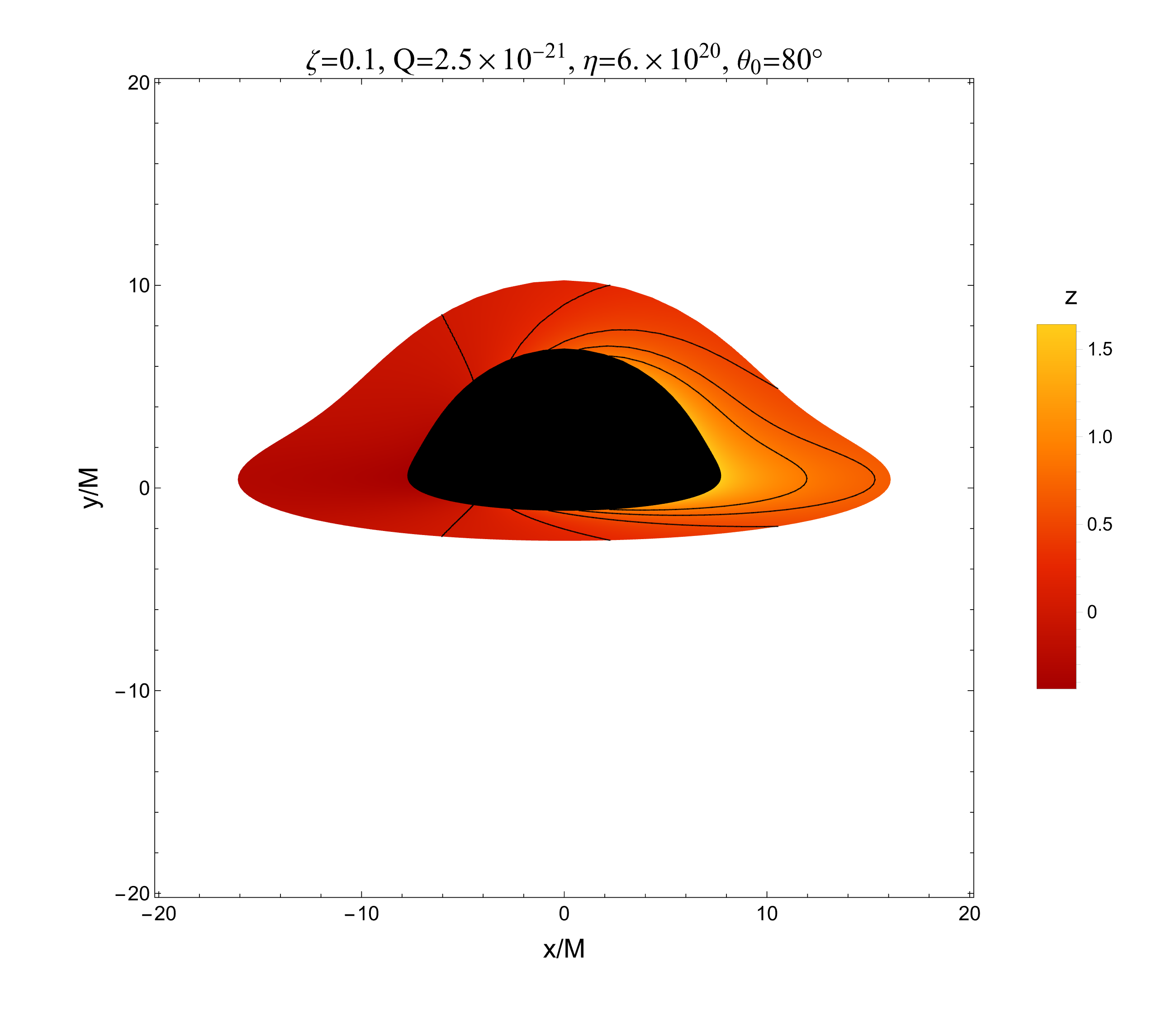}
	\end{subfigure}
	\hfill
	\begin{subfigure}[b]{0.31\textwidth}
		\centering
		\includegraphics[width=\textwidth]{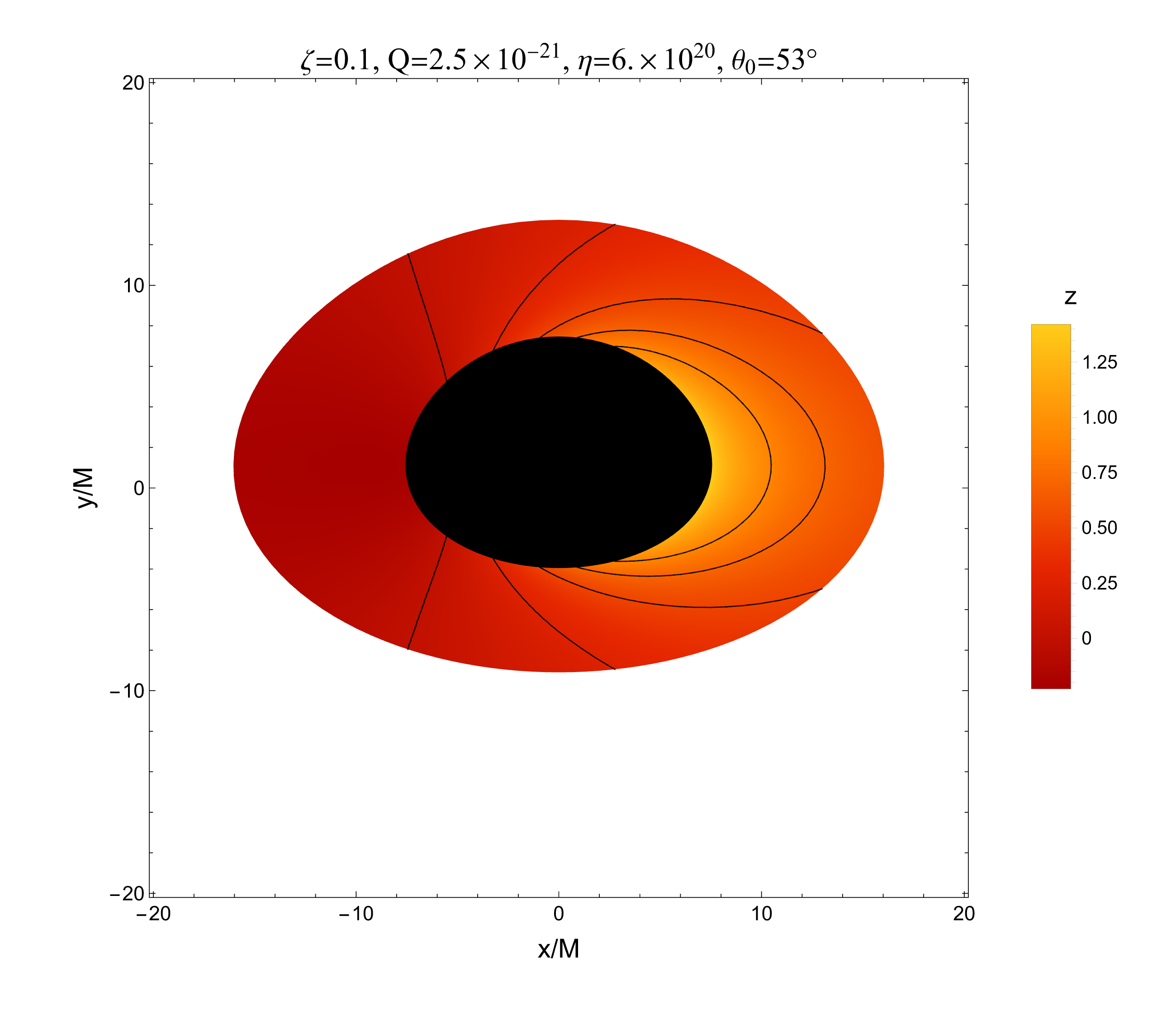}
	\end{subfigure}
	\hfill
	\begin{subfigure}[b]{0.31\textwidth}
		\centering
		\includegraphics[width=\textwidth]{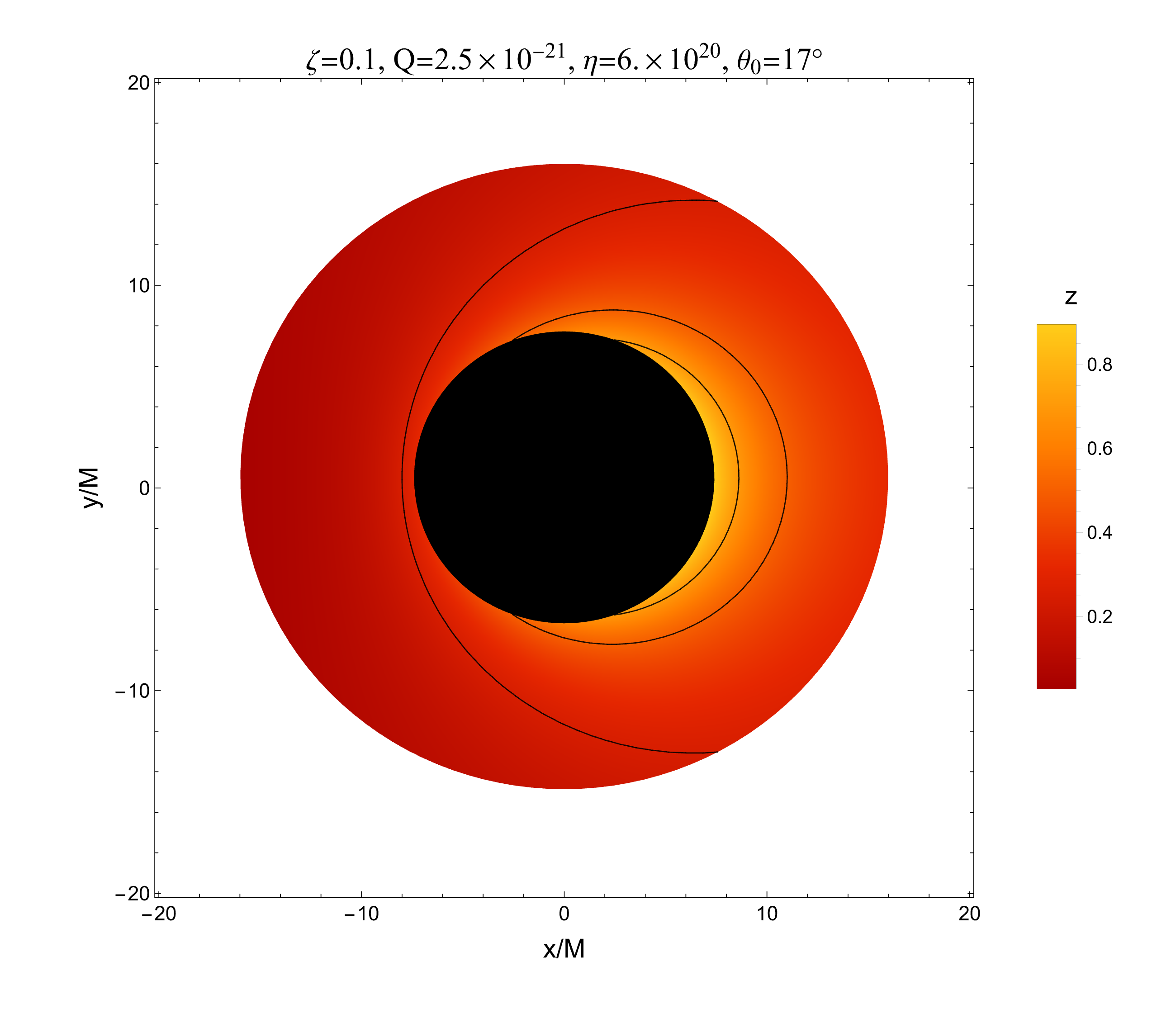}
	\end{subfigure}
	
	\caption{Influence of $\eta$ on the direct images of the redshift factor $z$ at different observation angles. The $\kappa>0$ case.
}
	\label{fig:Redshift diffeta-positive}
\end{figure}

\begin{figure}[htbp]
	\centering
	\begin{subfigure}[b]{0.31\textwidth}
		\centering
		\includegraphics[width=\textwidth]{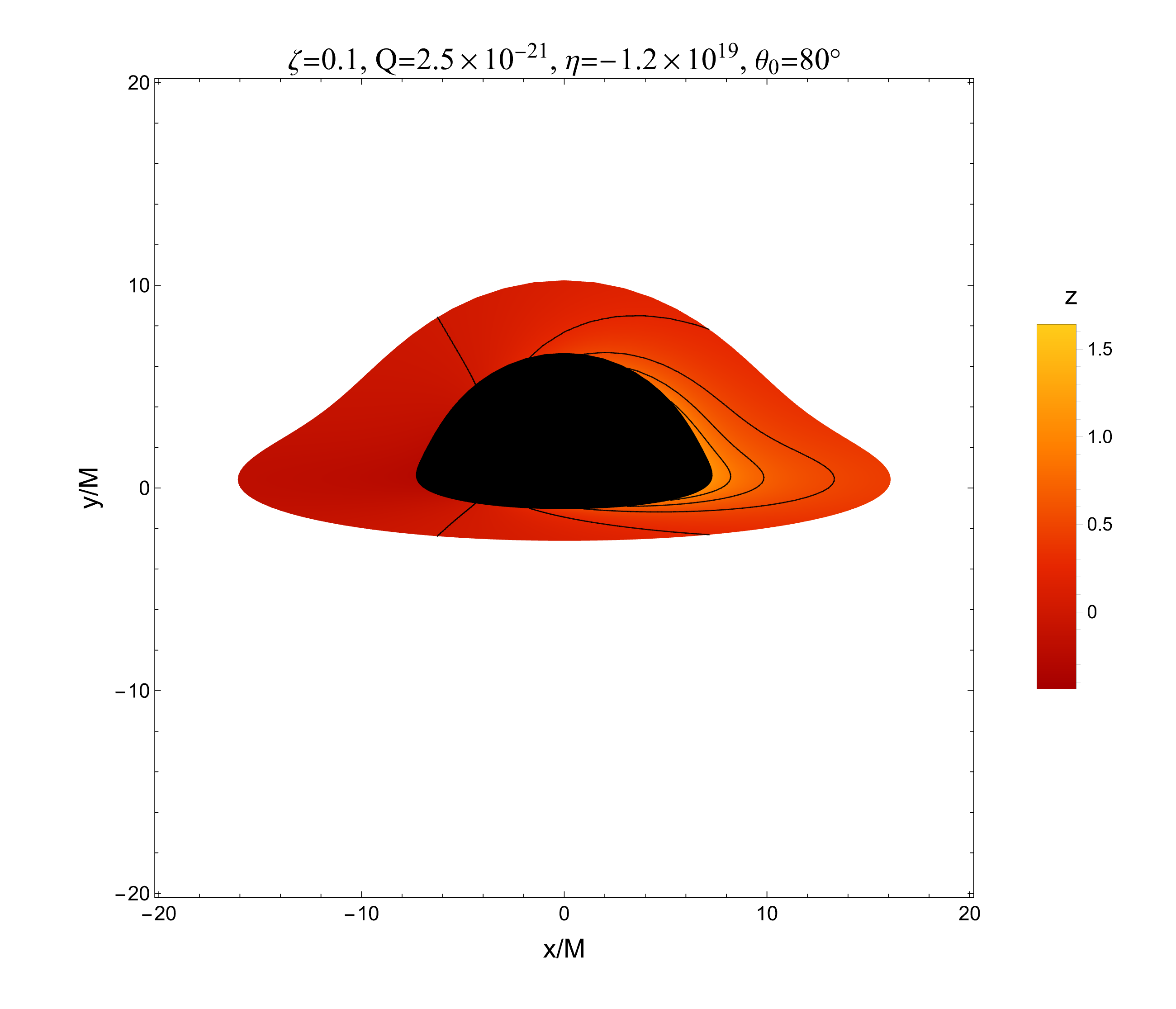}
	\end{subfigure}
	\hfill
	\begin{subfigure}[b]{0.31\textwidth}
		\centering
		\includegraphics[width=\textwidth]{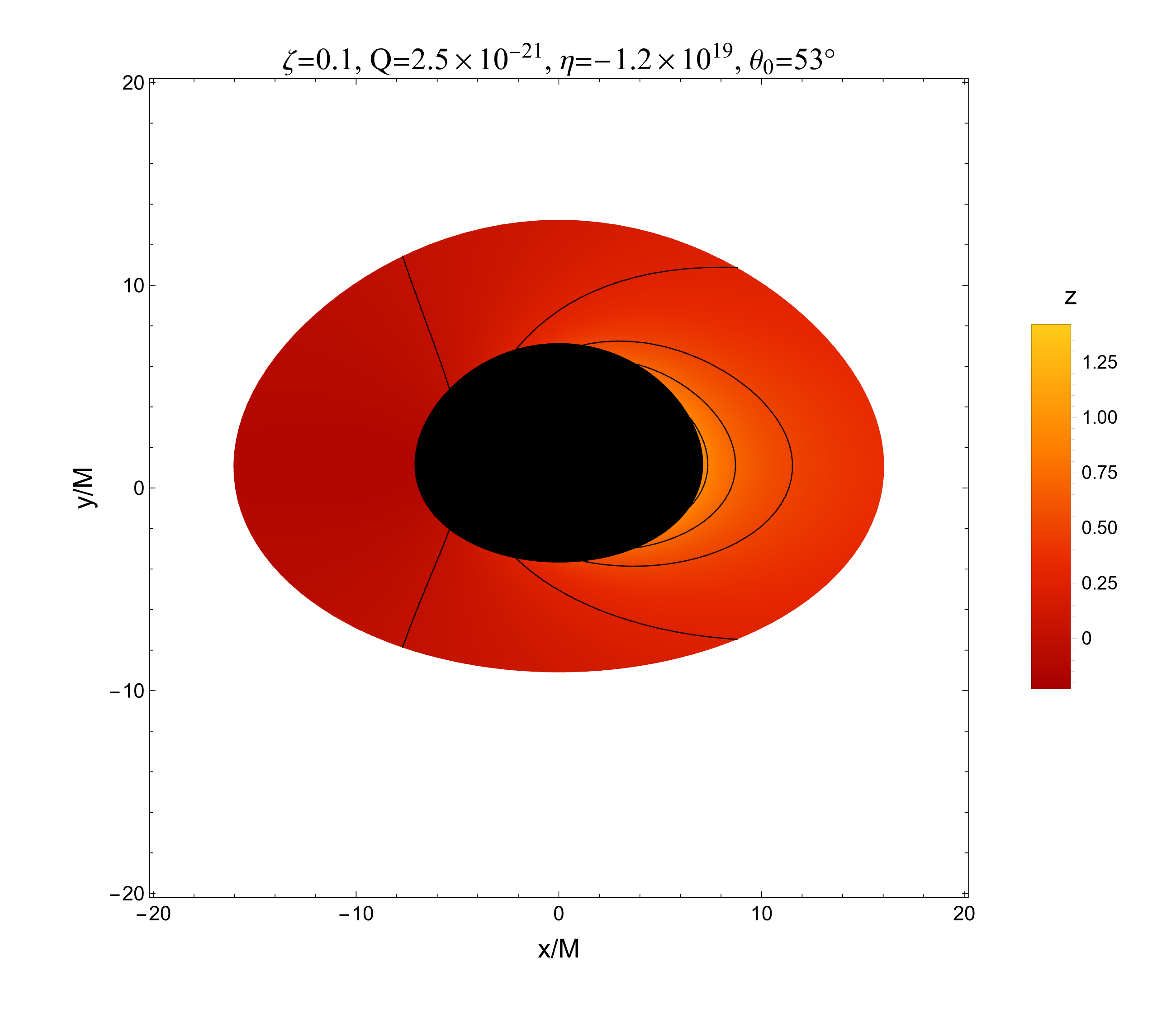}
	\end{subfigure}
	\hfill
	\begin{subfigure}[b]{0.31\textwidth}
		\centering
		\includegraphics[width=\textwidth]{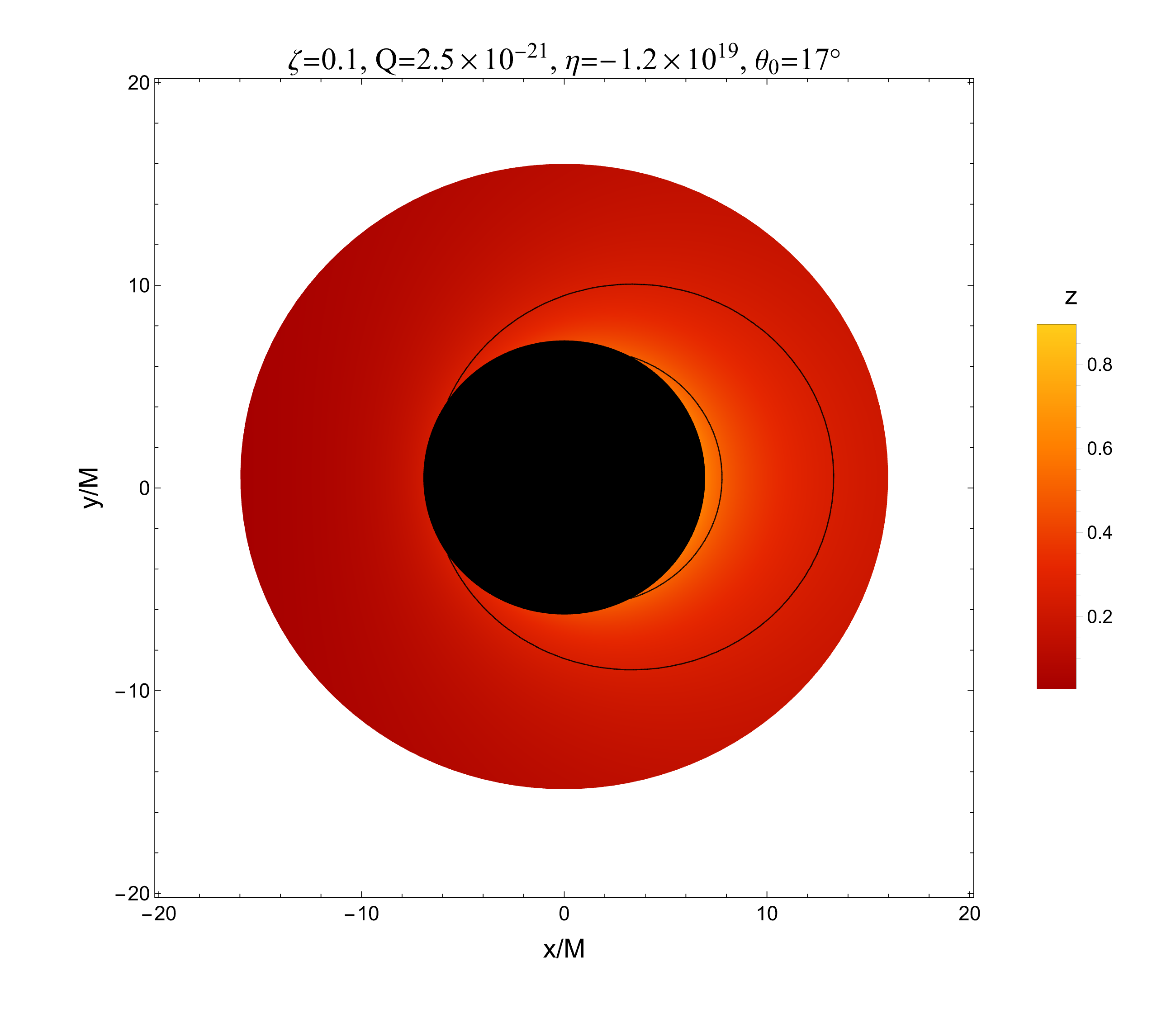}
	\end{subfigure}
	\par\medskip
	
	\begin{subfigure}[b]{0.31\textwidth}
		\centering
		\includegraphics[width=\textwidth]{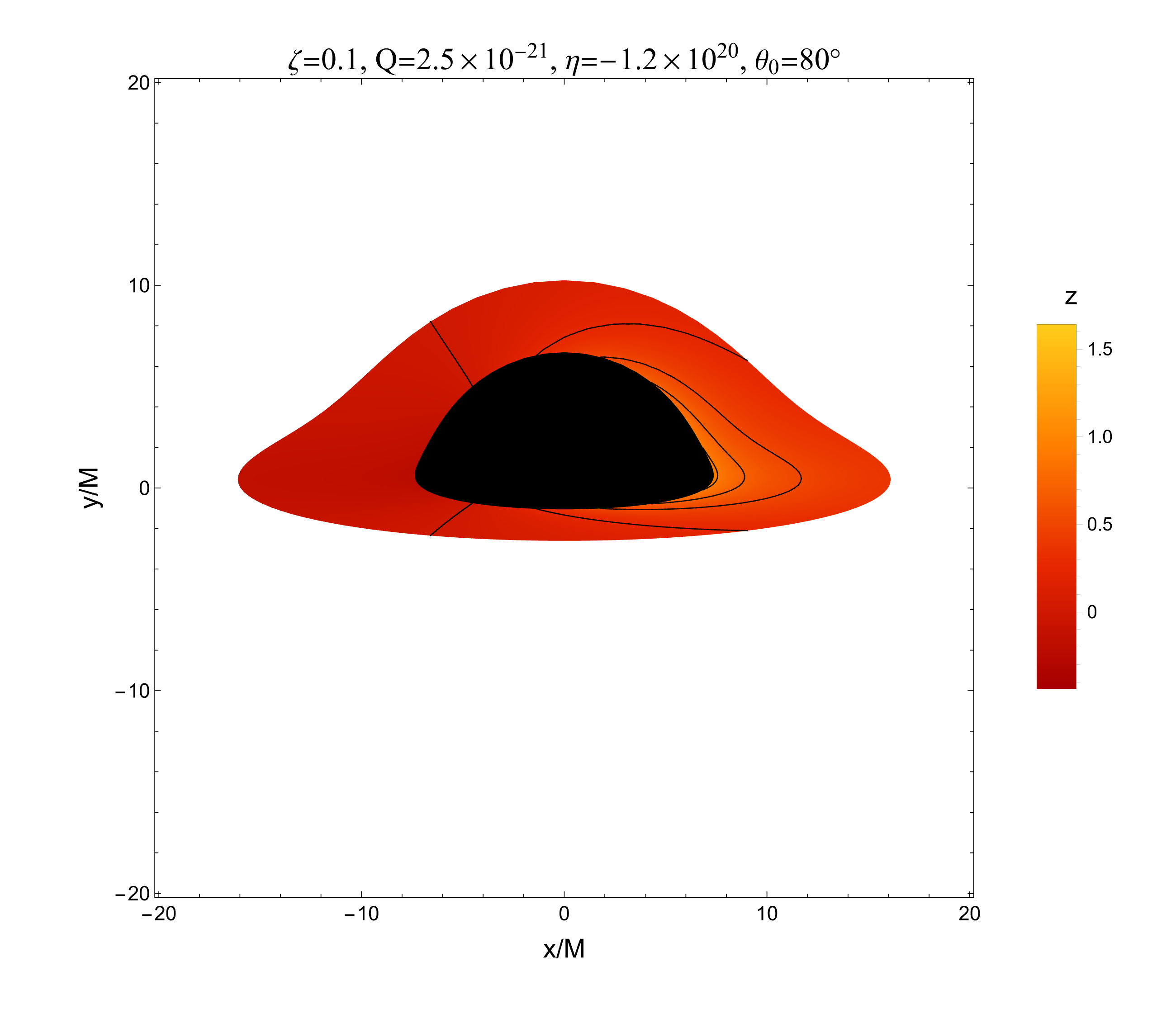}
	\end{subfigure}
	\hfill
	\begin{subfigure}[b]{0.31\textwidth}
		\centering
		\includegraphics[width=\textwidth]{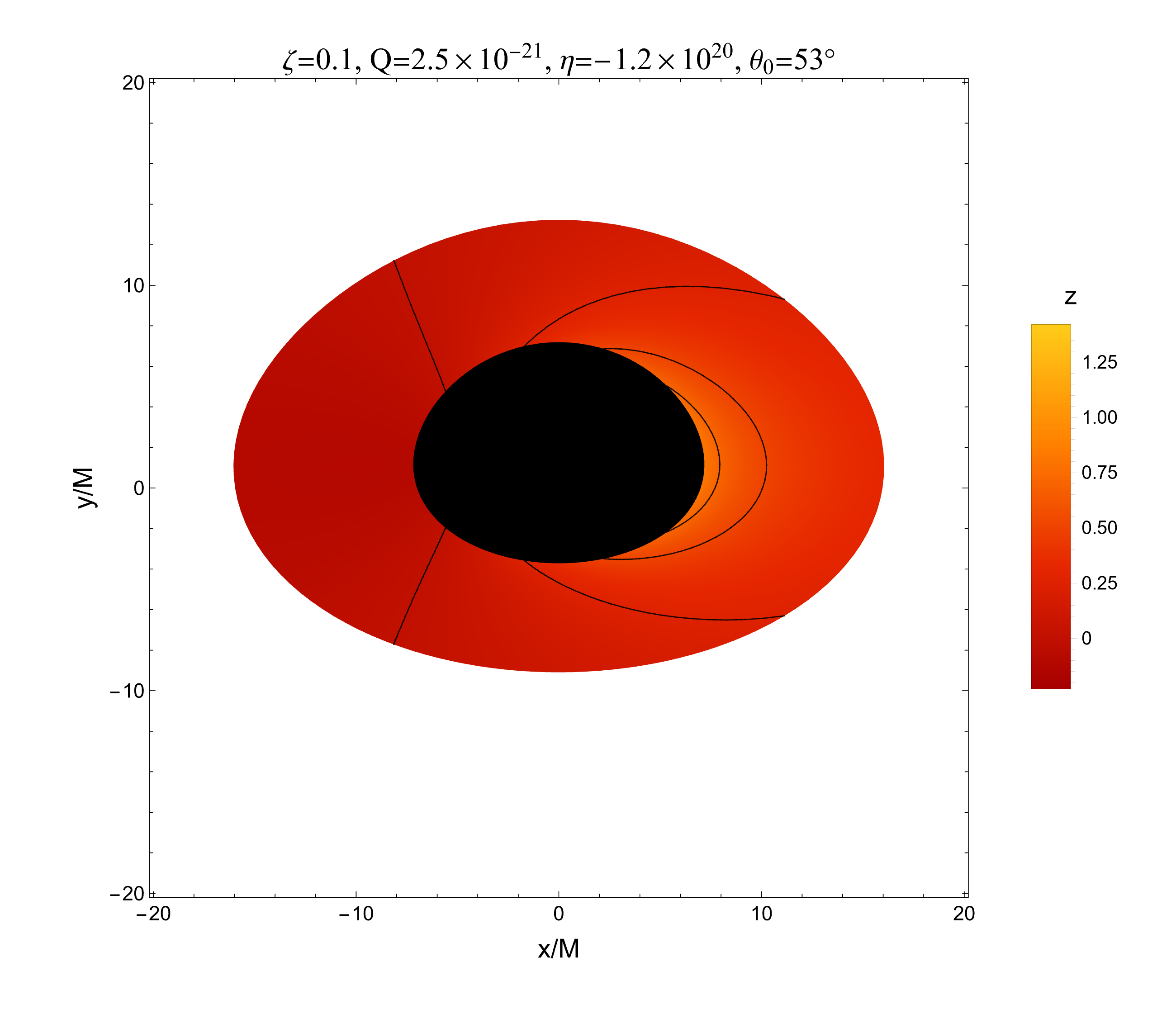}
	\end{subfigure}
	\hfill
	\begin{subfigure}[b]{0.31\textwidth}
		\centering
		\includegraphics[width=\textwidth]{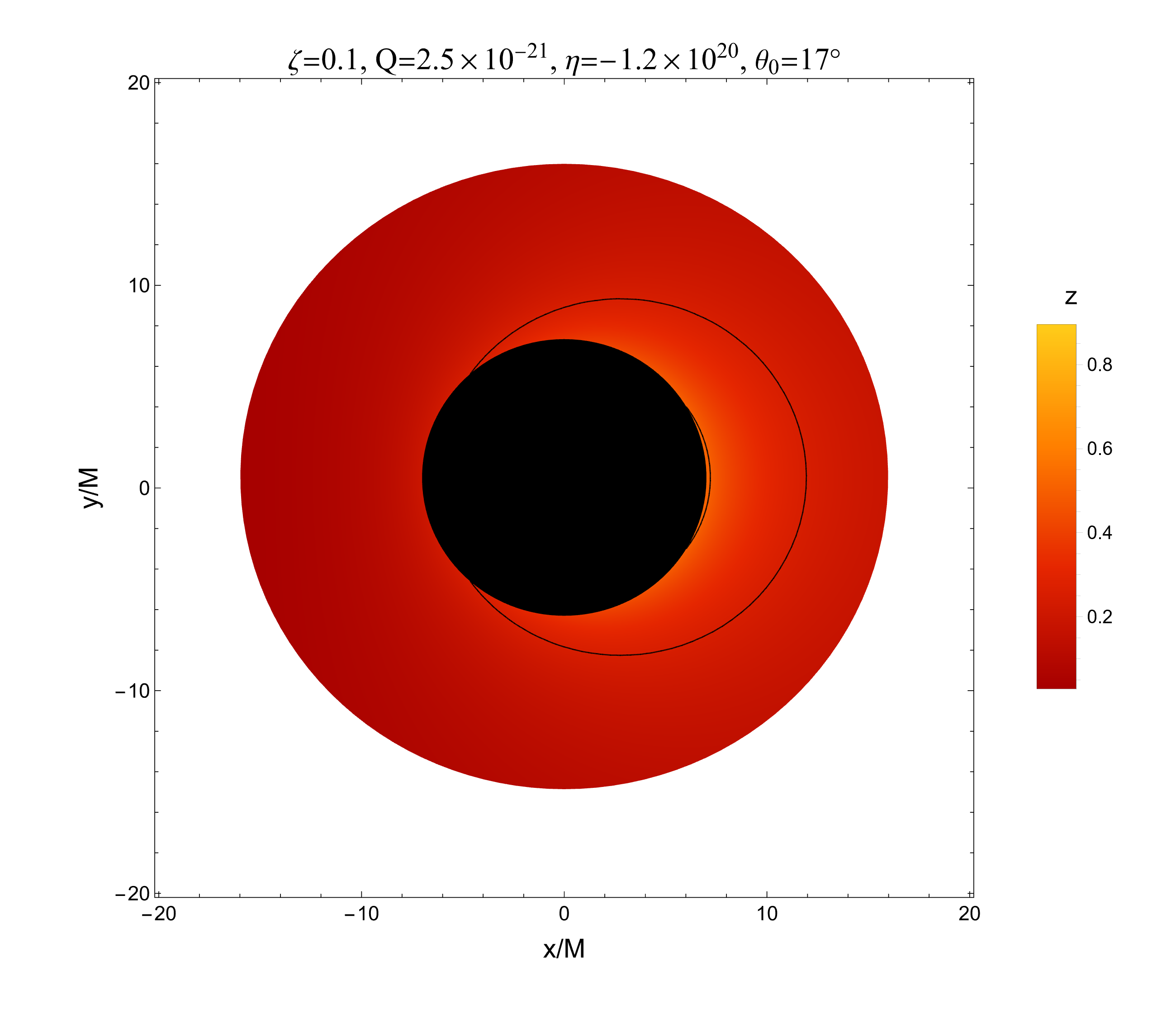}
	\end{subfigure}
	
	\caption{Influence of $\eta$ on the direct images of the redshift factor $z$ at different observation angles. The $\kappa<0$ case.}
	\label{fig:Redshift diffeta-negative}
\end{figure}

In Fig. \ref{fig:Redshift diffeta-positive} and Figs.\ref{fig:Redshift diffeta-negative}, we show the influence of positive $\eta$ and negative $\eta$ on the images of the redshift factors, respectively. The same contours are also plotted, and we can see that the redshift distribution is sensitive to both the sign and the magnitude of $\eta$ (thus $\kappa$). For the positive $\eta$ case, the coupling is $0.5$ for the first row and is $1.5$ for the second row. 
We can observe that the increase of $\eta$ broadens the high-redshift region, increases the maximum value of the redshift factor, and makes the $z$ contours move outward. 
For the negative $\eta$ case, the coupling is $-0.03$ for the first row and is $-0.3$ for the second row.  The increase of $|\eta|$ decreases the maximum value of the redshift factor, and makes the $z$ contours move inward. In both cases, the influence becomes more pronounced at higher inclination angles.


Then, we consider the direct images of the observed fluxes $F_{\mathrm{obs}}=F/(1+z)^4$. The relevant model constants are taken as: $M=2\times10^6M_\odot$, $\dot M=2\times10^{-3}M_\odot\,\mathrm{yr}^{-1}$, and the radius of outer edge of the disk is $15M$. The unit of the flux is $\mathrm{erg\,cm^{-2}\,s^{-1}}$.
In order to illustrate the influence of the parameters clearly, we also plot flux contours in each image. The values of the contour lines from the inside to the outside in sequence are taken as:
$F_{\mathrm{obs}}$=1.987, 1.288, 0.830, 0.528, 0.324, 0.181, 0.072 for all images at $\theta_0=17^\circ$; $F_{\mathrm{obs}}$=8.282, 5.368, 3.460, 2.199, 1.351, 0.755, 0.300 for all images at $\theta_0=53^\circ$; $F_{\mathrm{obs}}$=18.245, 11.827, 7.622, 4.845, 2.975, 1.662, 0.661 for all images at $\theta_0=80^\circ$.

\begin{figure}[htbp]
	\centering
	\begin{subfigure}[b]{0.31\textwidth}
		\centering
		\includegraphics[width=\textwidth]{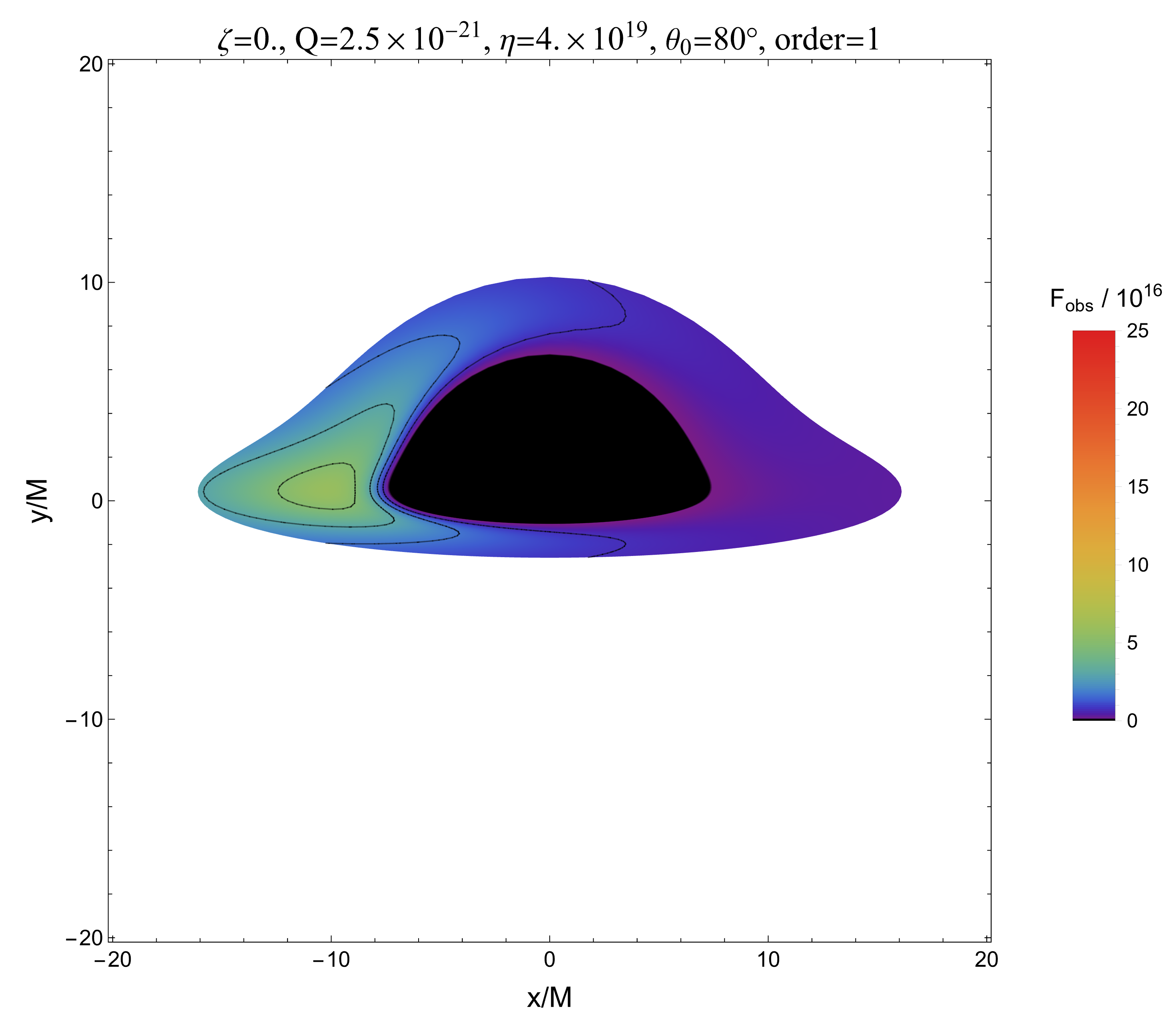}
		\label{fig:80fluxz0Q05}
	\end{subfigure}
	\hfill
	\begin{subfigure}[b]{0.31\textwidth}
		\centering
		\includegraphics[width=\textwidth]{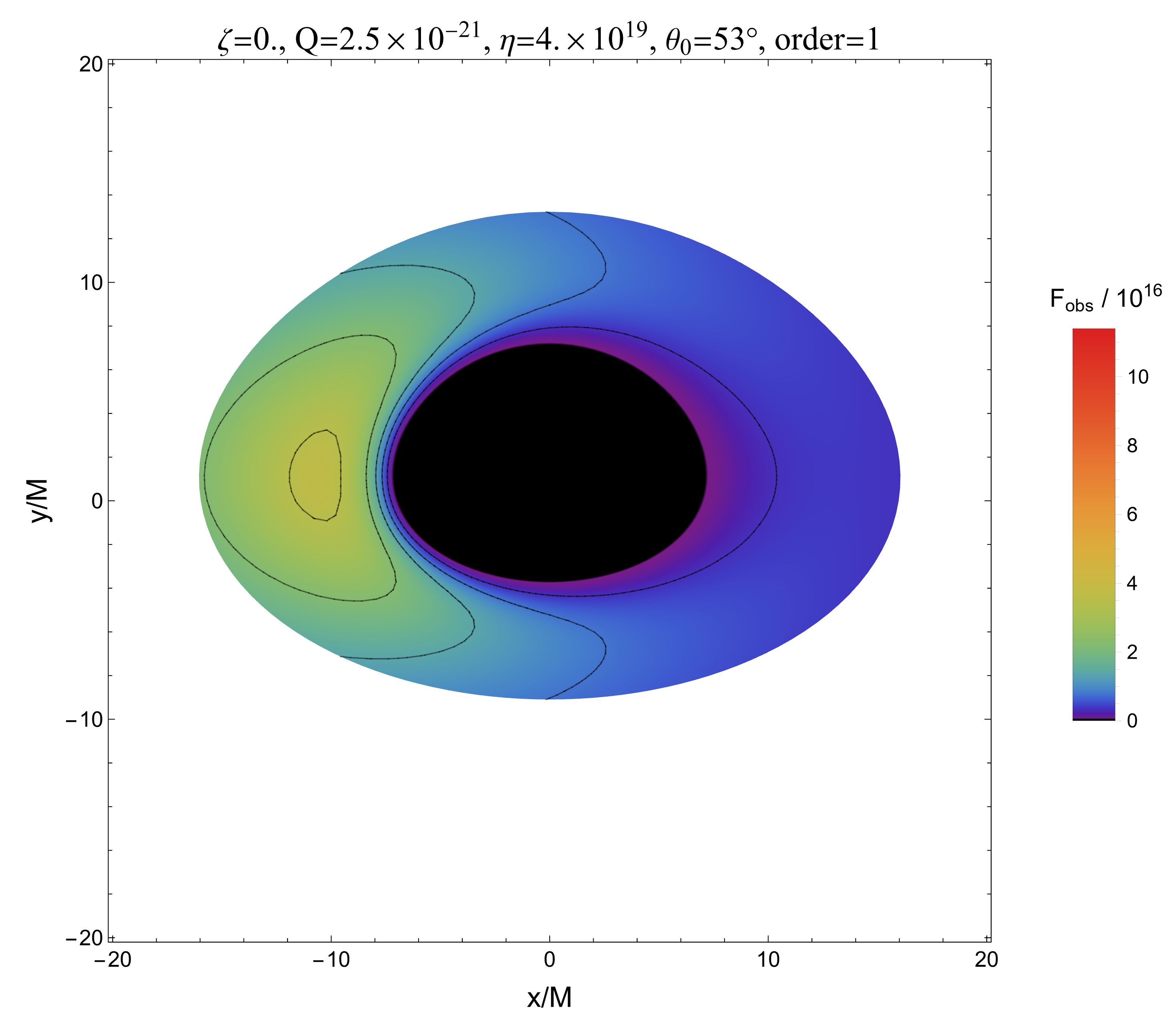}
		\label{fig:53fluxz0Q03}
	\end{subfigure}
	\hfill
	\begin{subfigure}[b]{0.31\textwidth}
		\centering
		\includegraphics[width=\textwidth]{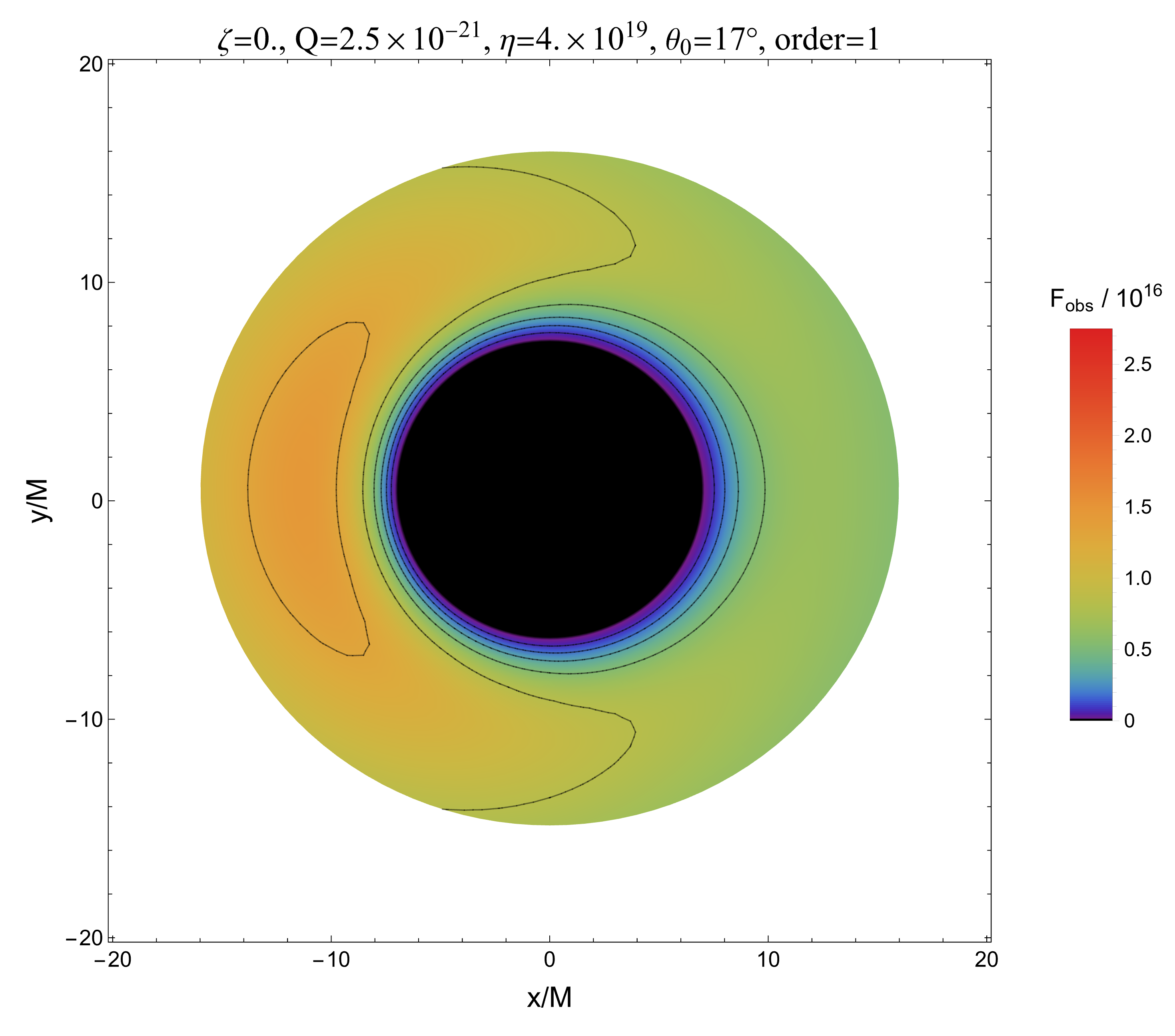}
		\label{fig:17fluxz0Q03}
	\end{subfigure}
	\par\medskip
	\begin{subfigure}[b]{0.31\textwidth}
		\centering
		\includegraphics[width=\textwidth]{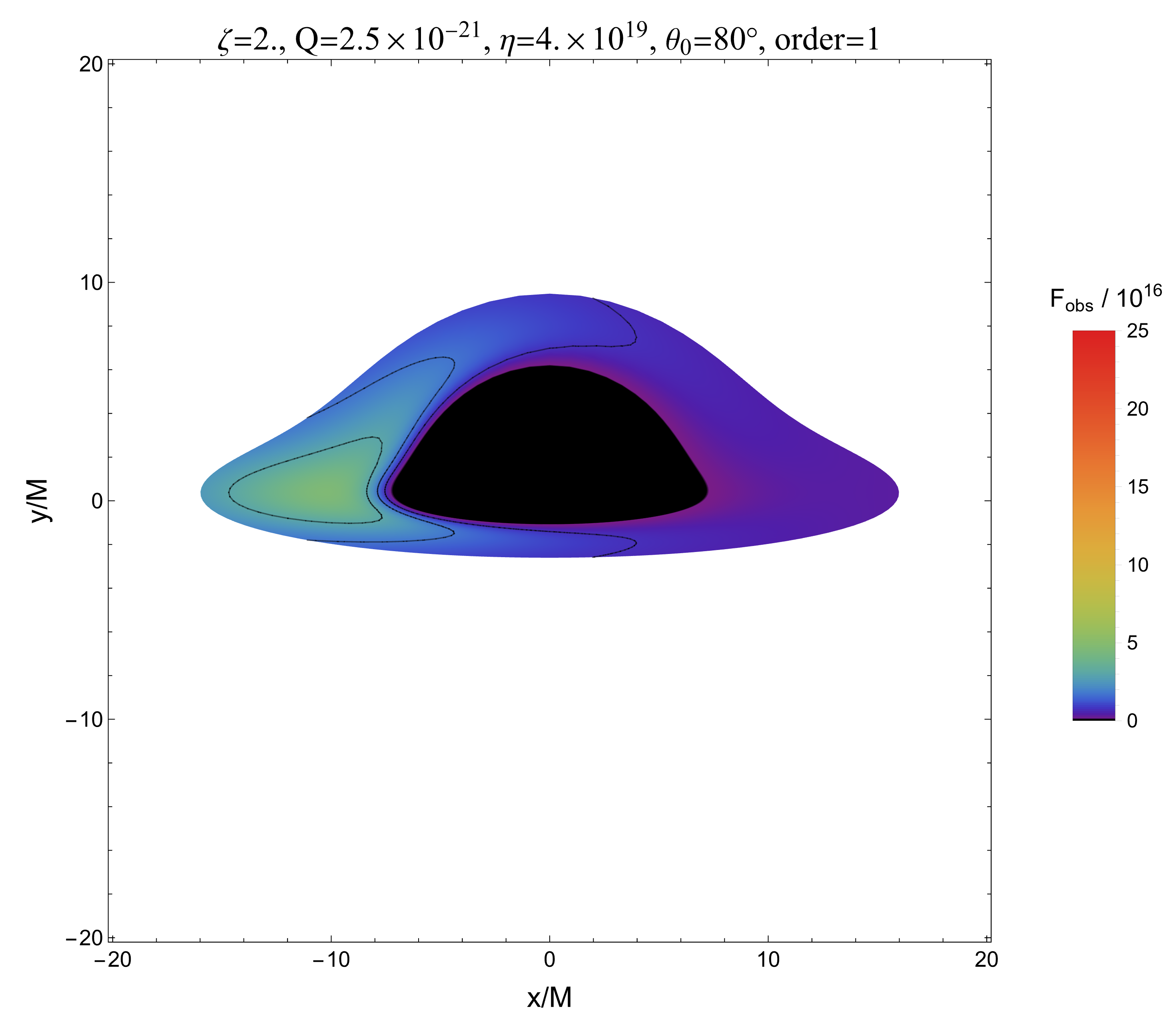}
		\label{fig:80fluxz2Q05}
	\end{subfigure}
	\hfill
	\begin{subfigure}[b]{0.31\textwidth}
		\centering
		\includegraphics[width=\textwidth]{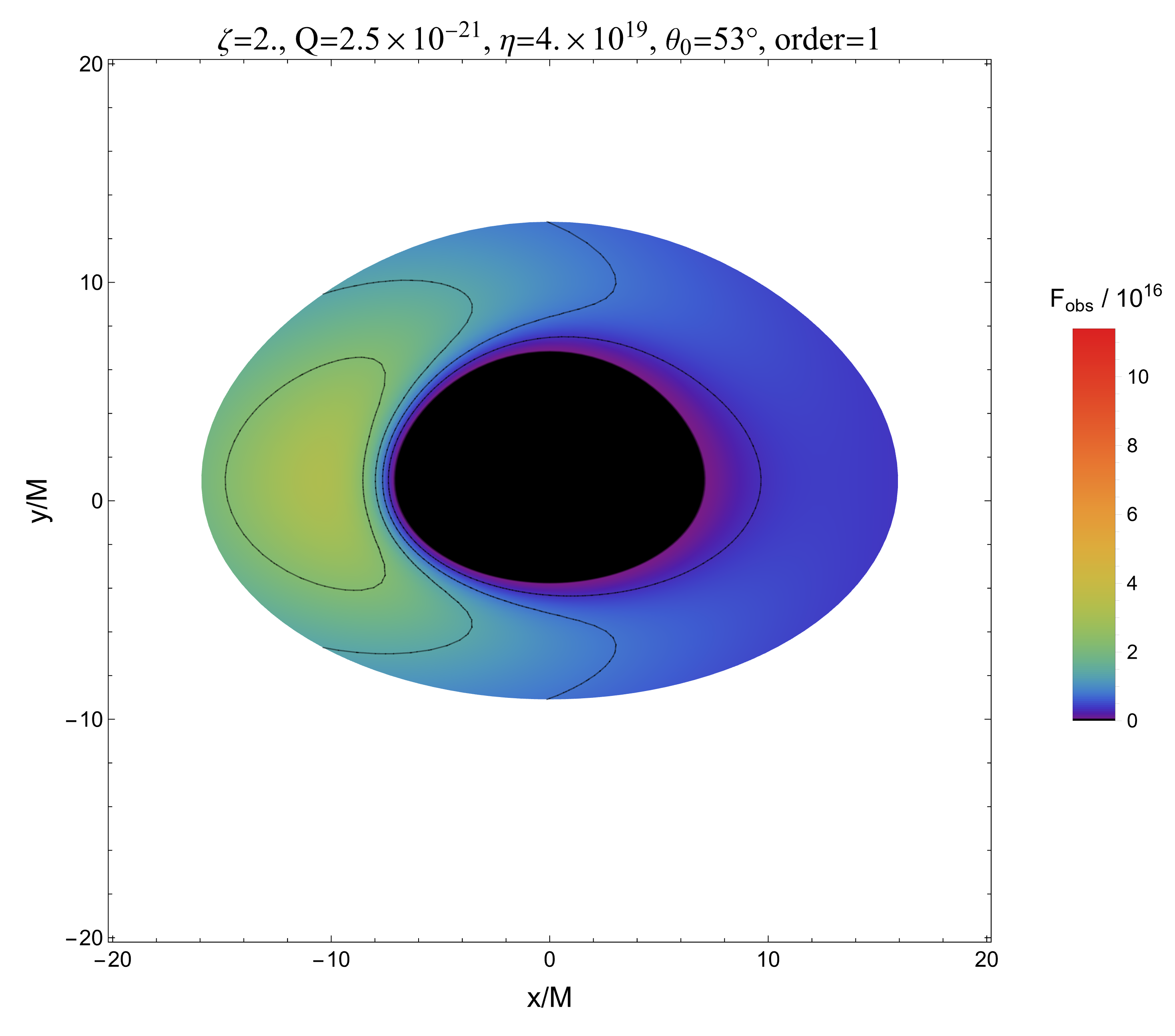}
		\label{fig:53fluxz2Q05}
	\end{subfigure}
	\hfill
	\begin{subfigure}[b]{0.31\textwidth}
		\centering
		\includegraphics[width=\textwidth]{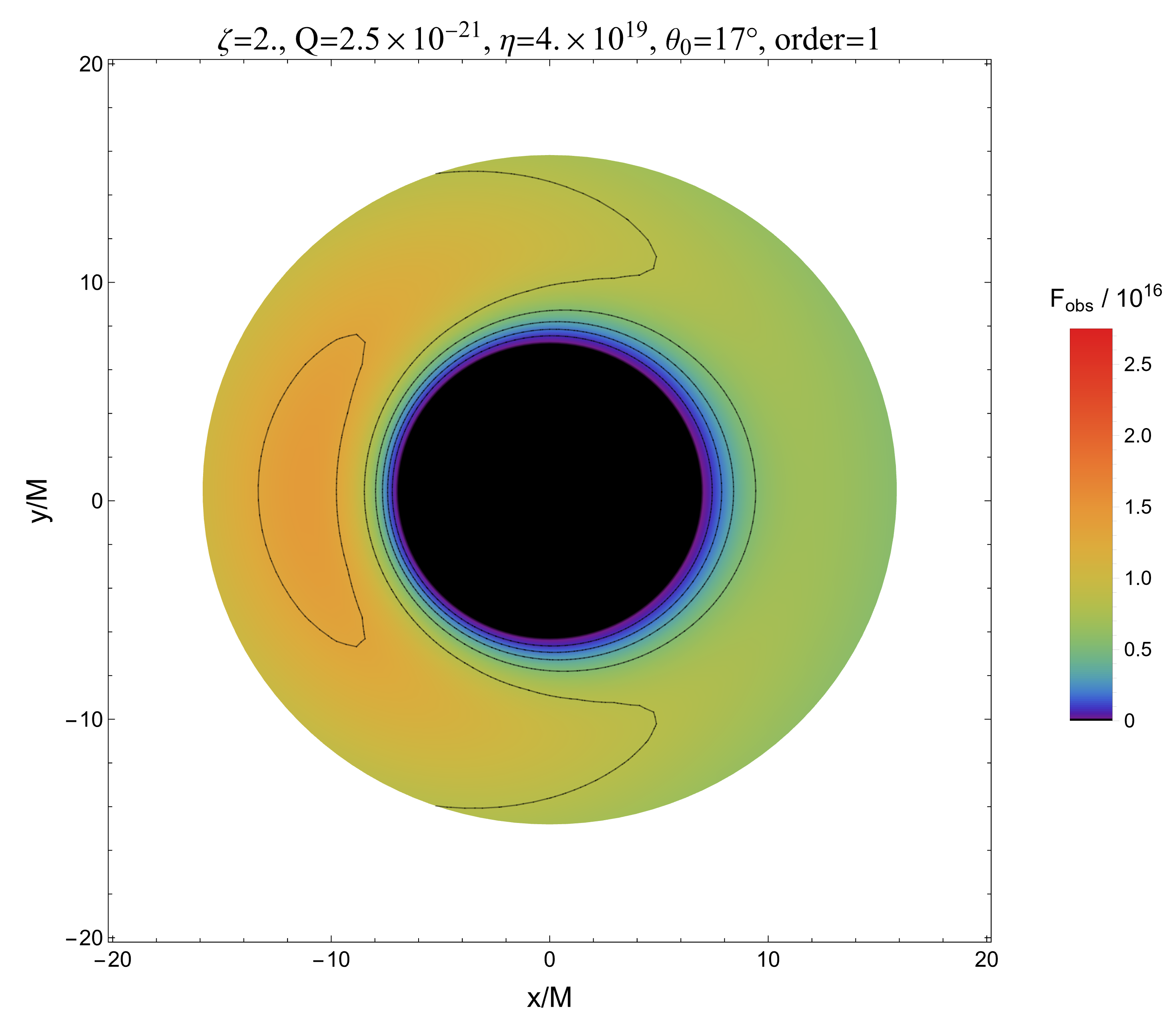}
		\label{fig:17fluxz2Q05}    
	\end{subfigure}
	
	\caption{Influence of $\zeta$ on the direct images of the observed flux $F_{\mathrm{obs}}$ at different observation angles when $\kappa=0.1$.}
	\label{fig:Flux diffz}
\end{figure}

In Fig.\ref{fig:Flux diffz}, we show the influence of the quantum parameter $\zeta$ on the direct images of the observed radiation flux at different observation angles. The electromagnetic coupling is fixed at $\kappa=0.1$. $\zeta=0$ for the images in the first row and  $\zeta=2$ for the images in the second row. 
By comparing the two direct images at each observation angle, we can see that the quantum parameter $\zeta$ reduces the maximum values of the observed flux.
The flux contours move inward and the coverage of every contour also reduces. The influence of the quantum parameter $\zeta$ on the direct images of the observed flux is more remarkable for the lager observation angle.

\begin{figure}[htbp]
	\centering
	\begin{subfigure}[b]{0.31\textwidth}
		\centering
		\includegraphics[width=\textwidth]{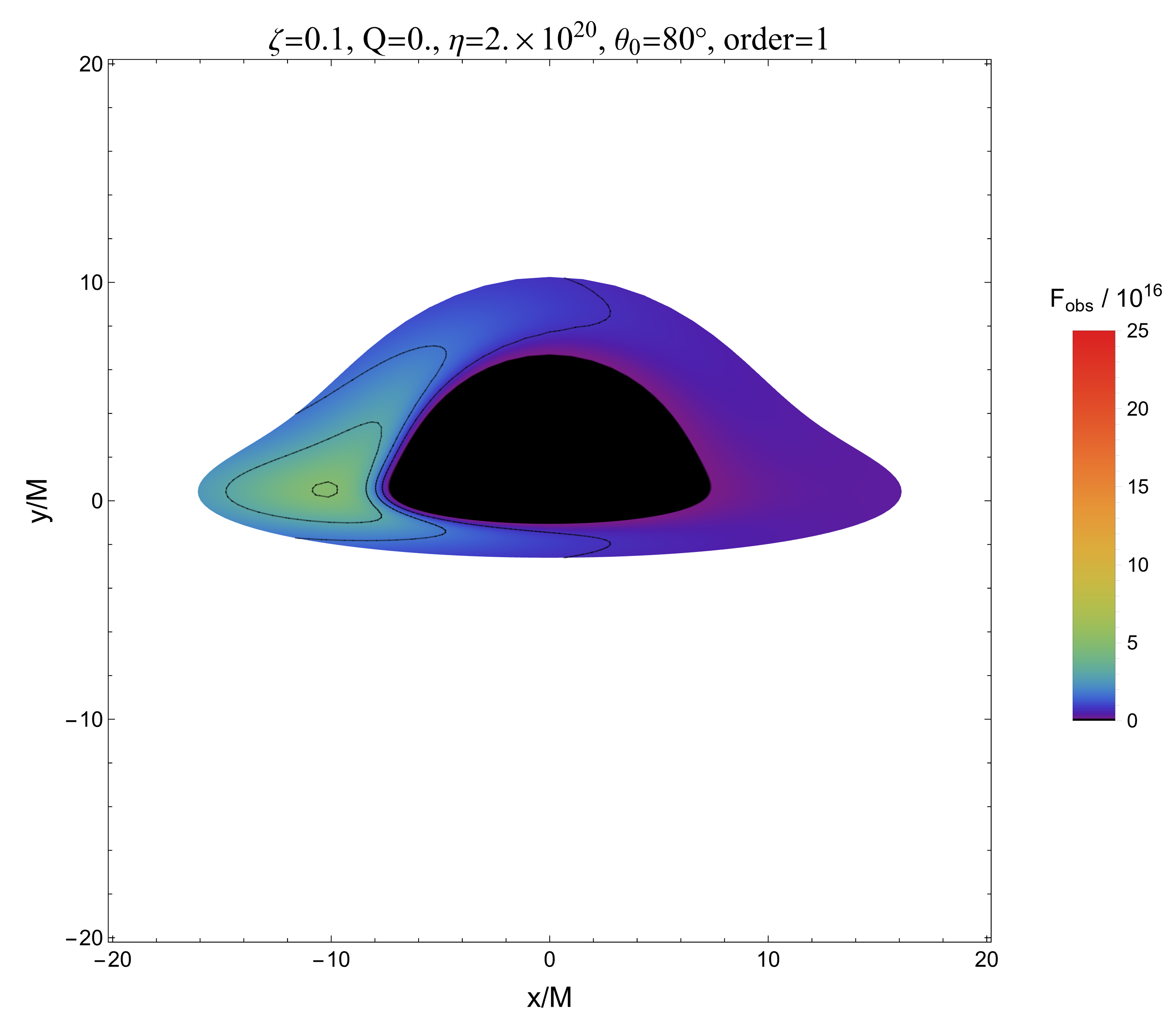}
		\label{fig:80flux15Q03}
	\end{subfigure}
	\hfill
	\begin{subfigure}[b]{0.31\textwidth}
		\centering
		\includegraphics[width=\textwidth]{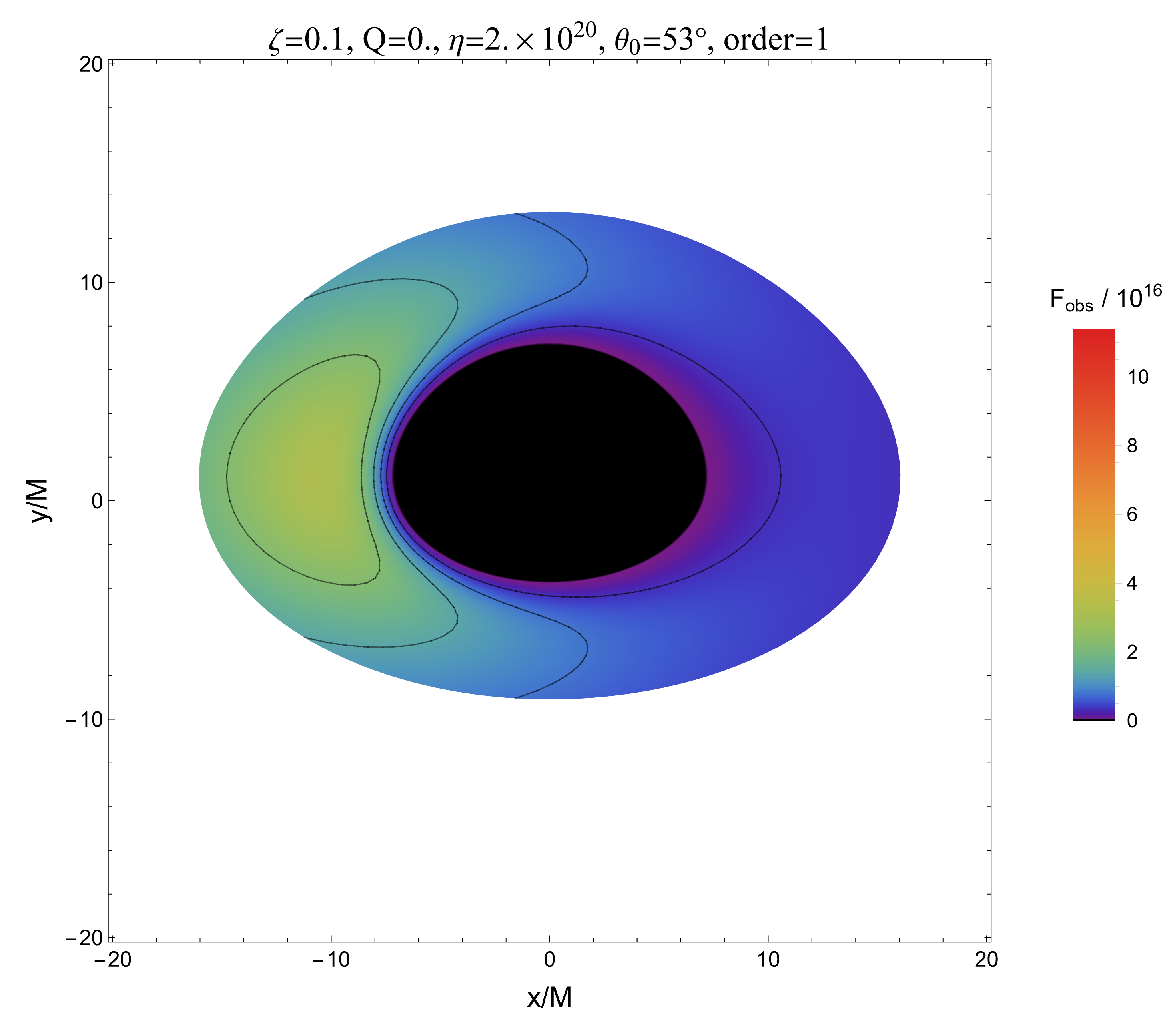}
		\label{fig:53flux15Q03}
	\end{subfigure}
	\hfill
	\begin{subfigure}[b]{0.31\textwidth}
		\centering
		\includegraphics[width=\textwidth]{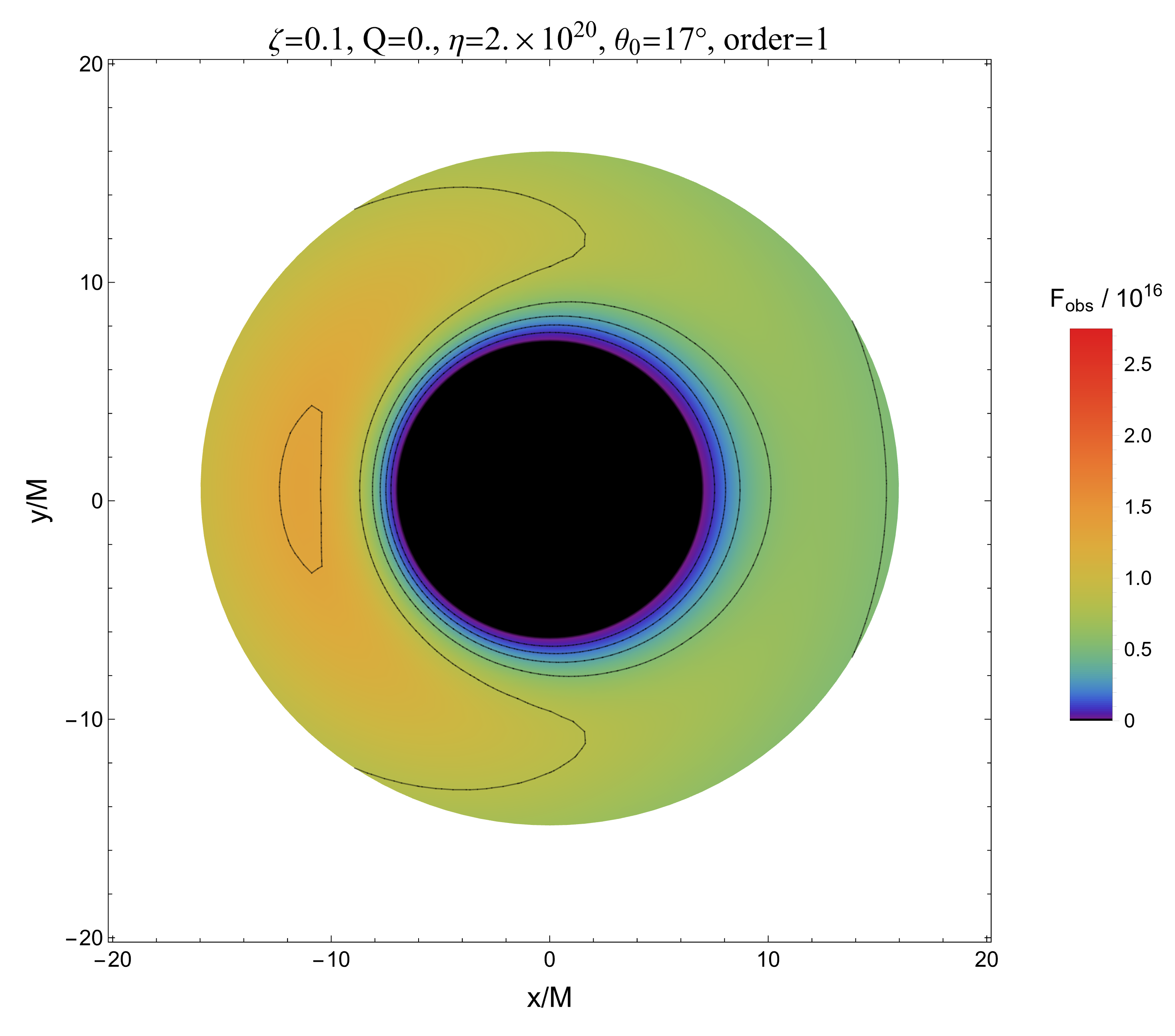}
		\label{fig:17flux15Q03}
	\end{subfigure}
	\par\medskip
	
	\begin{subfigure}[b]{0.31\textwidth}
		\centering
		\includegraphics[width=\textwidth]{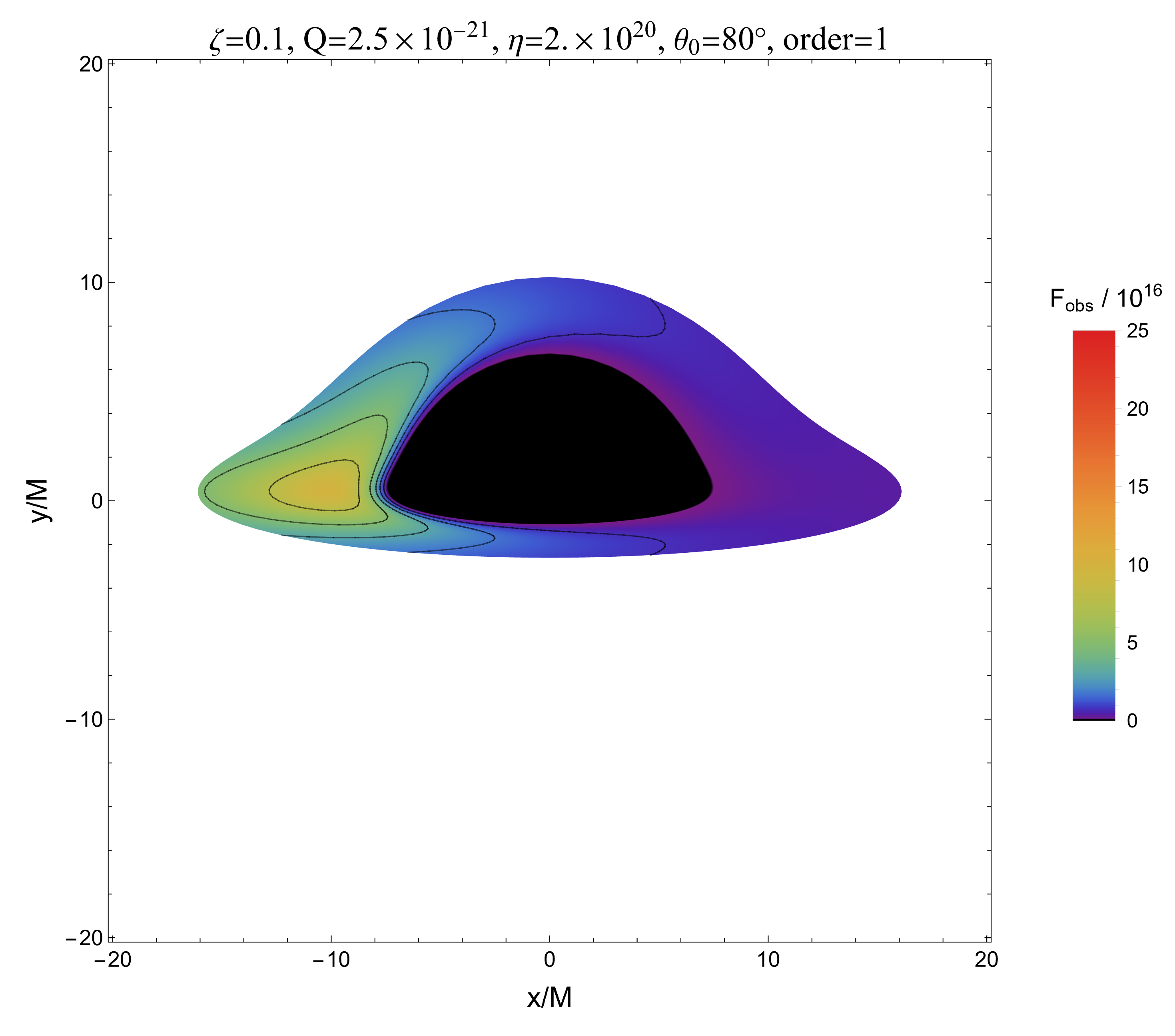}
		\label{fig:80flux15Q05}
	\end{subfigure}
	\hfill
	\begin{subfigure}[b]{0.31\textwidth}
		\centering
		\includegraphics[width=\textwidth]{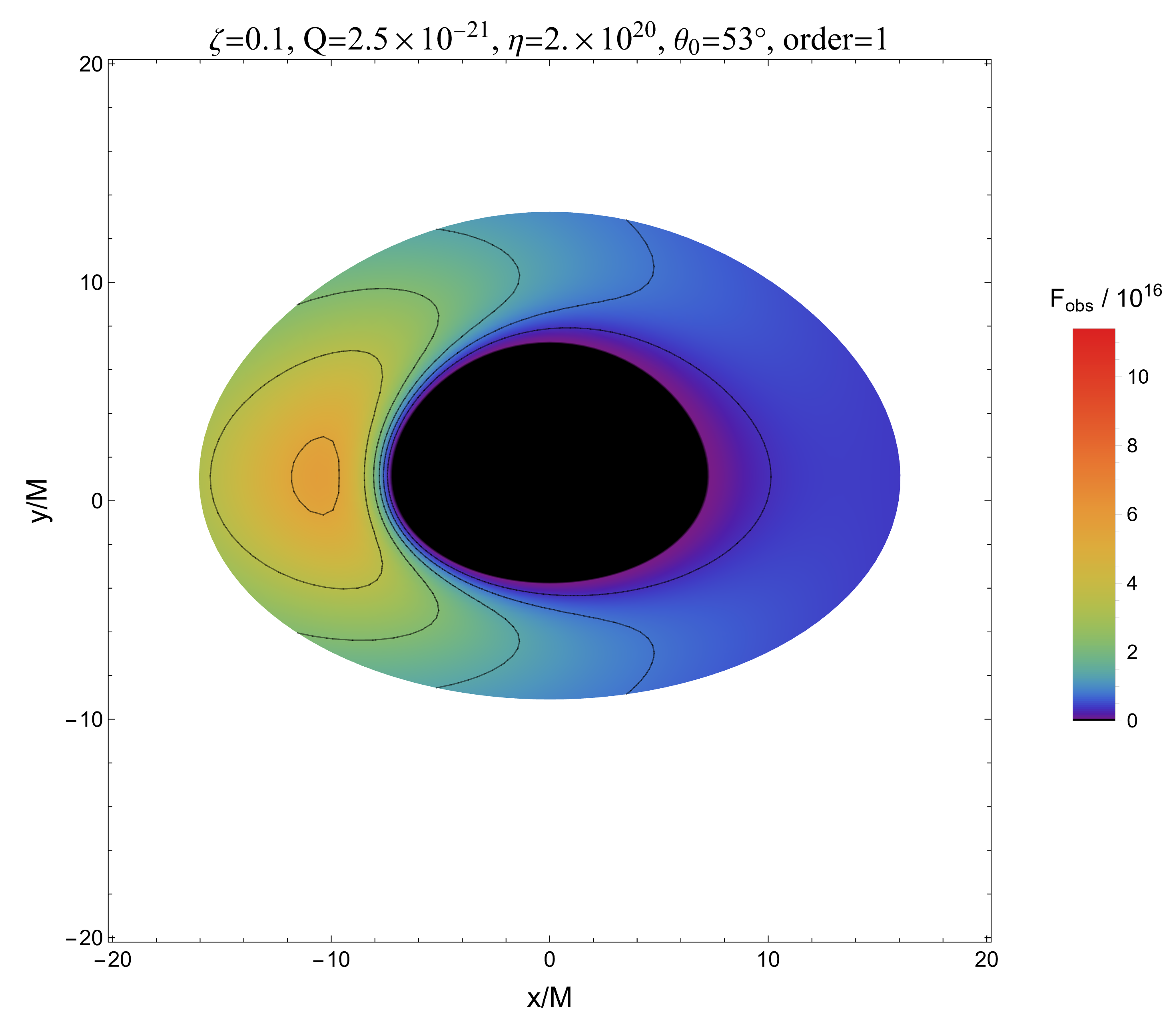}
		\label{fig:53flux15Q05}
	\end{subfigure}
	\hfill
	\begin{subfigure}[b]{0.31\textwidth}
		\centering
		\includegraphics[width=\textwidth]{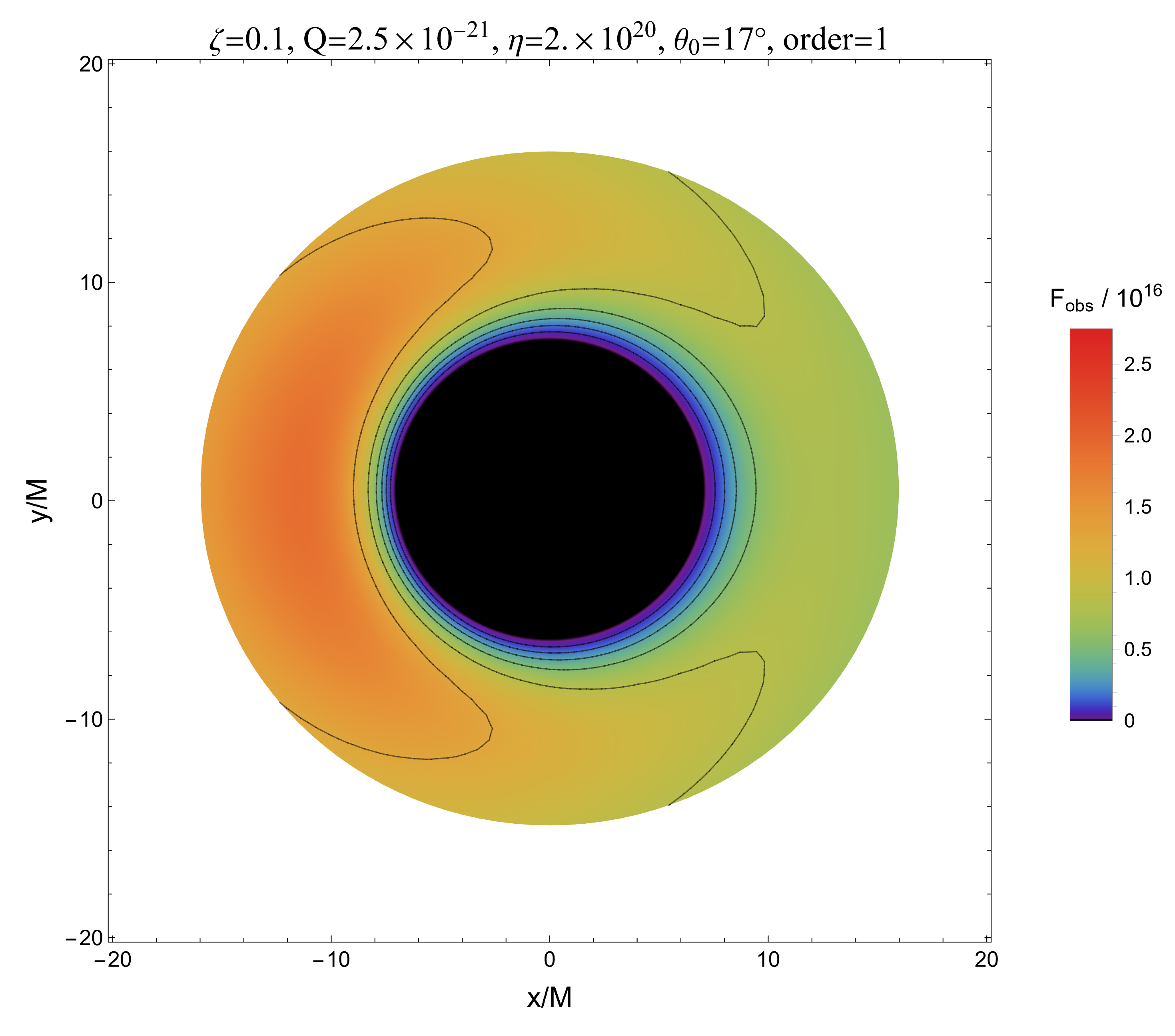}
		\label{fig:17flux15Q05}    
	\end{subfigure}
	
	\caption{Influence of $Q$ on the direct images of the observed flux $F_{\mathrm{obs}}$ at different observation angles when $\kappa>0$. }
	\label{fig:Flux diffQ}
\end{figure}

In Fig.\ref{fig:Flux diffQ}, we show the direct images of the observed fluxes at different observation angles and the influence of charge parameter $Q$ on them. $Q=0$ ($\kappa=0$) for the images in the first row and  $Q=2.5\times10^{-21}$ ( $\kappa=0.5$) for the images in the second row. 
By comparing the two direct images at each observation angle, we can see that the charge parameter $Q$ significantly increases the maximum value of the observed flux even when the black hole carries a little amount of electric charges. The flux contours move outward remarkably, and the observed flux increases everywhere as the increase of $Q$. These can be understood from the fact that when the electromagnetic coupling $\kappa>0$, the charge parameter increases the radiative efficiency.  We can also see that at small observation angle (e.g. $17^{\circ}$), the flux distribution is fairly symmetric, and at large observation angle (e.g. $80^{\circ}$), the asymmetry of the flux distribution increases. 

\begin{figure}[htbp]
	\centering
	\begin{subfigure}[b]{0.31\textwidth}
		\centering
		\includegraphics[width=\textwidth]{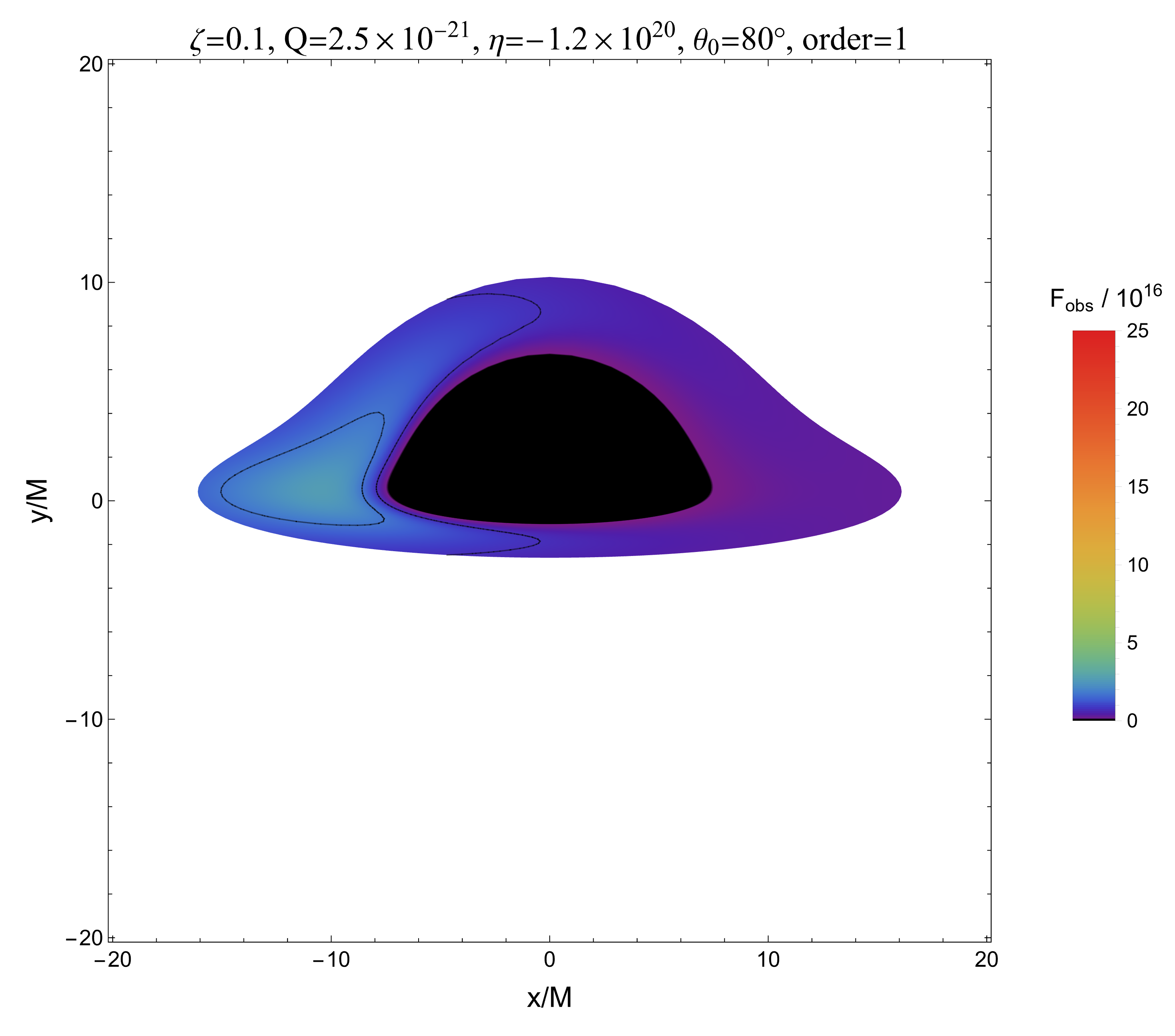}
	\end{subfigure}
	\hfill
	\begin{subfigure}[b]{0.31\textwidth}
		\centering
		\includegraphics[width=\textwidth]{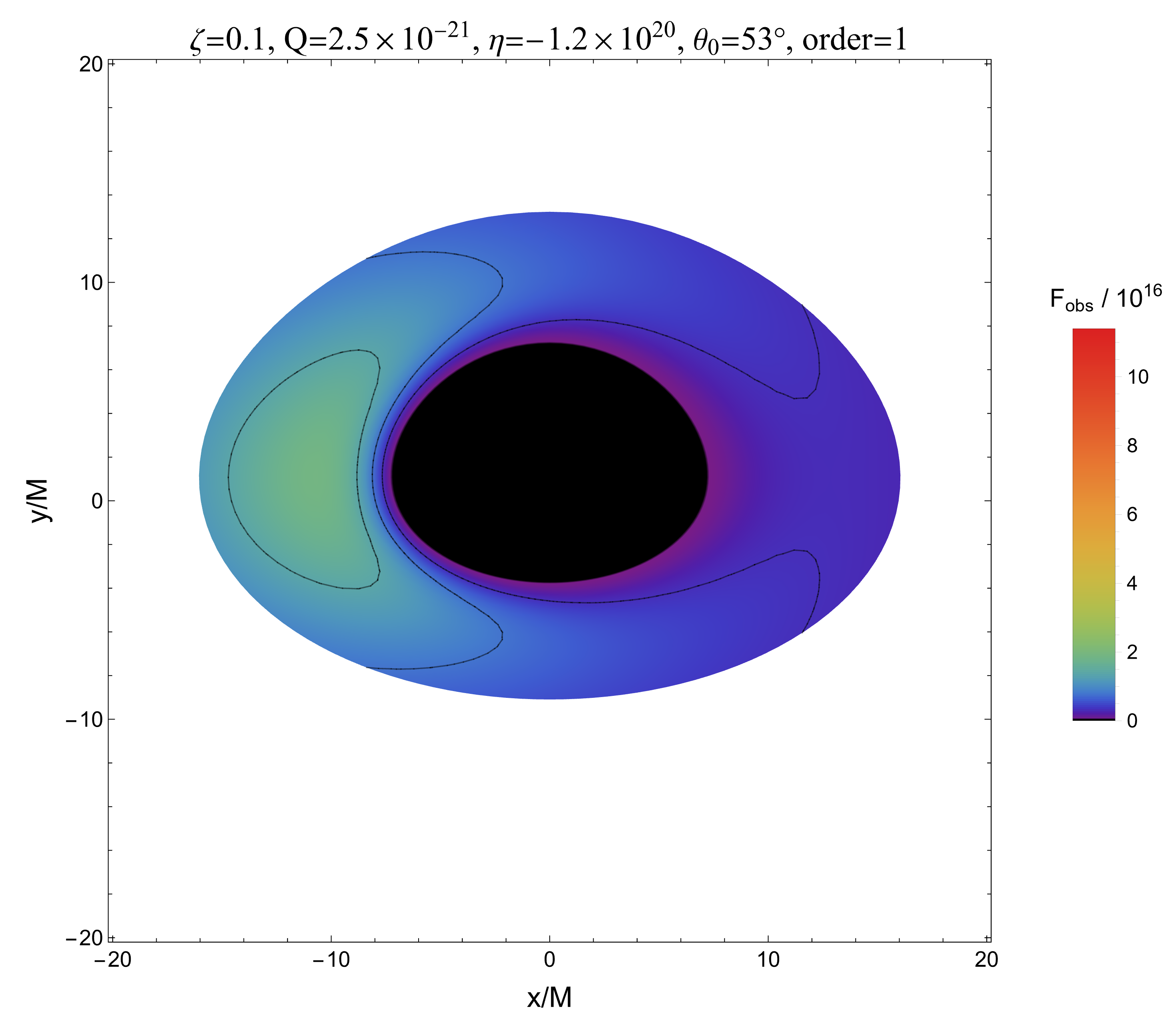}
	\end{subfigure}
	\hfill
	\begin{subfigure}[b]{0.31\textwidth}
		\centering
		\includegraphics[width=\textwidth]{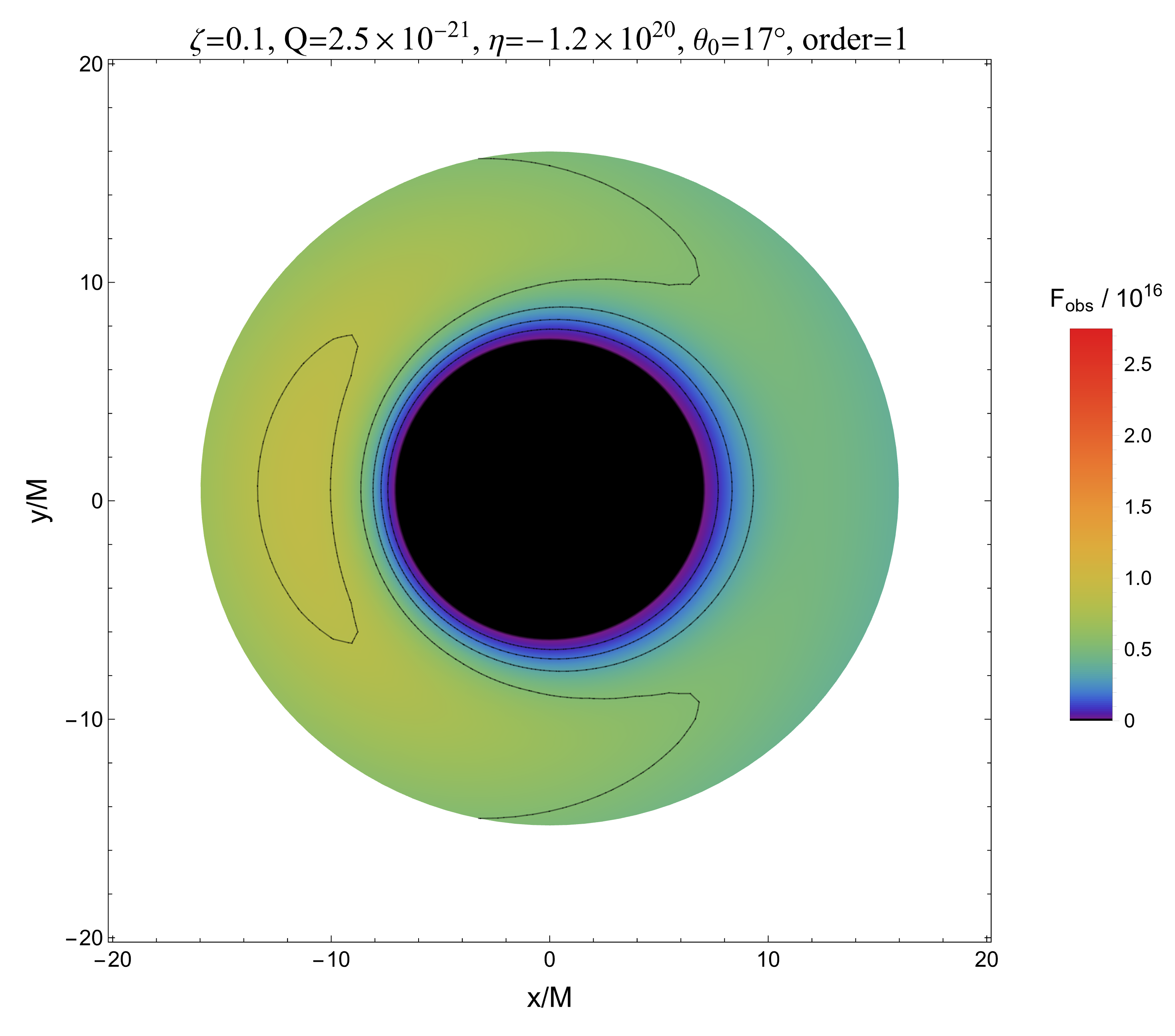}
	\end{subfigure}
	\par\medskip
	
	\begin{subfigure}[b]{0.31\textwidth}
		\centering
		\includegraphics[width=\textwidth]{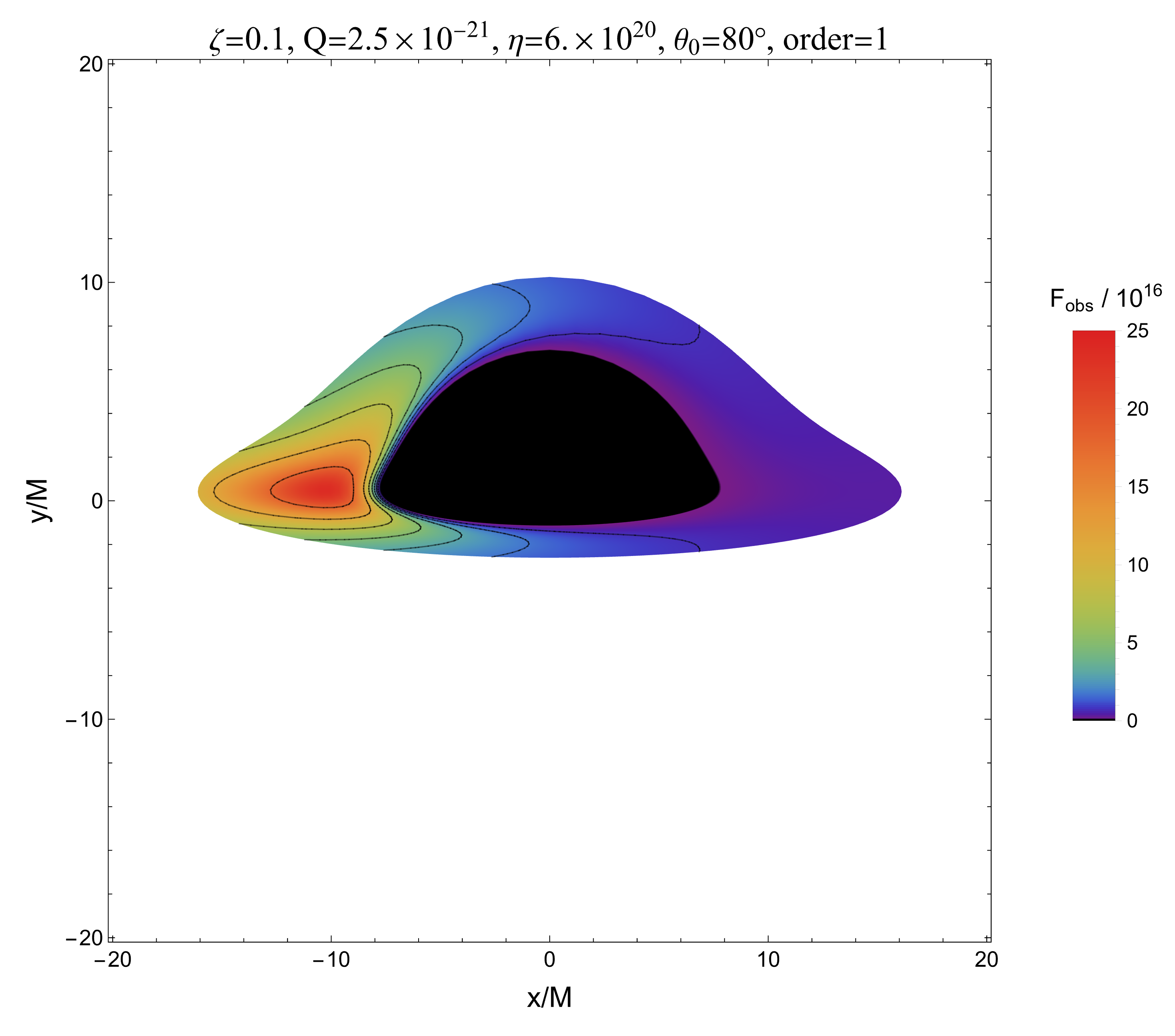}
	\end{subfigure}
	\hfill
	\begin{subfigure}[b]{0.31\textwidth}
		\centering
		\includegraphics[width=\textwidth]{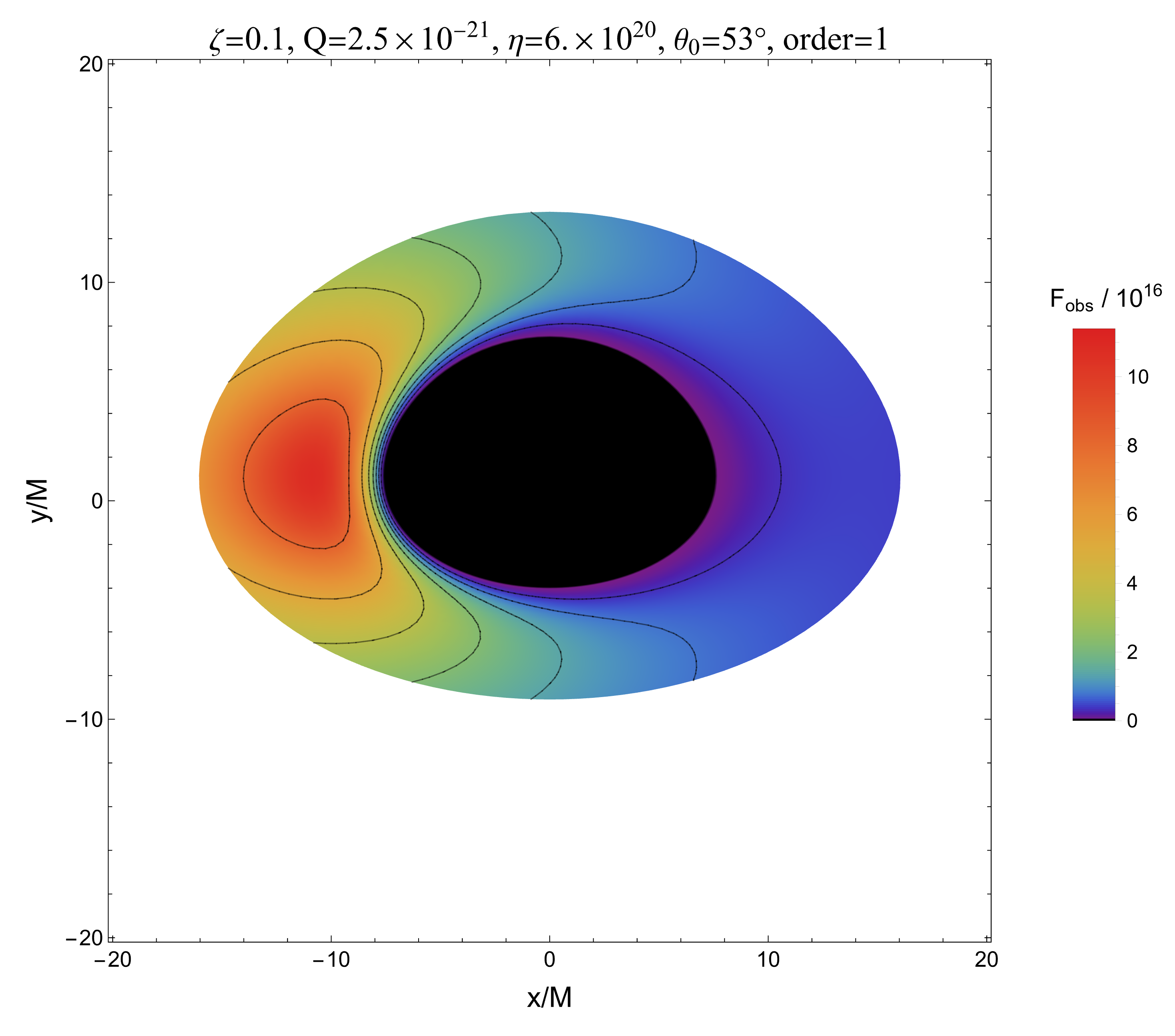}
	\end{subfigure}
	\hfill
	\begin{subfigure}[b]{0.31\textwidth}
		\centering
		\includegraphics[width=\textwidth]{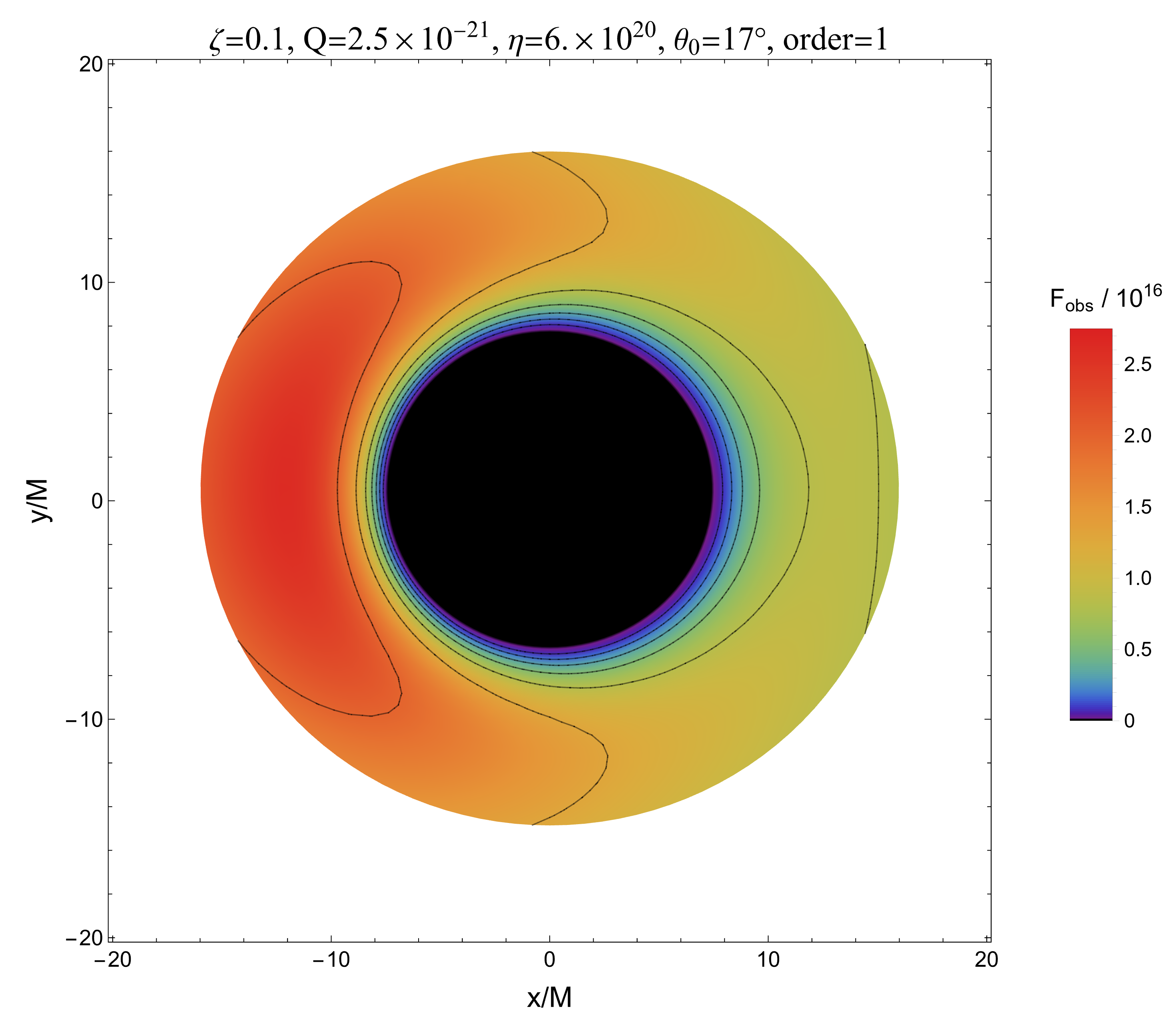}
	\end{subfigure}
	
	\caption{The direct images of the observed flux $F_{\mathrm{obs}}$ for two values of $\eta$ at different observation angles.}
	\label{fig:Flux diffeta}
\end{figure}

In Fig.\ref{fig:Flux diffeta}, we show the direct images of the observed fluxes at different observation angles and the influence of $\eta$ on them. $\eta=-1.2\times10^{20}$ ($\kappa=-0.3$) for the images in the first row and  $\eta=6\times10^{20}$ ($\kappa=1.5$) for the images in the second row. 
By comparing the two direct images at each observation angle, we can see that the observed flux for $\kappa>0$ is much stronger than that for $\kappa<0$. Taking the images with $Q=0$
in Fig.\ref{fig:Flux diffQ} as reference, we can also observe that a positive $\kappa$ increases the observed flux while a negative $\kappa$ decreases it. These can also be understood from the influence of $\kappa$ on the radiative efficiency.

\begin{figure}[htbp]
	\centering
\begin{subfigure}[b]{0.31\textwidth}
		\centering
		\includegraphics[width=\textwidth]{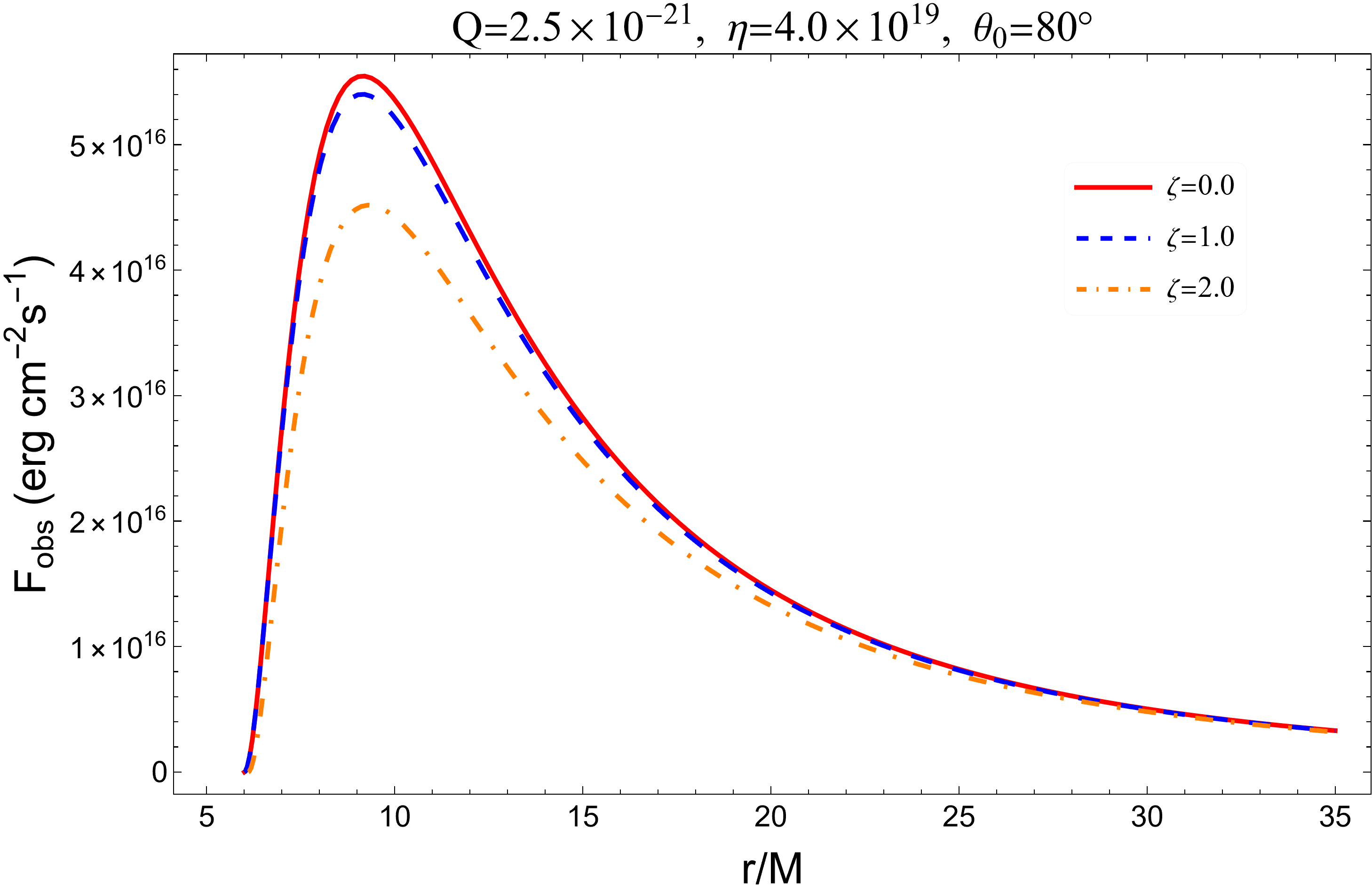}
	\end{subfigure}
	\hfill
	\begin{subfigure}[b]{0.31\textwidth}
		\centering
		\includegraphics[width=\textwidth]{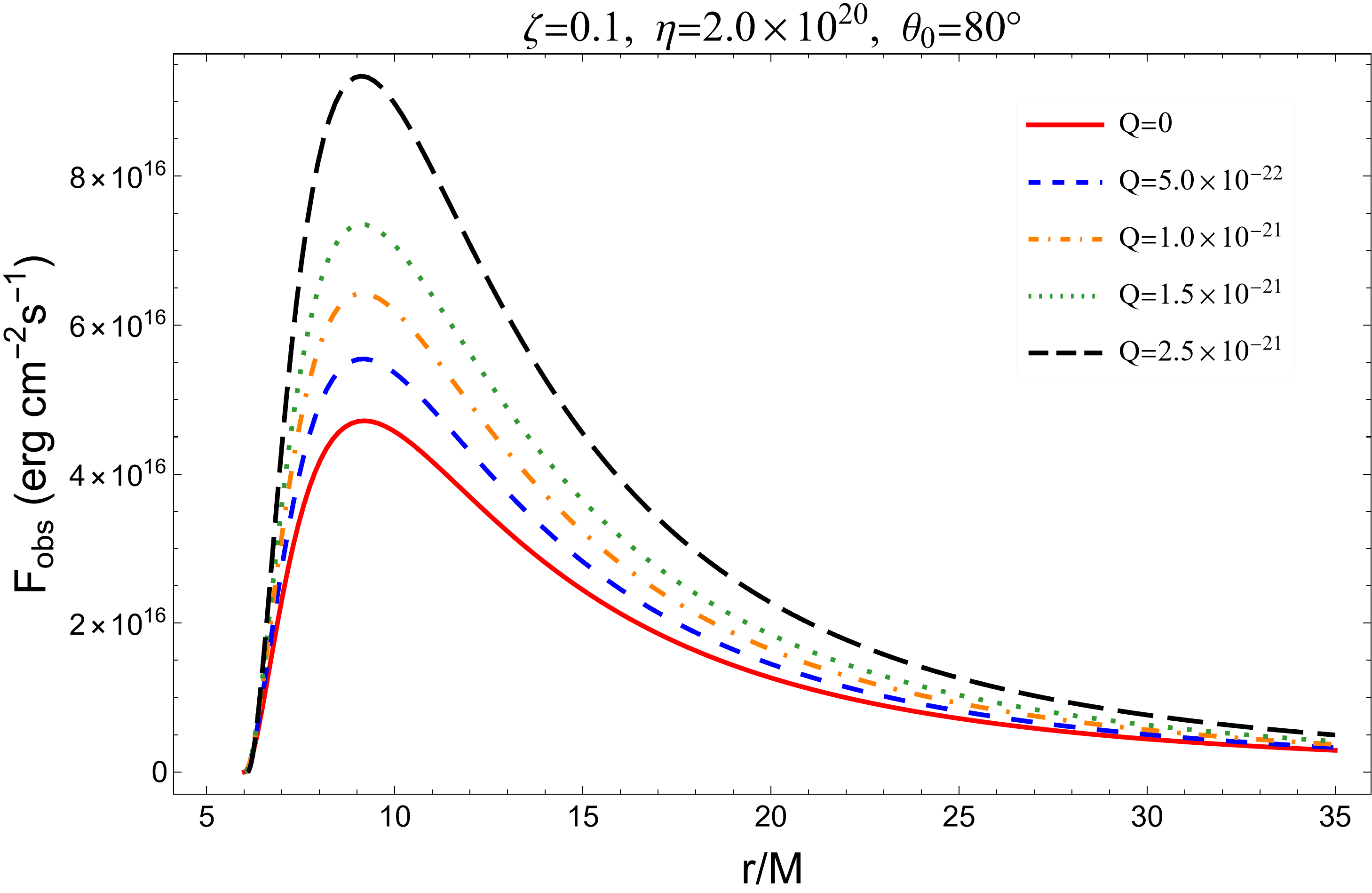}
	\end{subfigure}
	\hfill	
	\begin{subfigure}[b]{0.31\textwidth}
		\centering
		\includegraphics[width=\textwidth]{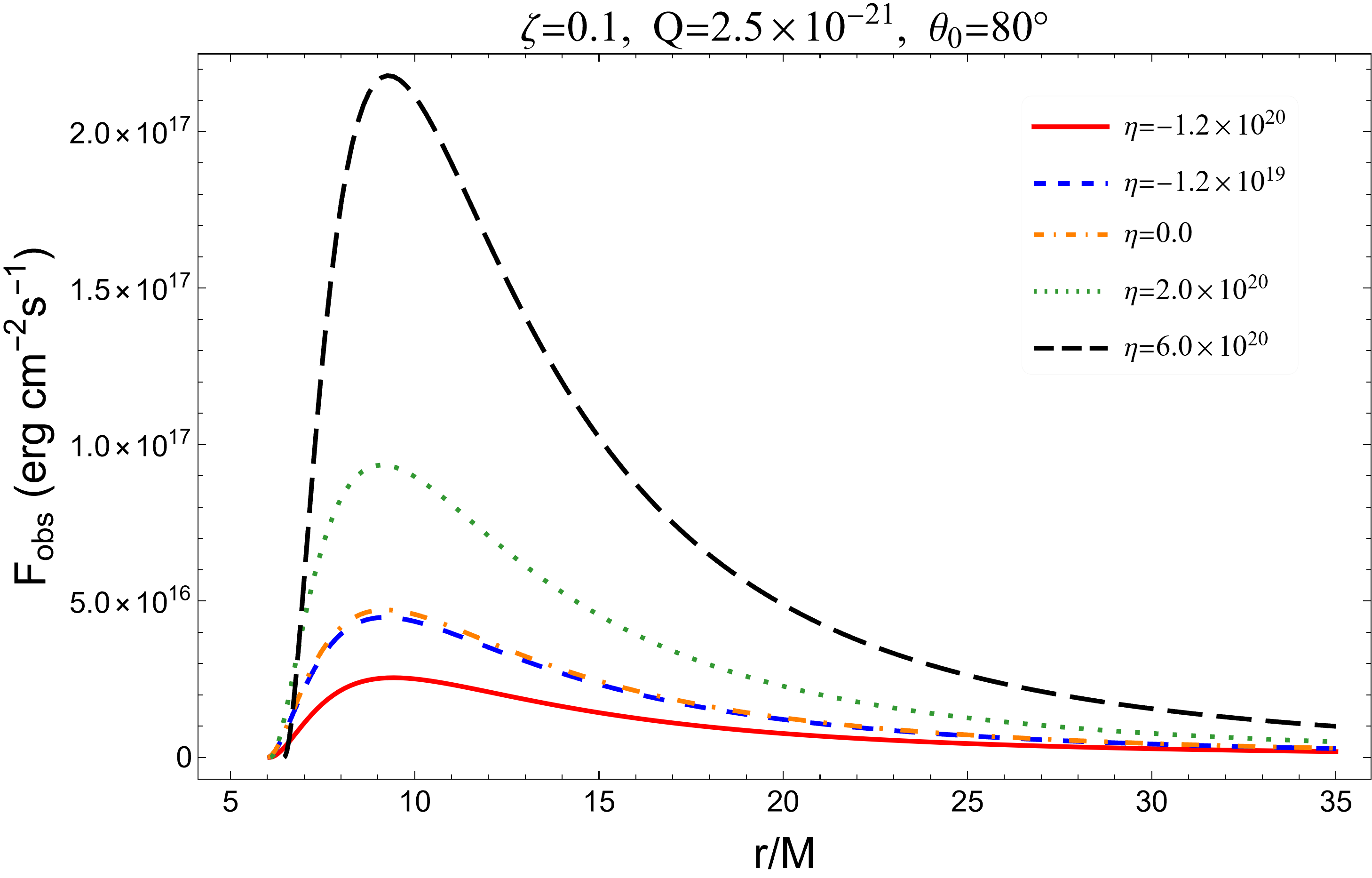}
	\end{subfigure}
	
\begin{subfigure}[b]{0.31\textwidth}
		\centering
		\includegraphics[width=\textwidth]{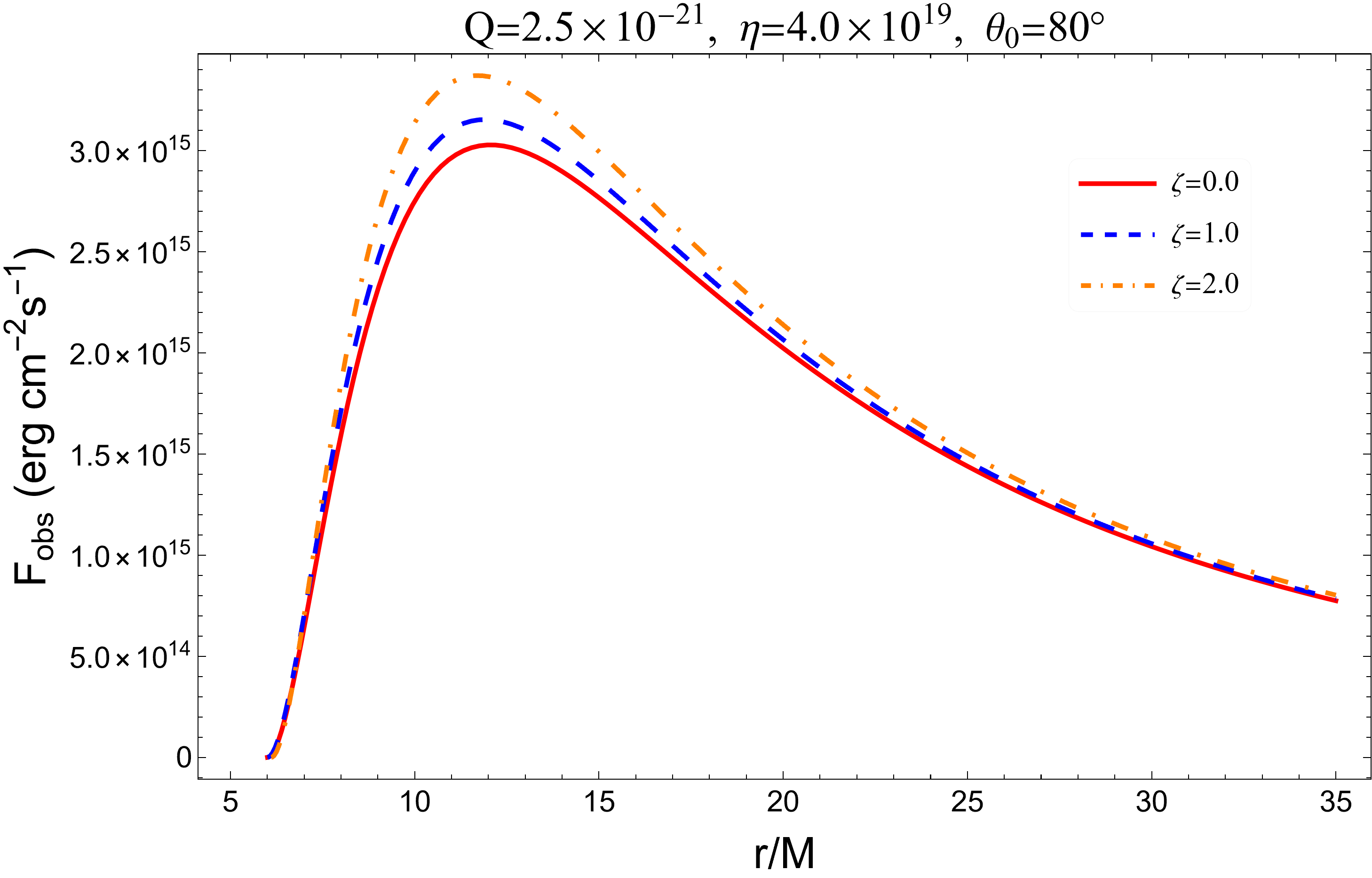}
	\end{subfigure}
	\hfill
	\begin{subfigure}[b]{0.31\textwidth}
		\centering
		\includegraphics[width=\textwidth]{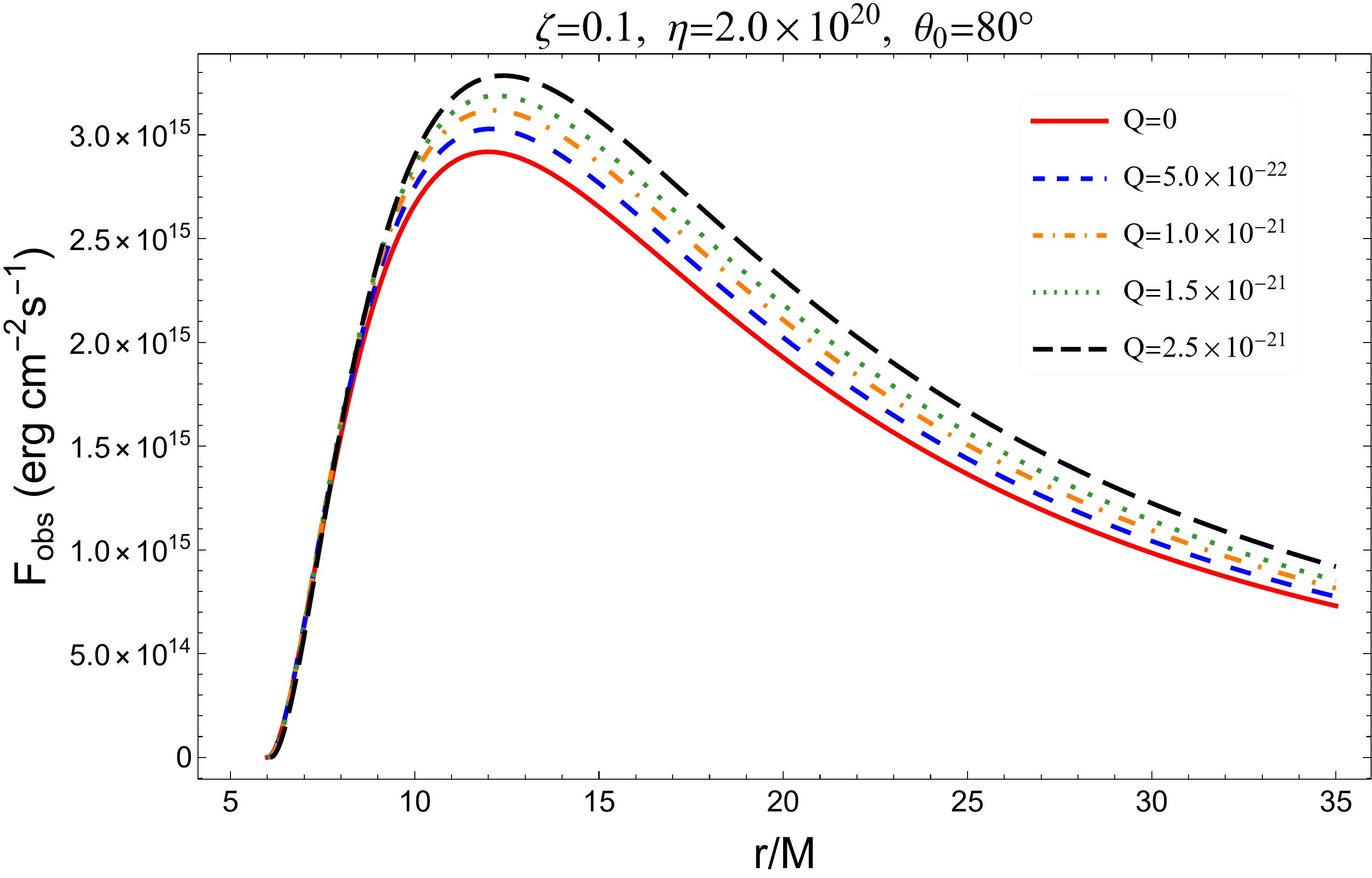}
	\end{subfigure}
	\hfill
	\begin{subfigure}[b]{0.31\textwidth}
		\centering
		\includegraphics[width=\textwidth]{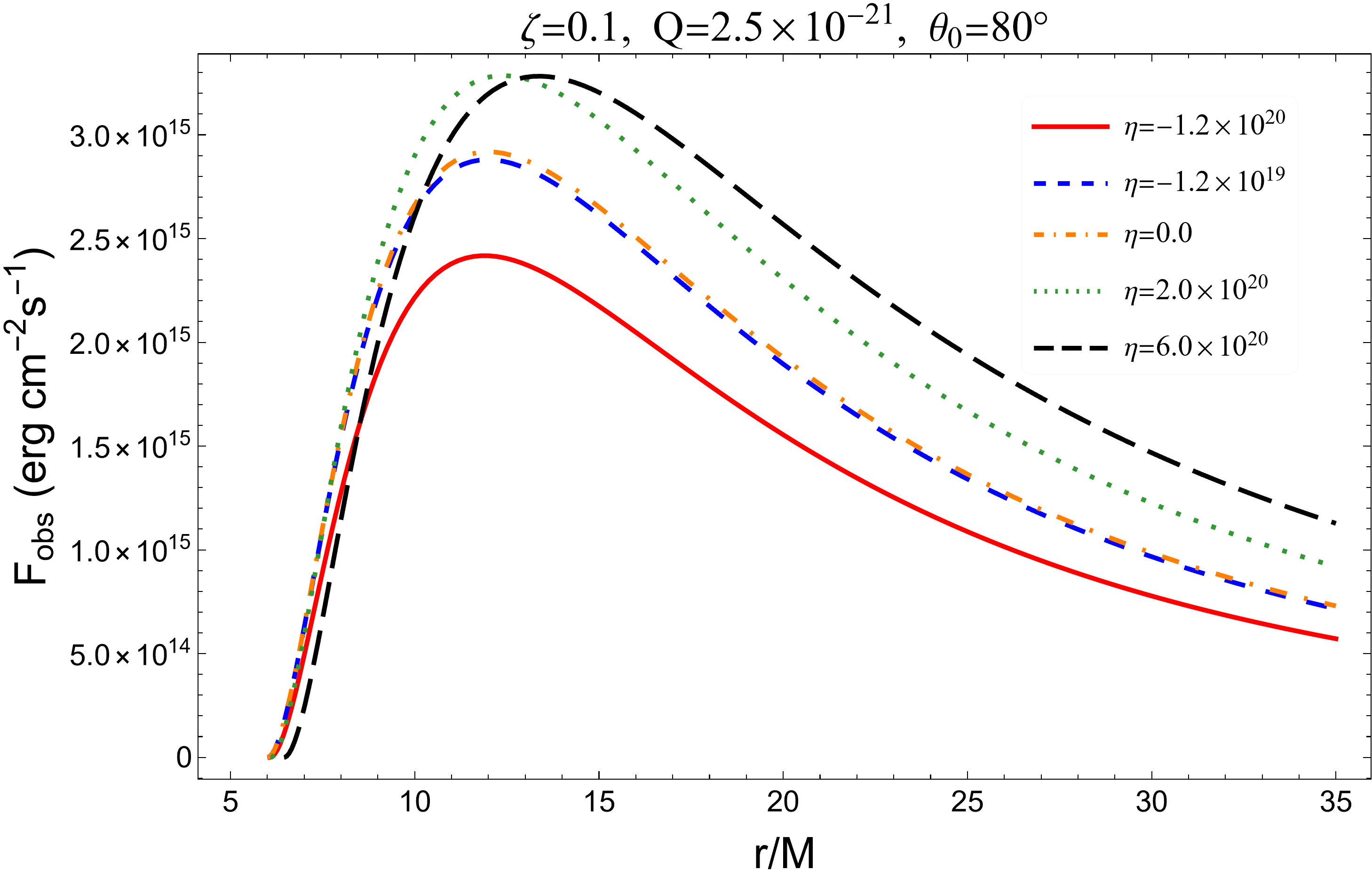}
	\end{subfigure}
	
	\caption{Radial distributions of the observed flux $F_{\mathrm{obs}}$ on the observation plane for various model parameters. 
 $\alpha=3\pi/2$ for the upper row and $\alpha=\pi/2$ for the lower row. }
	\label{fig:sideflux}
\end{figure}

From the various direct images of the observed fluxes, it is easy to see that on the observation plane, the most bright region is around $\alpha=\frac{3\pi}{2}$ corresponding to 
 particles in the disk approaching the observer while the most dark region is around $\alpha=\frac{\pi}{2}$ corresponding to particles in the disk moving away from the observer.
In Fig. \ref{fig:sideflux}, we show the radial distribution of the observed flux along these two special angles and the influence of model parameters on the distribution.
$\alpha=3\pi/2$ for the first row and  $\alpha=\pi/2$ for the second row. 
 From the first column, we see that when $\kappa>0$, the quantum parameter $\zeta$ lowers the observed flux in the bright region while strengthening the observed flux in the dark region. In the previous discussion, it is found that the parameter $\zeta$ increases the local radiation flux of the accretion disk. However, the redshift factor is also affected by the parameter $\zeta$, which leads to the different influence of $\zeta$ on the observed fluxes in the bright and dark regions. 
 From the second and third columns, we can see that when $\kappa>0$, the increase of $\kappa$ leads to significant enhancement of the observed flux in the bright region and to only slight enhancement of the flux in the dark region; when $\kappa<0$, the increase of $|\kappa|$ leads to remarkable reduction of the observed flux in both the bright and dark regions.

\section{Conclusions}

In this study, we investigate the shadow, circular orbital dynamics, radiation properties and the images of the LQRNBH surrounded by a thin accretion disk, and systematically analyze the influence of the loop quantum parameter $\zeta$ and the electromagnetic interaction on the these quantities.   
We focus on the case where the black hole has the same horizon structure as the corresponding classical Reissner-Nordström black hole.   
The black hole shadow of LQRNBH is derived, and constraints on the quantum parameter $\zeta$ and charge $Q$ of the LQRNBH are analyzed with the observational data of M87* and Sgr A*, which provide the parameter space of our subsequent analyses.

The circular motion of charged massive particles around the LQRNBH is considered. The sign and magnitude of the electromagnetic coupling $\kappa$ are both important for the circular motion. 
 The influence of model parameters on the particles' energy $E$, angular momentum $L$, and angular velocity $\Omega$ is analyzed. The influence of $\zeta$ is noticeable only when $\zeta$ is of order 1, and compared with the $\zeta=0$ case, the energy decreases slightly when $\kappa=3$ while increase slightly when $\kappa=-0.6$ in the $\zeta=1$ case.   
And, the angular momentum always decreases for both $\kappa=3$ and $\kappa=-0.6$. When $\kappa>0$, the increase of $\kappa$ significantly decreases the energy $E$ while increases the angular momentum $L$. When $\kappa<0$, the increase of $|\kappa|$ increases the energy $E$ while decreases the angular momentum $L$. The influence of the model parameters on the angular momentum of the particles on circular orbits remains obvious in the far region of the disk, providing a potential avenue for testing quantum gravity effect through measurement of the accretion disk.

The ISCO radius is important for the accretion disk model considered here. It is found that an ISCO exists only when $\kappa>-1$. In other words, the existence of ISCO needs a weak electromagnetic attractive force between the black hole and charged particles in the accretion disk. If we assume the ISCO is not far from the black hole (e.g. $<10M$), then $\kappa\lesssim5$. It is found that the ISCO radius increases as the increase of $|\kappa|$; as the increase of the quantum parameter $\zeta$, the radius of ISCO decreases when $\kappa>0$, while increases when $\kappa<0$. The ISCO energy is closely related to the radiation efficiency $\epsilon$ of the black hole.
The influence of the loop quantum parameter $\zeta$ on the radiation efficiency $\epsilon$ is very small. However, the electromagnetic coupling $\kappa$ has significant impact on $\epsilon$. As the increase of $|\kappa|$, $\epsilon$ shows a significant enhancement when $\kappa>0$ and decreases when $\kappa<0$.

The influence of the quantum parameter $\zeta$ on the local radiation flux of the accretion disk is small and depends on the sign of the electromagnetic coupling $\kappa$.  
As the increase of $\zeta$, the local radiation flux increases for $\kappa>0$ and decreases for $\kappa<0$. 
The influence of $\kappa$ on the local radiation flux is remarkable. 
As the increase of $|\kappa|$, the local radiation flux has significant enhancement for $\kappa>0$ and decreases remarkably for $\kappa<0$. 
 Based on the observational data of M87* and Sgr A*, we also provide two concrete examples to quantitatively illustrate the difference among LQRNBH, RNBH and Schwarzschild black hole in several typical quantities, such as the local radiation flux, temperature and location of their maximum values.

With the ray-tracing method, the direct images of the isoradial curves, redshift factors and observed fluxes are numerically studied. 
The impacts of the model parameters and observation angles on these images are illustrated. The maximum of the redshift factor decreases slightly as the increase of the quantum parameter $\zeta$. However, as the increase of $|\kappa|$, the redshift factor shows a significant increase when $\kappa>0$ and decreases when $\kappa<0$. 
For the observed fluxes, the quantum parameter $\zeta$ reduces its maximum value and increases its minimum value, which makes the distribution of the observed flux more uniform.
In contrast, the electromagnetic coupling $\kappa$ uniformly increases the observed flux for $\kappa>0$ and uniformly decreases the observed flux for $\kappa<0$.

The results of LQRNBH in this work focuses on the asymptotically flat LQRNBH surrounded by a thin disk model. Subsequent research may be extended to consider LQRNBH surrounded by dark matter halo or thick disk models \cite{Chen:2024nua,Li:2025ver,Li:2025knj,Aslam:2025hgl,Wang:2025qpv,Wang:2025gbj,Zeng:2025kyv,Yang:2025whw,Hou:2023bep,Battista:2023iyu,Wang:2025fmz}. 
The polarization images and patterns of the accretion flow around a black hole, especially the near horizon one,  can also be used to probe the black hole geometry \cite{Liang:2025bbn,Chen:2022scf,Hou:2024qqo,Chen:2024jkm,Zhang:2023bzv,Hou:2022eev}, then the study about how the loop quantum parameter affects these signatures and patterns could be future directions.  It is also interesting to consider the LQRNBH with nonzero cosmological constant and explore the influence of model parameters on the dynamics of Einstein ring \cite{Zeng:2025oef}. These further research may yield more insights on how to reveal the quantum gravity effects.

\clearpage
\bibliography{ref1}
	
\end{document}